\documentclass[
reprint,
superscriptaddress,
nofootinbib,
amsmath,amssymb,prd,showpacs
aps,
]{revtex4-2}

\usepackage{graphicx}
\usepackage{dcolumn}
\usepackage{bm}
\usepackage{siunitx}
\usepackage{xspace}
\usepackage{booktabs}
\usepackage{enumerate}
\usepackage[multiple]{footmisc}
\usepackage{hyperref}
\usepackage[svgnames]{xcolor}
\usepackage[capitalize]{cleveref}
\usepackage[caption=false]{subfig}
\makeatletter

\providecommand{\CCzeropi}{CC-$0\pi$\xspace}
\providecommand{\CConepi}{CC-$1\pi$\xspace}
\providecommand{\CConepiplus}{CC-$1\pi^+$\xspace}

\providecommand{\CCother}{CC-Other\xspace}

\providecommand{\minerva}{MINER$\nu$A\xspace}

\begin{document}

\title{First measurement of muon neutrino charged-current interactions on hydrocarbon\\ without pions in the final state using multiple detectors with correlated energy spectra at T2K}

\newcommand{\INSTHD}{\affiliation{University Autonoma Madrid, Department of Theoretical Physics, 28049 Madrid, Spain}}
\newcommand{\INSTEE}{\affiliation{University of Bern, Albert Einstein Center for Fundamental Physics, Laboratory for High Energy Physics (LHEP), Bern, Switzerland}}
\newcommand{\INSTFE}{\affiliation{Boston University, Department of Physics, Boston, Massachusetts, U.S.A.}}
\newcommand{\INSTGA}{\affiliation{University of California, Irvine, Department of Physics and Astronomy, Irvine, California, U.S.A.}}
\newcommand{\INSTI}{\affiliation{IRFU, CEA, Universit\'e Paris-Saclay, F-91191 Gif-sur-Yvette, France}}
\newcommand{\INSTGB}{\affiliation{University of Colorado at Boulder, Department of Physics, Boulder, Colorado, U.S.A.}}
\newcommand{\INSTFG}{\affiliation{Colorado State University, Department of Physics, Fort Collins, Colorado, U.S.A.}}
\newcommand{\INSTFH}{\affiliation{Duke University, Department of Physics, Durham, North Carolina, U.S.A.}}
\newcommand{\INSTJA}{\affiliation{E\"{o}tv\"{o}s Lor\'{a}nd University, Department of Atomic Physics, Budapest, Hungary}}
\newcommand{\INSTEF}{\affiliation{ETH Zurich, Institute for Particle Physics and Astrophysics, Zurich, Switzerland}}
\newcommand{\INSTIE}{\affiliation{CERN European Organization for Nuclear Research, CH-1211 Gen\'eve 23, Switzerland}}
\newcommand{\INSTEG}{\affiliation{University of Geneva, Section de Physique, DPNC, Geneva, Switzerland}}
\newcommand{\INSTHJ}{\affiliation{University of Glasgow, School of Physics and Astronomy, Glasgow, United Kingdom}}
\newcommand{\INSTDG}{\affiliation{H. Niewodniczanski Institute of Nuclear Physics PAN, Cracow, Poland}}
\newcommand{\INSTCB}{\affiliation{High Energy Accelerator Research Organization (KEK), Tsukuba, Ibaraki, Japan}}
\newcommand{\INSTIB}{\affiliation{University of Houston, Department of Physics, Houston, Texas, U.S.A.}}
\newcommand{\INSTED}{\affiliation{Institut de Fisica d'Altes Energies (IFAE) - The Barcelona Institute of Science and Technology, Campus UAB, Bellaterra (Barcelona) Spain}}
\newcommand{\INSTJC}{\affiliation{Institut f\"ur Physik, Johannes Gutenberg-Universit\"at Mainz, Staudingerweg 7, 55128 Mainz, Germany}}
\newcommand{\INSTEC}{\affiliation{IFIC (CSIC \& University of Valencia), Valencia, Spain}}
\newcommand{\INSTHH}{\affiliation{Institute For Interdisciplinary Research in Science and Education (IFIRSE), ICISE, Quy Nhon, Vietnam}}
\newcommand{\INSTEI}{\affiliation{Imperial College London, Department of Physics, London, United Kingdom}}
\newcommand{\INSTGF}{\affiliation{INFN Sezione di Bari and Universit\`a e Politecnico di Bari, Dipartimento Interuniversitario di Fisica, Bari, Italy}}
\newcommand{\INSTBE}{\affiliation{INFN Sezione di Napoli and Universit\`a di Napoli, Dipartimento di Fisica, Napoli, Italy}}
\newcommand{\INSTBF}{\affiliation{INFN Sezione di Padova and Universit\`a di Padova, Dipartimento di Fisica, Padova, Italy}}
\newcommand{\INSTBD}{\affiliation{INFN Sezione di Roma and Universit\`a di Roma ``La Sapienza'', Roma, Italy}}
\newcommand{\INSTEB}{\affiliation{Institute for Nuclear Research of the Russian Academy of Sciences, Moscow, Russia}}
\newcommand{\INSTHI}{\affiliation{International Centre of Physics, Institute of Physics (IOP), Vietnam Academy of Science and Technology (VAST), 10 Dao Tan, Ba Dinh, Hanoi, Vietnam}}
\newcommand{\INSTJD}{\affiliation{ILANCE, CNRS – University of Tokyo International Research Laboratory, Kashiwa, Chiba 277-8582, Japan}}
\newcommand{\INSTHA}{\affiliation{Kavli Institute for the Physics and Mathematics of the Universe (WPI), The University of Tokyo Institutes for Advanced Study, University of Tokyo, Kashiwa, Chiba, Japan}}
\newcommand{\INSTID}{\affiliation{Keio University, Department of Physics, Kanagawa, Japan}}
\newcommand{\INSTIF}{\affiliation{King's College London, Department of Physics, Strand, London WC2R 2LS, United Kingdom}}
\newcommand{\INSTCC}{\affiliation{Kobe University, Kobe, Japan}}
\newcommand{\INSTCD}{\affiliation{Kyoto University, Department of Physics, Kyoto, Japan}}
\newcommand{\INSTEJ}{\affiliation{Lancaster University, Physics Department, Lancaster, United Kingdom}}
\newcommand{\INSTII}{\affiliation{Lawrence Berkeley National Laboratory, Berkeley, CA 94720, USA}}
\newcommand{\INSTBA}{\affiliation{Ecole Polytechnique, IN2P3-CNRS, Laboratoire Leprince-Ringuet, Palaiseau, France}}
\newcommand{\INSTFC}{\affiliation{University of Liverpool, Department of Physics, Liverpool, United Kingdom}}
\newcommand{\INSTFI}{\affiliation{Louisiana State University, Department of Physics and Astronomy, Baton Rouge, Louisiana, U.S.A.}}
\newcommand{\INSTIH}{\affiliation{Joint Institute for Nuclear Research, Dubna, Moscow Region, Russia}}
\newcommand{\INSTHB}{\affiliation{Michigan State University, Department of Physics and Astronomy,  East Lansing, Michigan, U.S.A.}}
\newcommand{\INSTCE}{\affiliation{Miyagi University of Education, Department of Physics, Sendai, Japan}}
\newcommand{\INSTDF}{\affiliation{National Centre for Nuclear Research, Warsaw, Poland}}
\newcommand{\INSTFJ}{\affiliation{State University of New York at Stony Brook, Department of Physics and Astronomy, Stony Brook, New York, U.S.A.}}
\newcommand{\INSTGJ}{\affiliation{Okayama University, Department of Physics, Okayama, Japan}}
\newcommand{\INSTCF}{\affiliation{Osaka Metropolitan University, Department of Physics, Osaka, Japan}}
\newcommand{\INSTGG}{\affiliation{Oxford University, Department of Physics, Oxford, United Kingdom}}
\newcommand{\INSTIC}{\affiliation{University of Pennsylvania, Department of Physics and Astronomy,  Philadelphia, PA, 19104, USA.}}
\newcommand{\INSTGC}{\affiliation{University of Pittsburgh, Department of Physics and Astronomy, Pittsburgh, Pennsylvania, U.S.A.}}
\newcommand{\INSTFA}{\affiliation{Queen Mary University of London, School of Physics and Astronomy, London, United Kingdom}}
\newcommand{\INSTE}{\affiliation{University of Regina, Department of Physics, Regina, Saskatchewan, Canada}}
\newcommand{\INSTGD}{\affiliation{University of Rochester, Department of Physics and Astronomy, Rochester, New York, U.S.A.}}
\newcommand{\INSTHC}{\affiliation{Royal Holloway University of London, Department of Physics, Egham, Surrey, United Kingdom}}
\newcommand{\INSTBC}{\affiliation{RWTH Aachen University, III. Physikalisches Institut, Aachen, Germany}}
\newcommand{\INSTJB}{\affiliation{Departamento de F\'isica At\'omica, Molecular y Nuclear, Universidad de Sevilla, 41080 Sevilla, Spain}}
\newcommand{\INSTFB}{\affiliation{University of Sheffield, Department of Physics and Astronomy, Sheffield, United Kingdom}}
\newcommand{\INSTDI}{\affiliation{University of Silesia, Institute of Physics, Katowice, Poland}}
\newcommand{\INSTBB}{\affiliation{Sorbonne Universit\'e, CNRS/IN2P3, Laboratoire de Physique Nucl\'eaire et de Hautes Energies (LPNHE), Paris, France}}
\newcommand{\INSTEH}{\affiliation{STFC, Rutherford Appleton Laboratory, Harwell Oxford,  and  Daresbury Laboratory, Warrington, United Kingdom}}
\newcommand{\INSTCH}{\affiliation{University of Tokyo, Department of Physics, Tokyo, Japan}}
\newcommand{\INSTBJ}{\affiliation{University of Tokyo, Institute for Cosmic Ray Research, Kamioka Observatory, Kamioka, Japan}}
\newcommand{\INSTCG}{\affiliation{University of Tokyo, Institute for Cosmic Ray Research, Research Center for Cosmic Neutrinos, Kashiwa, Japan}}
\newcommand{\INSTHF}{\affiliation{Tokyo Institute of Technology, Department of Physics, Tokyo, Japan}}
\newcommand{\INSTGI}{\affiliation{Tokyo Metropolitan University, Department of Physics, Tokyo, Japan}}
\newcommand{\INSTHG}{\affiliation{Tokyo University of Science, Faculty of Science and Technology, Department of Physics, Noda, Chiba, Japan}}
\newcommand{\INSTF}{\affiliation{University of Toronto, Department of Physics, Toronto, Ontario, Canada}}
\newcommand{\INSTB}{\affiliation{TRIUMF, Vancouver, British Columbia, Canada}}
\newcommand{\INSTDJ}{\affiliation{University of Warsaw, Faculty of Physics, Warsaw, Poland}}
\newcommand{\INSTDH}{\affiliation{Warsaw University of Technology, Institute of Radioelectronics and Multimedia Technology, Warsaw, Poland}}
\newcommand{\INSTIJ}{\affiliation{Tohoku University, Faculty of Science, Department of Physics, Miyagi, Japan}}
\newcommand{\INSTFD}{\affiliation{University of Warwick, Department of Physics, Coventry, United Kingdom}}
\newcommand{\INSTGH}{\affiliation{University of Winnipeg, Department of Physics, Winnipeg, Manitoba, Canada}}
\newcommand{\INSTEA}{\affiliation{Wroclaw University, Faculty of Physics and Astronomy, Wroclaw, Poland}}
\newcommand{\INSTHE}{\affiliation{Yokohama National University, Department of Physics, Yokohama, Japan}}
\newcommand{\INSTH}{\affiliation{York University, Department of Physics and Astronomy, Toronto, Ontario, Canada}}

\INSTHD
\INSTEE
\INSTFE
\INSTGA
\INSTI
\INSTGB
\INSTFG
\INSTFH
\INSTJA
\INSTEF
\INSTIE
\INSTEG
\INSTHJ
\INSTDG
\INSTCB
\INSTIB
\INSTED
\INSTJC
\INSTEC
\INSTHH
\INSTEI
\INSTGF
\INSTBE
\INSTBF
\INSTBD
\INSTEB
\INSTHI
\INSTJD
\INSTHA
\INSTID
\INSTIF
\INSTCC
\INSTCD
\INSTEJ
\INSTII
\INSTBA
\INSTFC
\INSTFI
\INSTIH
\INSTHB
\INSTCE
\INSTDF
\INSTFJ
\INSTGJ
\INSTCF
\INSTGG
\INSTIC
\INSTGC
\INSTFA
\INSTE
\INSTGD
\INSTHC
\INSTBC
\INSTJB
\INSTFB
\INSTDI
\INSTBB
\INSTEH
\INSTCH
\INSTBJ
\INSTCG
\INSTHF
\INSTGI
\INSTHG
\INSTF
\INSTB
\INSTDJ
\INSTDH
\INSTIJ
\INSTFD
\INSTGH
\INSTEA
\INSTHE
\INSTH

\author{K.\,Abe}\INSTBJ
\author{N.\,Akhlaq}\INSTFA
\author{R.\,Akutsu}\INSTCB
\author{H.\,Alarakia-Charles}\INSTEJ
\author{A.\,Ali}\INSTGH\INSTB
\author{Y.I.\,Alj Hakim}\INSTEI
\author{S.\,Alonso Monsalve}\INSTEF
\author{C.\,Alt}\INSTEF
\author{C.\,Andreopoulos}\INSTFC
\author{M.\,Antonova}\INSTEC
% ...

\collaboration{The T2K Collaboration}\noaffiliation

\begin{abstract}
This paper reports the first measurement of muon neutrino charged-current interactions without pions in the final state using multiple detectors with correlated energy spectra at T2K. The data was collected on hydrocarbon targets using the off-axis T2K near detector (ND280) and the on-axis T2K near detector (INGRID) with neutrino energy spectra peaked at 0.6 GeV and 1.1 GeV respectively. The correlated neutrino flux presents an opportunity to reduce the impact of the flux uncertainty and to study the energy dependence of neutrino interactions. The extracted double-differential cross sections are compared to several Monte Carlo neutrino-nucleus interaction event generators showing the agreement between both detectors individually and with the correlated result.
\end{abstract}

\maketitle
\section{Introduction}

In the last decade, neutrino oscillations experiments have continued to collect more statistics and reduce their systematic uncertainties, moving forward to the precision era of neutrino oscillation physics \cite{PhysRevLett.124.161802, PhysRevD.97.072001, PhysRevLett.123.151803, PhysRevLett.121.241805, cite-key, PhysRevD.98.012002, PhysRevLett.120.071801}. 
For this reason, the next generation of long-baseline (LBL) neutrino oscillations experiments, DUNE (Deep Underground Neutrino Experiment) \cite{dune} and Hyper-Kamiokande \cite{2015}, require systematic errors reduced to a few percent to achieve their physics goals, including precise measurements of the neutrino mass hierarchy and leptonic CP-violation \cite{10.1143/PTP.28.870, pontecorvo1968neutrino}. In order to reach this unprecedented reduction of systematic uncertainties, our knowledge of neutrino--nucleus interaction cross sections must be improved. In the range of energies used in current LBL neutrino oscillation experiments, a precise knowledge of neutrino interactions with nucleons is crucial for the extrapolation from the near to the far detector. Incorrect modeling of neutrino interactions can affect the reconstructed neutrino energy, which can introduce bias in the measurement of neutrino oscillation parameters. This reduction of interaction uncertainty in part will be accomplished through measurements using the planned near detectors for DUNE and Hyper-Kamiokande, but can also be achieved through performing measurements and exploring new techniques with current generation neutrino experiments.

Neutrino charged-current quasielastic (CCQE) interactions, also referred to as one-particle-one-hole (1p1h) excitations, can be written as:
\begin{equation}
    \nu_{\ell} + n \rightarrow \ell^{-} + p,
\end{equation}
where $\nu_\ell$ is the incident neutrino of flavor $\ell$, $n$ and $p$ are the struck neutron and outgoing proton respectively, and $\ell$ is the charged lepton. CCQE interactions are the dominant reaction at the T2K (T\={o}kai-to-Kamioka) neutrino beam energy (peaked at 0.6 GeV). Modelling these interactions with a bound nucleon inside a nucleus is complex, requiring treatment of the Fermi motion, removal energy, and nucleon-nucleon correlations. Various models exist for predicting the initial state nucleon momentum and removal energy, such as Fermi gas models or relativistic mean field models, and for modelling correlations between nucleons, for example the random phase approximation method \cite{SINGH1992587,GIL1997543, PhysRevC.70.055503, PhysRevC.73.025504, PhysRevC.80.065501}. Interactions with correlated pairs of nucleons, referred to as multi-nucleon or two-particle-two-hole (2p2h) excitations, are possible due to meson-exchange currents or short range correlations in the nucleus \cite{PhysRevC.80.065501, DELORME1985263, MARTEAU200076, PhysRevC.81.045502, PhysRevC.83.045501, NIEVES201272, PhysRevC.84.055502}. These multi-nucleon interactions enhance the neutrino cross section in the energy range of T2K and can easily be confused for CCQE interactions, which can then bias the oscillation analysis if not considered. The global picture of neutrino cross-section data is still complicated as many results are in tension with each other, and the available models and Monte Carlo (MC) generators cannot accurately describe many different results across experiments \cite{BETANCOURT20181, PhysRevD.105.092004}. In recent T2K oscillation results \cite{t2knaturepaper}, the dominant systematic uncertainty is from the nucleon removal energy on charged current quasielastic interactions, showcasing the need for further study of neutrino cross sections and cross-section modelling.

\begin{figure}[hbt]
    \centering
    \includegraphics[width=0.40\textwidth,angle=0]{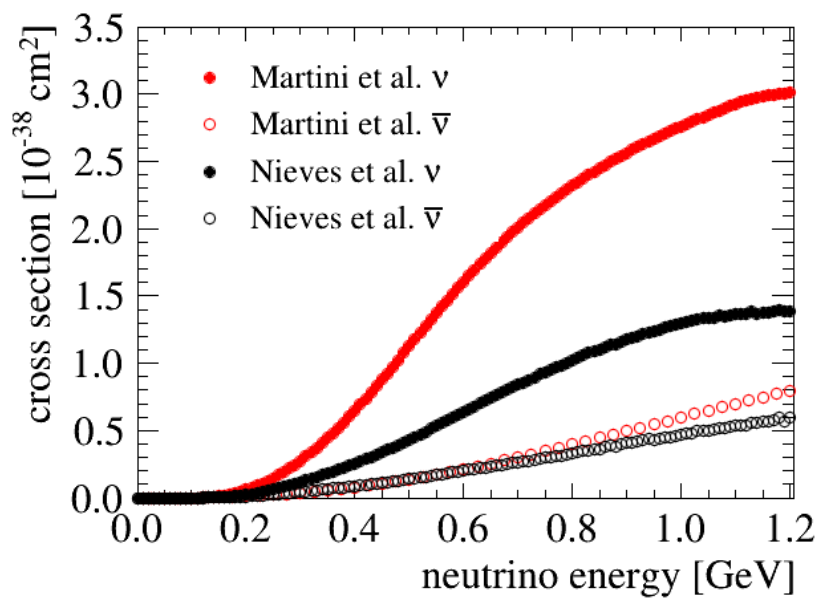}
    \caption{Multi-nucleon cross section on ${}^{12}$C as a function of energy for the Nieves \textit{et al.} and the Martini \textit{et al.} models.}
    \label{fig:multinucleon_xsec_enu}
\end{figure}

The near detectors (close to the neutrino source) used by T2K provide a unique opportunity to perform a combined measurement using the same neutrino beam with two detectors exposed to different but correlated spectra of incident neutrinos, and is the subject of the analysis presented in this paper. 
Neutrino detectors measure the rate of neutrino interactions, which is primarily a product of the neutrino flux and neutrino cross section.
Changes in both the flux and cross-section models can cause the observed event rate to change, often in similar ways, and this degeneracy limits the ability to separate individual effects due to either the flux or cross section. The correlation between the different fluxes at the near detectors provides additional information that can be used to constrain the flux uncertainty and break some of the degeneracy between flux and cross section.
The different neutrino energy spectra seen at each detector also presents an opportunity to study the energy dependence of neutrino interactions within the same analysis framework.
An example is the energy dependence of multi-nucleon interactions, which comprise a non-negligible fraction of the samples used to measure the cross section presented in this paper (on average 10\% across all samples).
The multi-nucleon cross-section prediction as a function of energy for the Nieves \textit{et al.} model \cite{PhysRevC.83.045501}, which is the default multi-nucleon model used in T2K, and the Martini \textit{et al}. model \cite{PhysRevC.80.065501} shows differences mainly related to normalization of by about a factor of two to three across the neutrino energy range used at T2K (shown in Fig. \ref{fig:multinucleon_xsec_enu} from \cite{PhysRevD.96.092006}). This variation between models motivated an additional systematic uncertainty for the T2K oscillation analysis \cite{PhysRevLett.124.161802}.
The analysis presented in this paper takes advantage of the T2K near detector setup to perform the first measurement using multiple detectors with different neutrino energy spectra.

\section{The T2K experiment}
T2K \cite{ABE2011106} is a second-generation long-baseline neutrino oscillation experiment based in Japan, which is able to measure neutrino oscillations with a $\nu_\mu$ ($\bar\nu_\mu$) beam. The neutrino beam is produced at the Japan Proton Accelerator Research Complex (J-PARC) in T\={o}kai. It is first detected 280 m downstream from the source at the near detector complex, where the flavor composition of the incoming neutrino flux is not expected to be affected by oscillations, and then it travels 295 km to the Super-Kamiokande (SK) far detector~\cite{FUKUDA2003418, ABE2014253}, located in Hida, where oscillations significantly affect the flavor composition. The near detector complex houses the two detectors of primary interest for the analysis presented in this paper: a detector on the axis of the neutrino beam, called INGRID~\cite{OTANI2010368} (Interactive Neutrino GRID), and a detector located 2.5 degrees off-axis, called ND280 \cite{KARLEN200691} (Near Detector at 280 meters). INGRID primarily serves as a neutrino beam and flux monitor, measuring the total rate of neutrino interactions and the beam direction. ND280 is dedicated to the study of the un-oscillated spectrum of neutrinos at 280 meters from the production point and neutrino interaction cross-section properties.

The Super-Kamiokande far detector is a deep underground 50 kton water Cherenkov detector. The SK detector, as with ND280, is situated at 2.5 degrees off-axis, meaning that it is exposed to the same relatively narrow energy band neutrino flux, peaked at the oscillation maximum, around 0.6 GeV.

\subsection{Neutrino beam}

\begin{figure}[hbt]
    \centering
    \includegraphics[width=0.40\textwidth,angle=0]{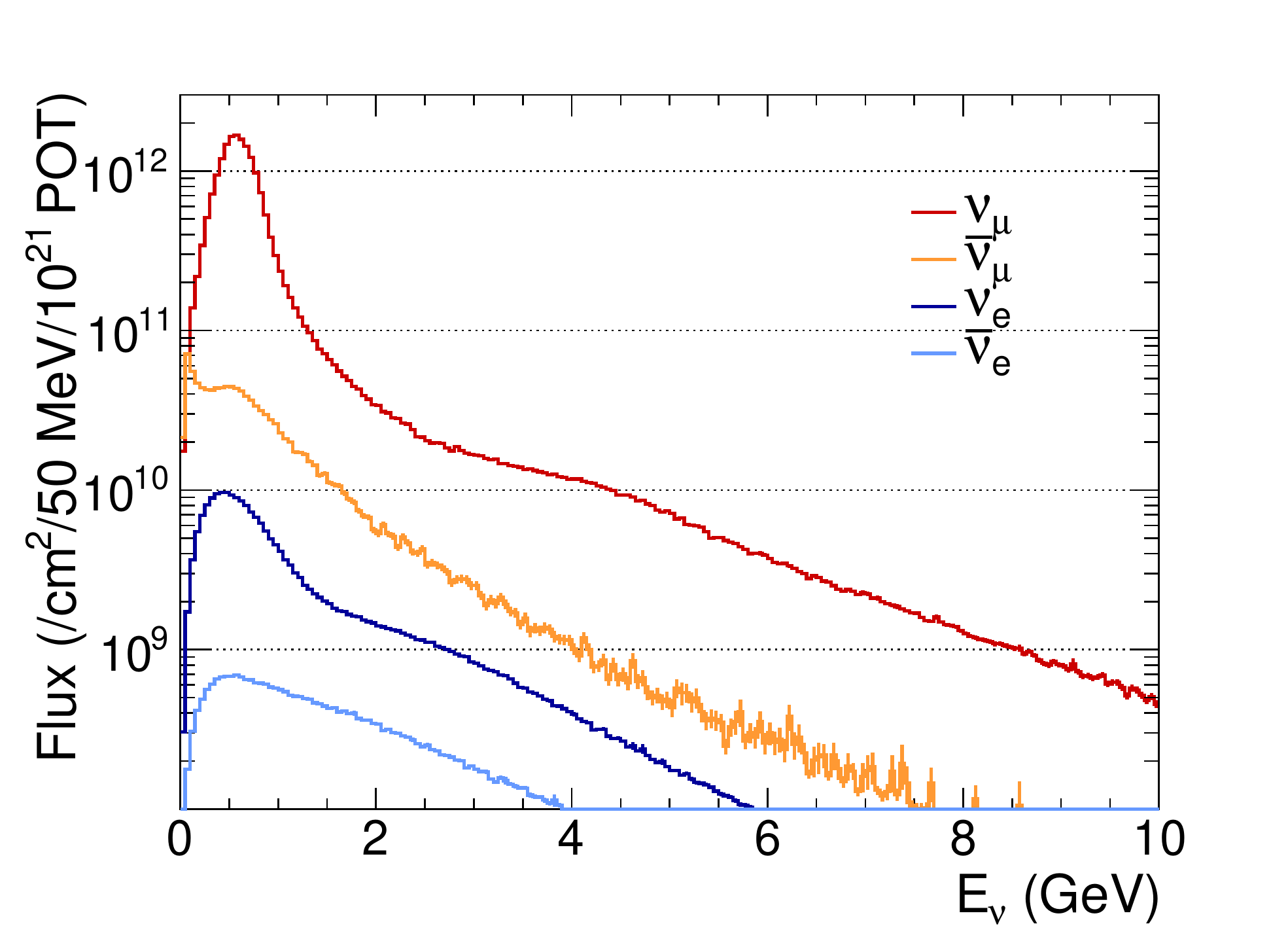}
    \includegraphics[width=0.40\textwidth,angle=0]{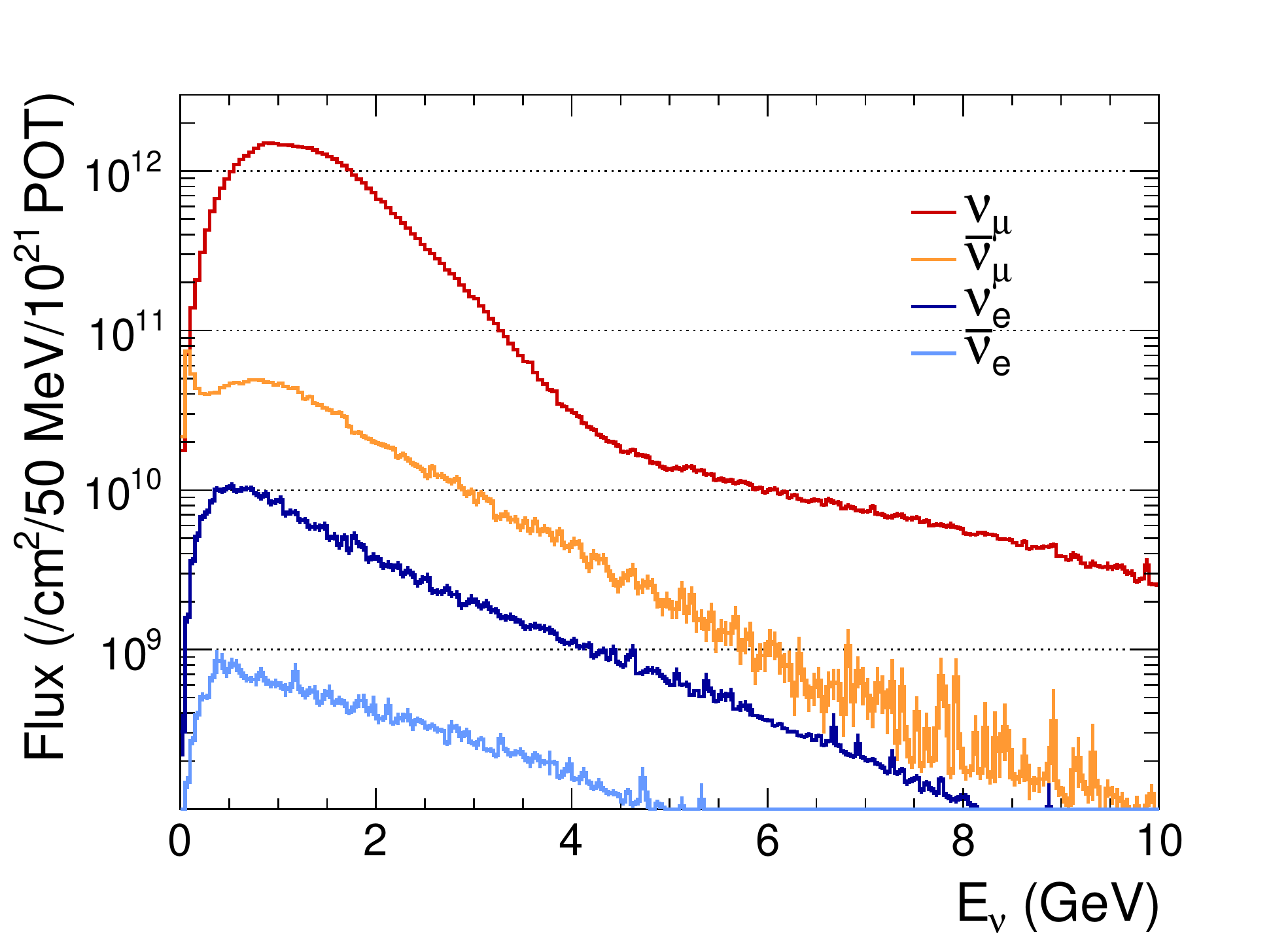}
    \caption{Nominal neutrino mode flux prediction at ND280 (top) and INGRID (bottom) separated by neutrino flavor.}
    \label{fig:nominal_flux}
\end{figure}

T2K neutrinos come from in-flight decays of focused hadrons emitted from an extended, 91.4 cm long, monolithic graphite target. The target is bombarded with a 30 GeV proton beam produced at J-PARC. Interactions of beam protons inside the target initiate a chain of hadronic interactions, the charged products of which are focused upon exit from the target using a series of three magnetic horns. The polarity of the horn current determines whether a $\nu_{\mu}$ (neutrino mode) or $\bar{\nu}_{\mu}$ (anti-neutrino mode) enhanced beam is produced, by focusing predominantly positively or negatively charged pions and kaons, respectively. These mesons are then left to decay, e.g. via $\pi^{\pm}\rightarrow \mu^{\pm}+\nu_{\mu}(\bar{\nu}_{\mu})$, in a 96 m long decay volume, capped with a concrete beam dump at the downstream end. Behind the beam dump, a muon monitor \cite{MATSUOKA2010385, MATSUOKA2010591} is used to measure the secondary beam stability. 
INGRID and ND280 are exposed to the same neutrino beam, but are placed at different angles relative to the beam center which gives a different integrated flux and energy spectrum for each detector. The neutrino flux peaks around 0.6 GeV at ND280 and around 1.1 GeV at INGRID, and the nominal $\nu_\mu$ fluxes are shown in Fig. \ref{fig:nominal_flux}. The beam composition at INGRID and ND280 when running in neutrino mode is shown in Tab. \ref{tab:beam_comp}.
\begin{table}[htb]
    \centering
    \begin{tabular}{c|c|c|c}
        & $\nu_\mu$ & $\bar\nu_\mu$ & $\nu_e$ + $\bar\nu_e$\\
        \hline
        INGRID & 95.3\%    & 3.9\%         & 0.8\%\\
        \hline
        ND280  & 92.9\%    & 5.9\%         & 1.2\%\\    
    \end{tabular}
    \caption{Neutrino beam composition at INGRID and ND280.}
    \label{tab:beam_comp}
\end{table}

\subsection{INGRID}

The INGRID detector is an on-axis neutrino detector located 280 m downstream of the proton target. It consists of 14 identical detector modules (referred to as standard modules) and an extra module called the Proton Module (PM).

\begin{figure}[hbt]
    \centering
    \includegraphics[width=0.3\textwidth,angle=0]{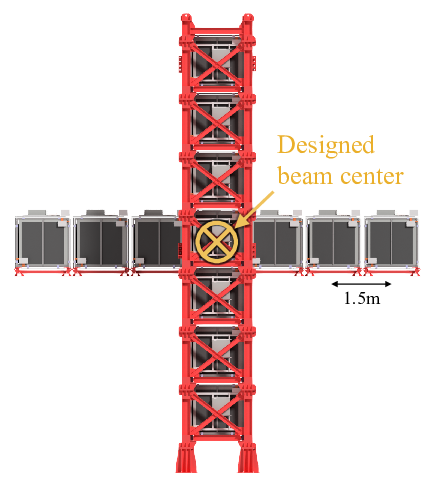}
    \caption{Overview of the 14 standard modules and cross configuration.}
    \label{fig:ingrid_standard_modules}
\end{figure}

\begin{figure}[hbt]
    \centering
    \includegraphics[width=0.4\textwidth,angle=0]{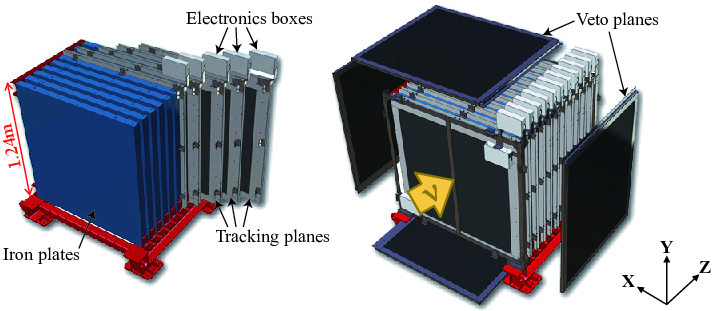}
    \caption{An exploded view of a standard module.}
    \label{fig:ingrid_standard_module}
\end{figure}

The main purpose of the standard modules is to monitor the neutrino beam direction.
The 14 identical standard modules are arranged in two identical groups along the horizontal and vertical axes, as shown in Fig. \ref{fig:ingrid_standard_modules}.
Each of the modules consists of nine iron target plates and eleven tracking scintillator planes surrounded by veto scintillator planes to reject charged particles coming from outside the modules~\cite{NIMA.694.211}, as shown in Fig. \ref{fig:ingrid_standard_module}.


\begin{figure}[hbt]
    \centering
    \includegraphics[width=0.3\textwidth,angle=0]{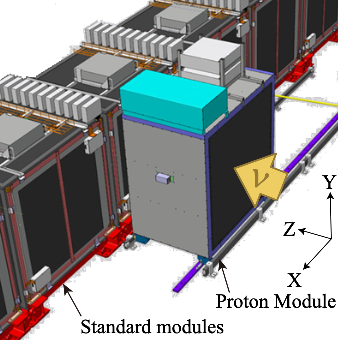}
    \caption{A schematic view of the Proton Module and the standard modules.}
    \label{fig:proton_mod_location}
\end{figure}

\begin{figure}[hbt]
    \centering
    \includegraphics[width=0.3\textwidth,angle=0]{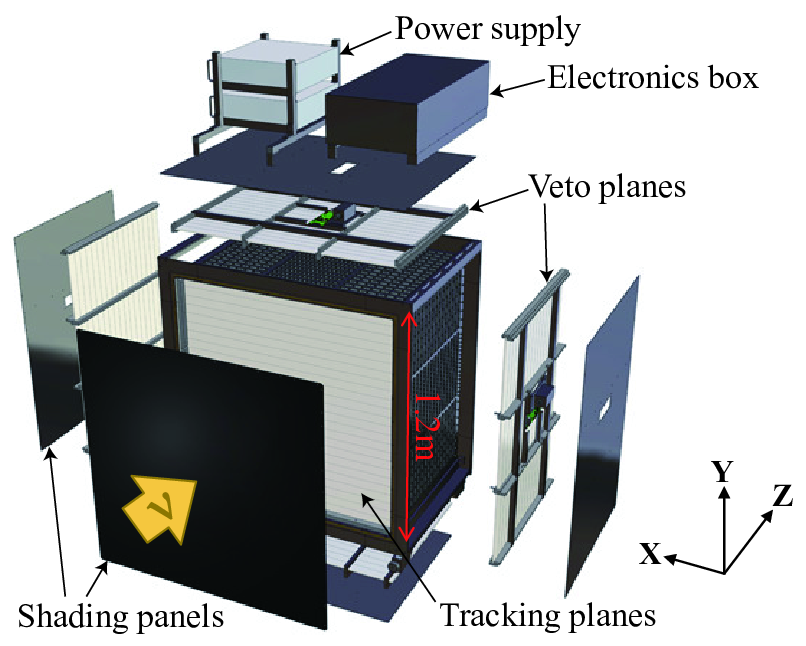}
    \caption{An exploded view of the Proton Module.}
    \label{fig:proton_mod}
\end{figure}

By contrast, the Proton Module was specifically developed for neutrino cross-section measurements.
It is located at the beam center between the horizontal and vertical standard modules as shown in Fig \ref{fig:proton_mod_location}. 
It is a fully-active tracking detector consisting of 36 tracking layers surrounded by veto planes (shown in Fig. \ref{fig:proton_mod}), where each tracking layer is an array of two types of scintillator bars~\cite{PRD.90.052010}. Each scintillator plane covers an area of $120\times120 \; \text{cm}^{2}$ transverse to the beam direction.
The tracking layers also serve as the neutrino interaction target, with the total target mass of the scintillator and fibers in the fiducial volume being 292.1 kg.

\subsection{ND280}

\begin{figure}[hbt]
    \centering
    \includegraphics[width=0.45\textwidth,angle=0]{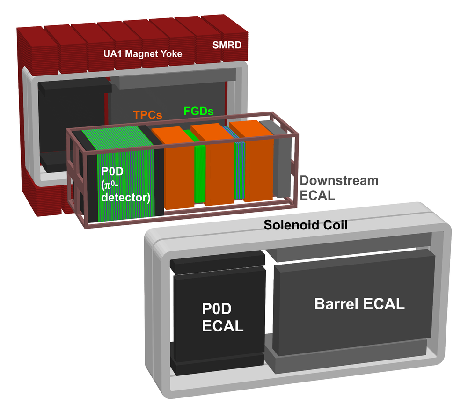}
    \caption{An exploded view of the ND280 off-axis detector.}
    \label{fig:nd280}
\end{figure}

The off-axis near detector ND280 (Fig. \ref{fig:nd280}), is a magnetized particle tracking device.
It consists of a number of sub-detectors installed inside the refurbished UA1/NOMAD magnet, which provides a 0.2 T field used to measure the charge and momentum of particles passing through ND280. Inside the UA1 magnet, the neutrino beam first passes through the  $\pi^0$ detector (P0D)~\cite{ASSYLBEKOV201248} and then the inner tracker, both of which are surrounded by an electromagnetic calorimeter (ECal)~\cite{Allan_2013}. Moreover the UA1 magnet yoke is instrumented with plastic scintillator to perform as a muon range detector (SMRD)~\cite{AOKI2013135} in order to track high angle muons and “sand muons” coming from neutrino interactions in the rock upstream of the detector. The tracker region of ND280 consists of three Time Projection Chambers (TPC1, 2, 3)~\cite{ABGRALL201125}, interleaved with two Fine-Grained Detectors (FGD1, 2)~\cite{AMAUDRUZ20121}. The upstream FGD1 detector is made of fifteen XY planes of polystyrene scintillator with each plane having 2 $\times$ 192 bars, while the downstream FGD2 contains seven polystyrene scintillator modules interleaved with six modules of water in between. The FGDs provide 1.1 tons target mass each for neutrino interactions and tracking of the charged particles coming from the interaction vertex, while the TPCs provide 3D tracking and determine the momentum and energy loss of each charged particle traversing them. The observed energy loss in the TPCs, combined with the measurement of the momentum, is used for particle identification (PID). The analysis presented here is focused on neutrino interactions on carbon, including only events occurring in FGD1.

\section{Event simulation and selection}
\label{sec:strategy}
The goal of this analysis was to perform a simultaneous fit to ND280 and INGRID data, extracting the muon neutrino flux-integrated differential cross section on hydrocarbon without pions in the final state as a function of the outgoing muon kinematics for both the off- and on-axis T2K flux. Signal events are defined by a neutrino interaction with an outgoing muon, zero pions, and any number of other hadrons in the final state and are referred to as \CCzeropi events (or topology). This signal definition is chosen because it is the most common interaction for the T2K oscillation analysis and to match what is accessible to the detectors: the outgoing final-state particles that exit the nucleus. Particles produced in the neutrino interaction can re-interact as they leave the nucleus, potentially producing new particles or being absorbed, referred to as final-state interactions (FSI). Defining the signal in terms of the final-state particles reduces the model dependence of attempting to correct for FSI effects. Similarly, the cross section is measured as a function of the outgoing muon kinematics as opposed to using the reconstructed neutrino energy or momentum transfer to avoid as much model dependence as possible.

\subsection{Event simulation}
\label{subsec:sim}

The T2K neutrino flux simulation \cite{Abe:2012av} is based on the modeling of proton interactions with the graphite target and propagating the produced particles through the target station, allowing for further interactions. Interactions within the target are simulated using the \textsc{fluka 2011} package \cite{Ferrari:2005zk, BOHLEN2014211} while out-of-target interactions and decays are handled by the \textsc{geant3} \cite{Brun:1994aa} and \textsc{gcalor} \cite{Zeitnitz:1992vw} packages. Hadronic interactions and multiplicities are tuned using NA61/SHINE thin-target data \cite{Abgrall:2011ae, Abgrall:2011ts, Abgrall:2016fs} and data from other experiments \cite{allaby, eichten, chemakin}. The proton beam conditions, horn current, and neutrino beam position are monitored and used as inputs to the flux simulation to provide additional constraints. Combined, this data-driven procedure gives an overall flux normalization prior uncertainty of about $8.5\%$ at ND280 and $9.9\%$ at INGRID for this analysis, which is dominated by hadron production and interaction uncertainties. The ND280 and INGRID flux predictions are produced simultaneously using the same input parameters, and this results in correlated uncertainties that are included in this analysis (and described further in Sec. \ref{subsec:uncertainties}).

Neutrino interactions in the detectors and the outgoing kinematics of the produced final-state particles are simulated using the NEUT neutrino event generator \cite{HAYATO2002171, Hayato:2009zz}. NEUT describes charged-current quasi-elastic (CCQE) neutrino-nucleon interactions using the spectral function (SF) approach from \cite{BENHAR1994493} with the quasi-elastic axial mass ($M^{QE}_{A}$) set to 1.21 $\mathrm{GeV}/c^2$ based on the K2K CCQE cross-section measurement in \cite{PhysRevD.74.052002}. Multi-nucleon correlations (also referred to as 2p2h interactions) are based on the model from Nieves \textit{et. al.} \cite{PhysRevC.83.045501}. Resonant pion production (RES) is described by the Rein--Sehgal model \cite{REIN198179} using updated nucleon form factors \cite{PhysRevD.77.053001} and the resonant axial mass ($M^{RES}_{A}$) set to 0.95 $\mathrm{GeV/}c^2$. Coherent pion production uses the updated Berger--Sehgal model \cite{PhysRevD.76.113004}. Deep inelastic scattering (DIS) interactions are modeled using the GRV98 parton distribution functions \cite{Gluck:1998xa} with corrections from Bodek and Yang \cite{doi:10.1063/1.1594324} to extend the validity of the treatment to lower four-momentum transfer ($Q^2 \lesssim 1.5 \, \mathrm{GeV}^2$). NEUT begins modeling DIS processes for interactions with hadronic invariant mass $W > 1.3 \, \mathrm{GeV}/c^2$. For interactions with $1.3 < W < 2.0 \, \mathrm{GeV}/c^2$ a custom hadronization model \cite{Aliaga:2020rqb} is used to interpolate between RES and DIS processes, while for $W > 2.0 \, \mathrm{GeV}/c^2$, \textsc{pythia/jetset} \cite{Sjostrand:1993yb} is used for the hadronization model. Hadrons produced in the primary neutrino--nucleon interaction must propagate through the nuclear remnant before they can be detected. Interactions before the hadrons leave are referred to as final-state interactions (FSI), and are simulated using a semi-classical intra-nuclear cascade model \cite{PhysRevC.6.631, OSET198513}. The MC productions for each detector use the same physics models in NEUT, but are based on slightly different versions, with ND280 using version 5.3.2 and INGRID using version 5.3.3, however this has a negligible impact on the analysis.

The propagation of the final-state particles through the detector medium after exiting the nucleus is performed using a \textsc{geant4} \cite{AGOSTINELLI2003250} simulation. Both detector simulations use \textsc{qgsp\_bert} for the hadronic physics list \cite{Allison:2016lfl}. The detector readout simulation is handled by a custom electronics simulation separately for ND280 and INGRID \cite{ABE2011106}.

\subsection{Data samples}

This analysis uses neutrino-mode data collected between 2010 and 2017 during T2K Runs 2 through 8. The ND280 sample corresponds to a total of $11.53 \times 10^{20}$ POT (protons on target), while the INGRID sample corresponds to a total of $6.04 \times 10^{20}$ POT. The breakdown of collected data by run period is listed in Tab. \ref{tab:pot_table}. The INGRID detector configuration was changed after Run 4 where the Proton Module was moved to a different location in the detector hall, which limits the usable data for this analysis and is the main reason for the difference in total POT between the ND280 and INGRID samples. T2K Run 3b used a lower horn current (205 kA instead of 250 kA) during data-taking, and is included in the ND280 data set but excluded from the INGRID data set. This is what was done in previous ND280 \cite{PhysRevD.93.112012,PhysRevD.98.032003,PhysRevD.101.112001} and INGRID analyses \cite{PhysRevD.91.112002}, and kept the same for this analysis for consistency and could be revisited for future versions.

\begin{table}[htb]
    \centering
    \begin{tabular}{l|r|r|l}
         T2K Run & ND280 & INGRID & Date Range \\
         \hline
         Run 2  & 0.792 & 1.115 & Nov. 2010 -- Mar. 2011 \\
         Run 3b & 0.217 & ----- & Mar. 2012 -- Mar. 2012 \\
         Run 3c & 1.364 & 1.373 & Apr. 2012 -- Jun. 2012 \\
         Run 4  & 3.426 & 3.551 & Oct. 2012 -- May  2013 \\
         Run 8  & 5.730 & ----- & Oct. 2016 -- Apr. 2017 \\
         \hline
         Total  &11.529 & 6.039 & \\
    \end{tabular}
    \caption{Recorded POT in units of $10^{20}$ after accounting for detector up-time separated by run period for ND280 and INGRID that are included in this analysis. The proton module was moved from its on-axis position before Run 8.}
    \label{tab:pot_table}
\end{table}

\subsection{Signal selection}

The signal selection for this analysis is designed to select muon neutrino events with no detected pions in the final state and any number of visible protons, referred to as the \CCzeropi topology. The target material is the plastic scintillator in either FGD1 (for ND280) or the Proton Module (for INGRID). The individual selections for ND280 and INGRID were developed for previous analyses, described in Refs. \cite{PhysRevD.93.112012,PhysRevD.98.032003,PhysRevD.101.112001} and Ref. \cite{PhysRevD.91.112002} for ND280 and INGRID respectively, and minor updates necessary for the joint fit and the addition of new data were made for the analysis presented in this paper.

\subsubsection{ND280}

The ND280 selection first requires passing a set of data quality cuts, and then requires the interaction vertex to be within the FGD1 fiducial volume (FV). The FV is defined to include events with a vertex at least five scintillator bars from the edge in the X and Y directions, and excludes the first XY module as an upstream veto. Events with a single negatively charged muon candidate and any number of proton candidates sharing a common vertex are identified and classified into different samples based on the detectors (FGD1 or TPCs) used to measure the momentum of the muon candidate and the proton candidate(s), if any. This sample separation by detector and particle content allows for a more precise treatment of the detector systematics due to the different detector responses. Tracks are identified by their energy deposition and curvature compared to the expected distributions for each particle hypothesis. The momentum of each reconstructed track is measured either by curvature in the TPCs or by range in FGD1 (and ECAL). Events with a detected associated decay electron in FGD1 are treated as background as these events are likely to have produced an untracked stopped pion decaying into a muon followed by a Michel electron from the muon decay. The signal events are classified into the following samples:

\begin{itemize}
    \item Sample I ($\mu$TPC): defined by a single muon candidate in the TPCs and no other tracks;
    \item Sample II ($\mu$TPC+pTPC): a muon candidate in the TPCs with one or more proton candidates in the TPC;
    \item Sample III ($\mu$TPC+pFGD): a muon candidate in the TPCs and a proton candidate in FGD1;
    \item Sample IV ($\mu$FGD+pTPC): a muon candidate in FGD1 (possibly reaching the ECAL) and a proton candidate in the TPC;
    \item Sample V ($\mu$FGD): a muon candidate in FGD1 (possibly reaching the ECAL) and no other tracks.
\end{itemize}

\begin{figure*}[hbt]
    \centering
    \subfloat{\includegraphics[width=0.40\textwidth,angle=0]{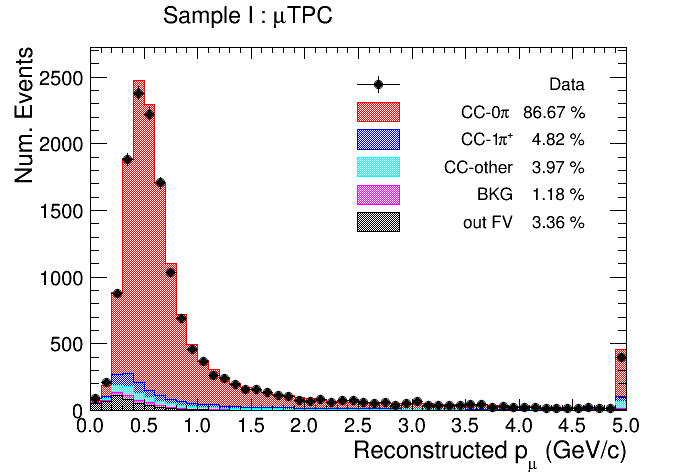}}
    \subfloat{\includegraphics[width=0.40\textwidth,angle=0]{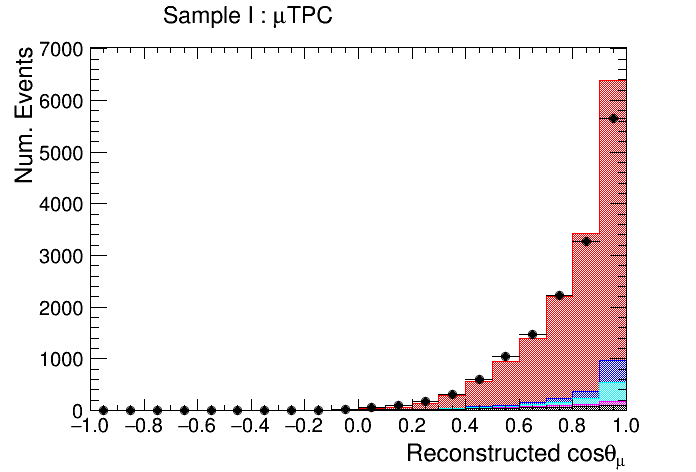}}
    \hfill
    \subfloat{\includegraphics[width=0.40\textwidth,angle=0]{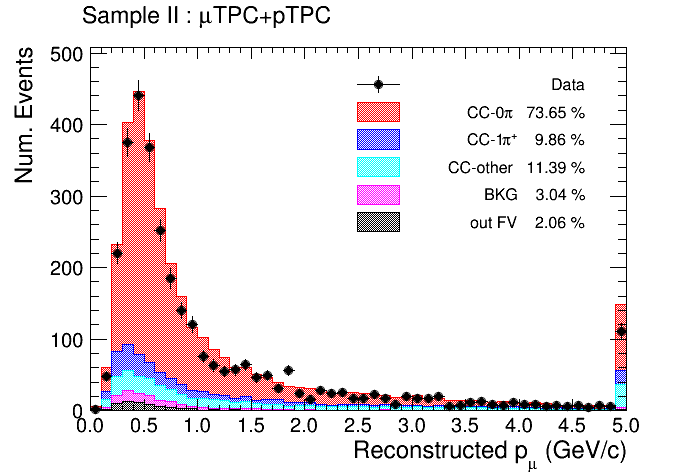}}
    \subfloat{\includegraphics[width=0.40\textwidth,angle=0]{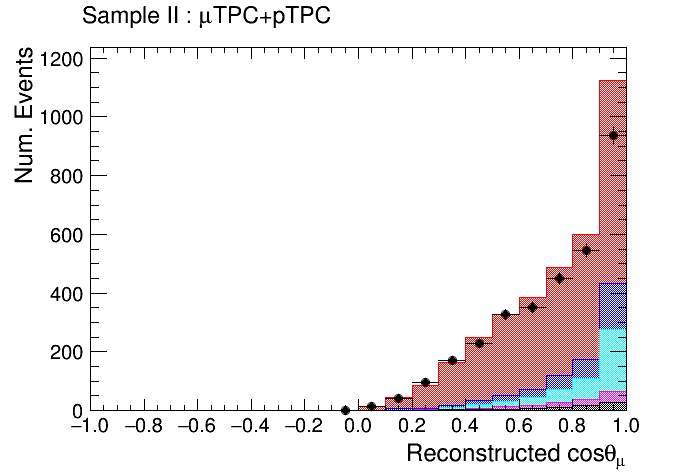}}
    \hfill
    \subfloat{\includegraphics[width=0.40\textwidth,angle=0]{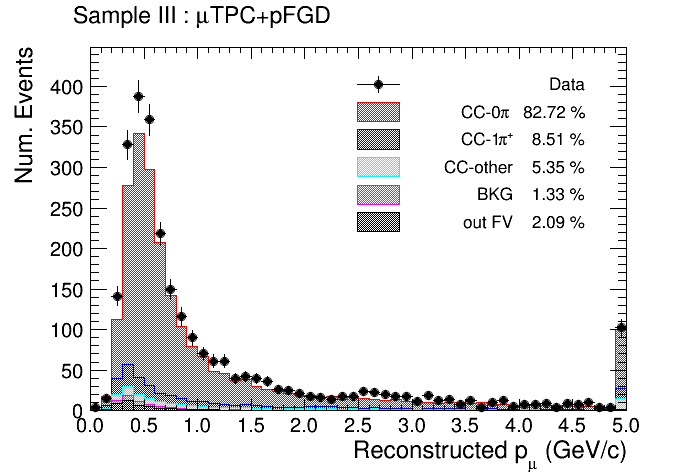}}
    \subfloat{\includegraphics[width=0.40\textwidth,angle=0]{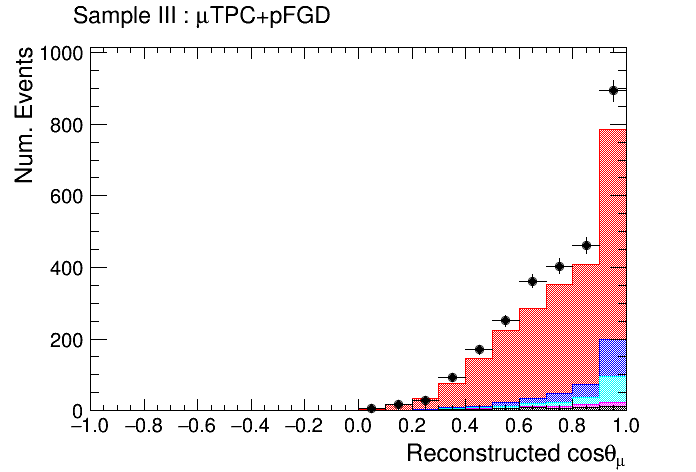}}
    \caption{Event distribution for measured data and MC prediction in reconstructed muon momentum and angle for the ND280 signal samples stacked by true topology. The purity of each topology is listed in the legend, and the last bin for muon momentum contains all events with momentum greater than 5 GeV/c.}
    \label{fig:nd280_signal_samples_tpc}
\end{figure*}

\begin{figure*}[hbt]
    \centering
    \subfloat{\includegraphics[width=0.40\textwidth,angle=0]{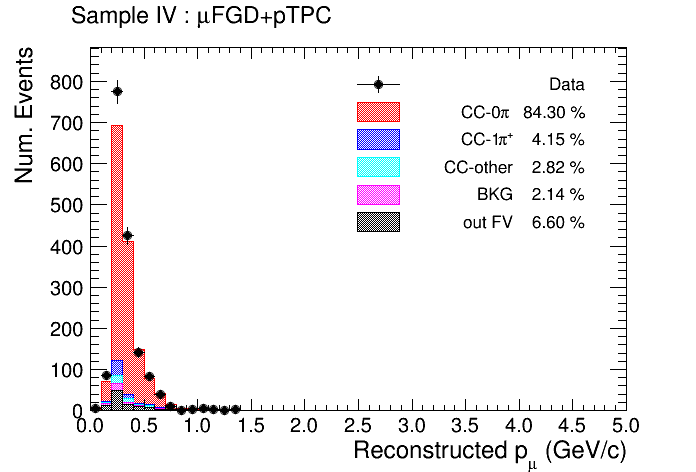}}
    \subfloat{\includegraphics[width=0.40\textwidth,angle=0]{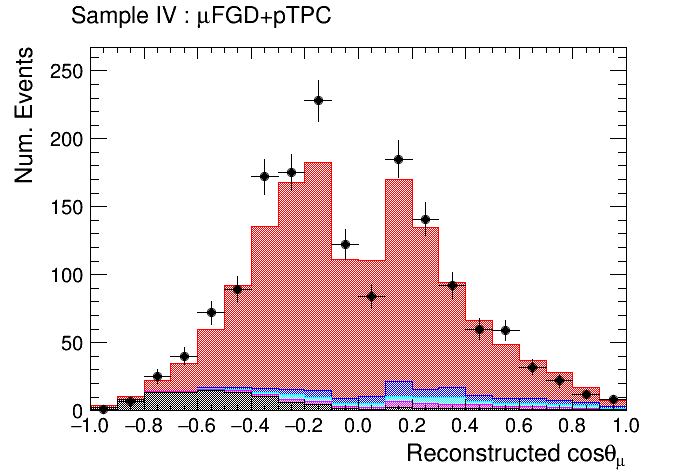}}
    \hfill
    \subfloat{\includegraphics[width=0.40\textwidth,angle=0]{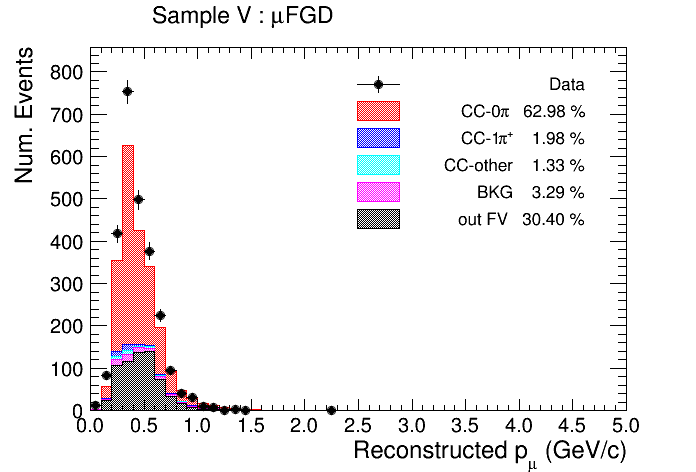}}
    \subfloat{\includegraphics[width=0.40\textwidth,angle=0]{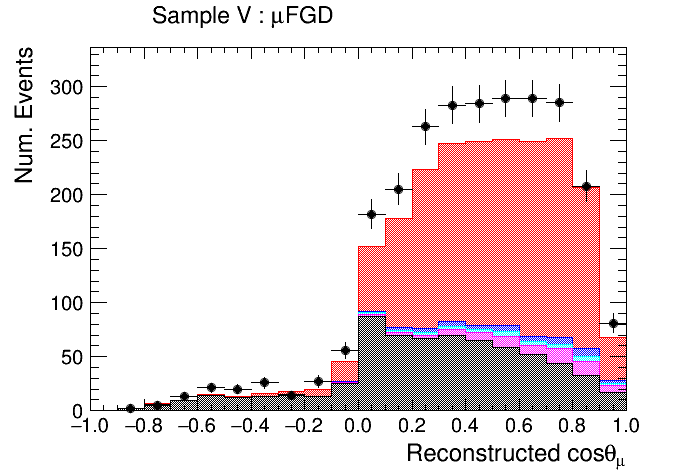}}
    \caption{Event distribution for measured data and MC prediction in reconstructed muon momentum and angle for the ND280 signal samples stacked by true topology. The purity of each topology is listed in the legend.}
    \label{fig:nd280_signal_samples_fgd}
\end{figure*}

The majority of events in the signal sample ($\sim 62\%$) are events with a single reconstructed muon and no other tracks, with most muons reaching the TPCs. The kinematic distributions of each sample separated by true topology are shown in Figs. \ref{fig:nd280_signal_samples_tpc} and \ref{fig:nd280_signal_samples_fgd} with data plus statistical errors overlaid. Most samples achieve a high purity of \CCzeropi events (approximately 82\% pure when integrated over all signal samples) with the main background coming from misidentified or unidentified pions from \CConepiplus (events with a muon and single positive pion track) or \CCother (events with a muon and multiple pion tracks) events. The normalization of the data and nominal MC is very similar when integrated across all the samples, but varies within 15\% per sample. A noticeable feature is the slight deficit of data events compared to the nominal MC at very forward angles for the combined $\mu$TPC sample (representing $\sim 85\%$ of the total event sample). The selected events are binned using the reconstructed muon momentum and cosine of the angle with respect to the beam direction (list of bin edges available in Appendix \ref{app:binning}). The binning scheme is designed such that bins are not finer than the detector resolution.

The cross section is extracted by adding the contributions from each sample, but the samples are kept separate in the analysis. This is important because events with and without protons and which subdetectors were used in the reconstruction are affected by different systematics and backgrounds.

\subsubsection{INGRID}

The INGRID selection first requires passing a set of data quality cuts, and then requires the interaction vertex to be within the Proton Module fiducial volume. The FV is defined to be the transverse central $\pm 50 \times \pm 50 \; \mathrm{cm}^2$ region of the Proton Module and excludes the first four scintillator layers as an upstream veto. Events with exactly one or two tracks sharing a common vertex are selected where one track must be minimum ionizing -- the muon candidate -- and the second track, if present, must be proton-like according to the PID. The INGRID PID algorithm is based on a boosted decision tree that uses the dE/dx along the track and the distribution of deposited energy with respect to distance from the end point of the track. Protons will tend to deposit more energy at the end of the track compared to muons or pions (referred to as a Bragg peak). The muon candidate track must either stop in the Proton Module or reach the standard module directly downstream, where it may also stop or traverse the entire module and escape. Events where the muon escapes out the side of the Proton Module are rejected. The momentum of a stopping muon candidate track is measured by calculating an equivalent distance traversed in iron, and muon candidate tracks that travel through the entire standard module and escape have a lower limit on their momentum. The momentum threshold for a muon to escape the standard module is approximately 1 GeV/\textit{c}. The selected events are binned using the reconstructed momentum and angle with respect to the beam direction (list of bin edges available in Appendix \ref{app:binning}). Stopping and escaping events are considered together as a single \CCzeropi sample in this analysis (Sample IX).

\begin{figure*}[hbt]
    \centering
    \subfloat{\includegraphics[width=0.40\textwidth,angle=0]{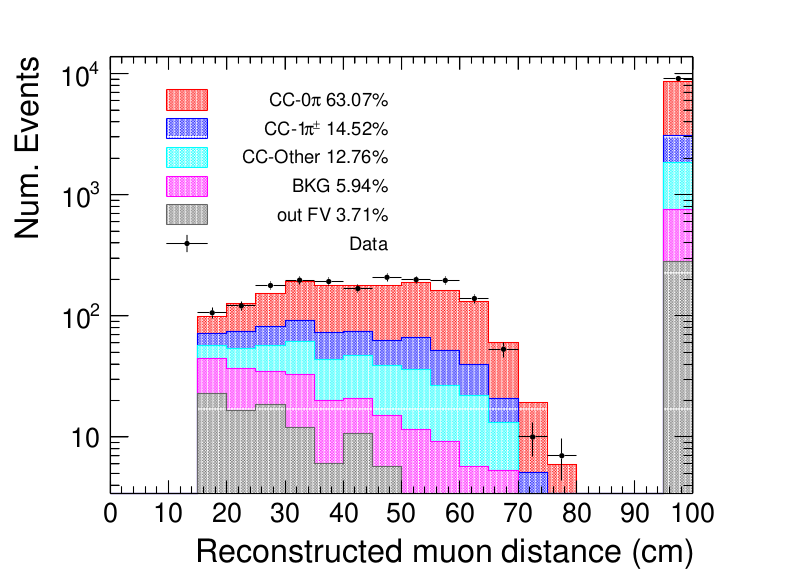}}
    \subfloat{\includegraphics[width=0.40\textwidth,angle=0]{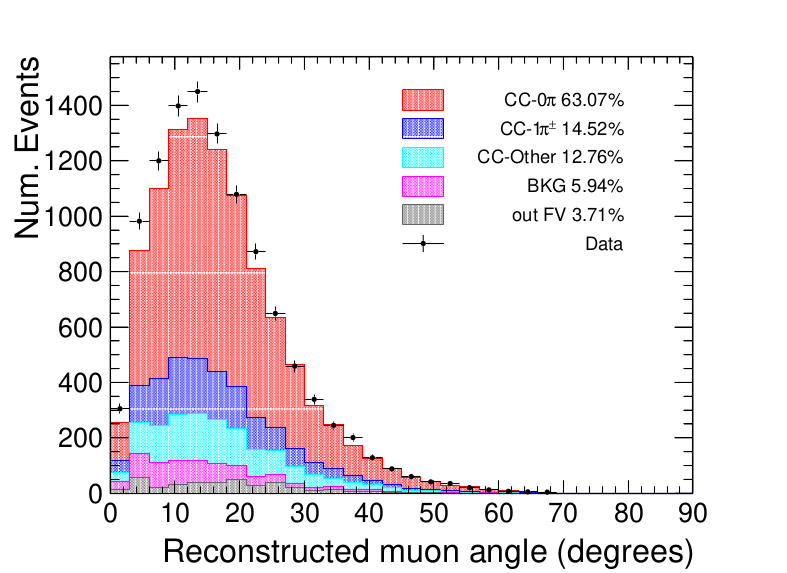}}
    \hfill
    \caption{Event distribution for measured data and MC prediction in reconstructed equivalent distance in iron and angle for the INGRID signal sample stacked by true topology. Through-going events are all placed in the final distance bin. The purity of each topology is listed in the legend.}
    \label{fig:ingrid_signal_samples}
\end{figure*}

The kinematic distribution of the signal sample separated by true topology is shown in Fig. \ref{fig:ingrid_signal_samples} with data plus statistical errors overlaid. The INGRID sample is notably less pure than the ND280 sample, with a much higher background primarily coming from pions being misidentified as muons. For the cross-section extraction, the kinematic regions of $p_\mu < 0.35 \; \mathrm{GeV}/c$ and $\cos(\theta_\mu) < 0.50$ are excluded from the analysis to remove regions of no acceptance due to the detector geometry.

\clearpage
\subsection{Control regions}

To provide a better constraint on the background contributions, a set of control samples are included in the analysis. As with the signal selections, the ND280 and INGRID control samples are designed to select similar types of events, however the additional capabilities of ND280 allow for more complicated event topologies.

\subsubsection{ND280}

The ND280 selection includes three control samples to select events with a pion, constraining the primary background contribution to the signal selection. These samples follow similar initial data quality cuts and criteria for identifying a muon candidate, but also require the identification of a pion candidate. A new addition for this analysis compared to previous ND280 analyses is the inclusion of a separate sample designed to identify low momentum pion events by detecting the presence of a Michel decay electron in FGD1. The control samples are categorized by the pion content as follows:

\begin{itemize}
    \item Sample VI (\CConepiplus): defined by a single muon candidate in the TPC and one $\pi^{+}$ candidate in the TPC;
    \item Sample VII (\CCother): a muon candidate, one $\pi^{+}$ candidate, and at least one additional track in the TPC;
    \item Sample VIII (CC-Michel): a muon candidate in the TPC and a delayed Michel electron in FGD1 indicating the presence of a low momentum $\pi^{+}$ below tracking threshold.
\end{itemize}

\begin{figure*}[hbt]
    \centering
    \subfloat{\includegraphics[width=0.40\textwidth,angle=0]{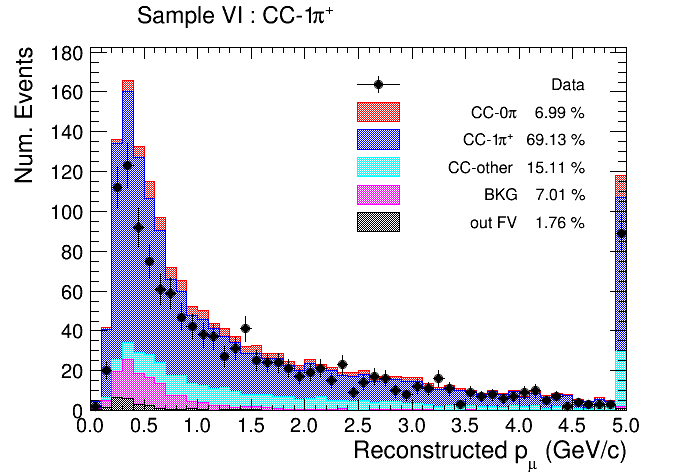}}
    \subfloat{\includegraphics[width=0.40\textwidth,angle=0]{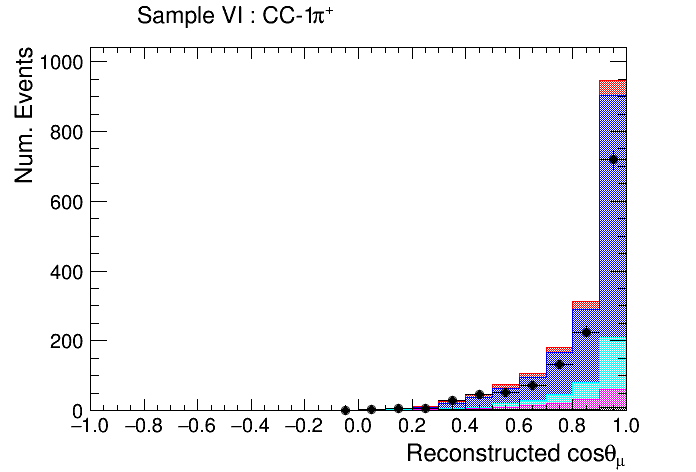}}
    \hfill
    \subfloat{\includegraphics[width=0.40\textwidth,angle=0]{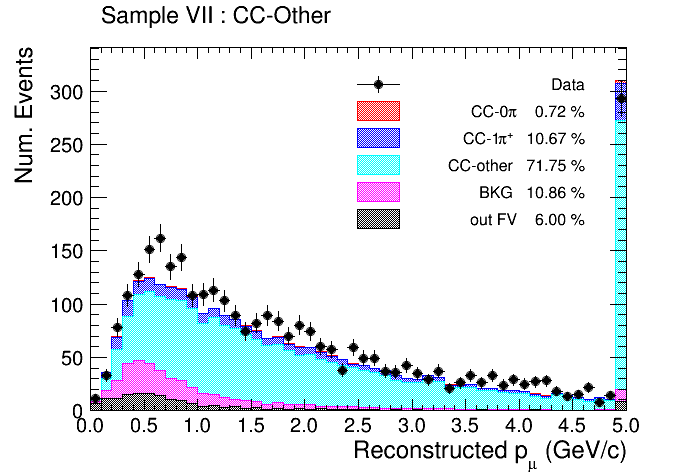}}
    \subfloat{\includegraphics[width=0.40\textwidth,angle=0]{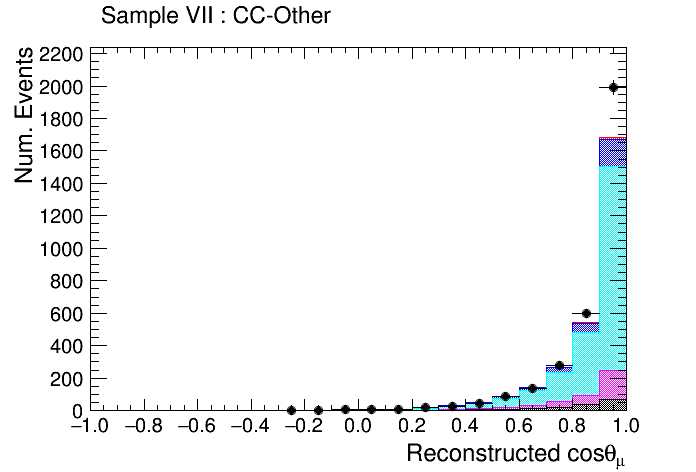}}
    \hfill
    \subfloat{\includegraphics[width=0.40\textwidth,angle=0]{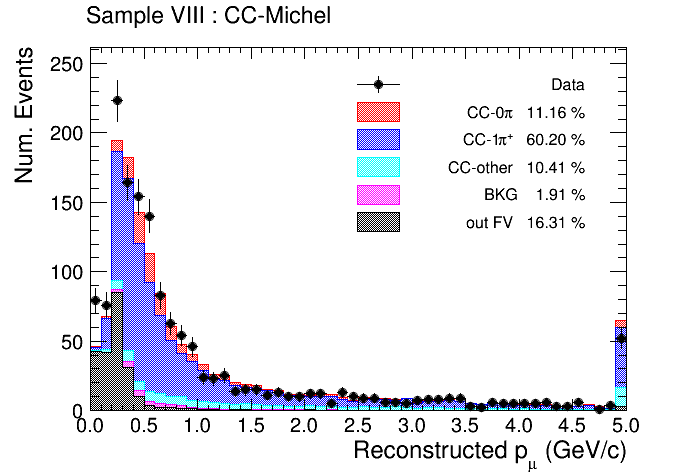}}
    \subfloat{\includegraphics[width=0.40\textwidth,angle=0]{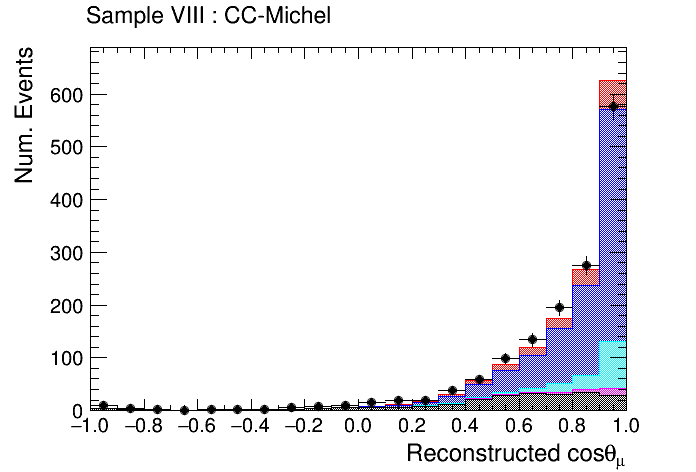}}
    \hfill
    \caption{Event distribution for measured data and MC prediction in reconstructed muon momentum and angle for the ND280 control samples stacked by true topology. The purity of each topology is listed in the legend, and the last bin for muon momentum contains all events with momentum greater than 5 GeV/c.}
    \label{fig:nd280_control_samples}
\end{figure*}

The kinematic distributions of each sample separated by true topology are shown in Fig. \ref{fig:nd280_control_samples}. The data clearly shows a deficit compared to the nominal MC prediction for the \CConepiplus sample while the opposite is seen in the \CCother sample, highlighting the need to include the control samples for a data driven background constraint. This deficit of \CConepiplus events has been observed in previous ND280 analyses \cite{PhysRevD.101.112001, PhysRevD.101.112004}. However in the CC-Michel sample, which contains mostly \CConepiplus events, the data has a similar overall normalization compared to the nominal MC prediction, and also shows an excess at the peak of the distribution. This tension between the \CConepiplus and CC-Michel samples and the impact on the analysis is discussed further in Sec. \ref{sec:results}.

\subsubsection{INGRID}

\begin{figure*}[hbt]
    \centering
    \subfloat{\includegraphics[width=0.40\textwidth,angle=0]{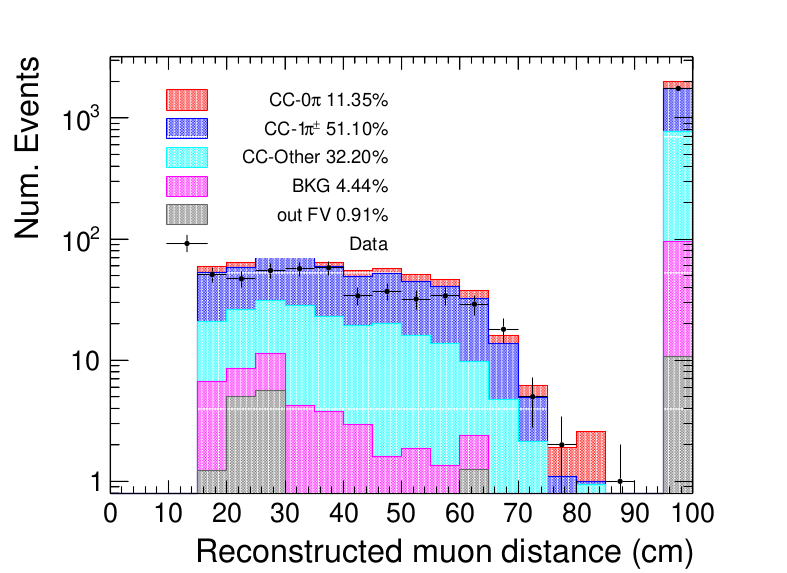}}
    \subfloat{\includegraphics[width=0.40\textwidth,angle=0]{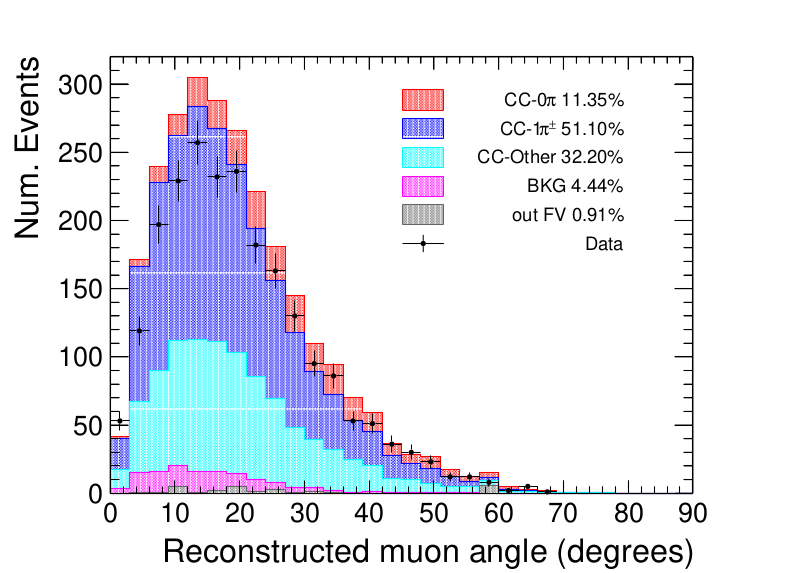}}
    \hfill
    \caption{Event distribution for measured data and MC prediction in reconstructed equivalent distance in iron and angle for the INGRID control sample stacked by true topology. Through-going events are all placed in the final distance bin. The purity of each topology is listed in the legend.}
    \label{fig:ingrid_control_samples}
\end{figure*}

The INGRID selection includes a single control sample to select events with a single pion candidate track. Events must contain exactly two or three tracks that share a common vertex, with the highest-momentum minimum-ionizing track labelled as the muon candidate, the other minimum-ionizing track as the pion candidate, and a third track that if present must be proton-like. The PID cuts have been tuned for this sample to have a higher efficiency for selecting pion tracks compared to selecting proton tracks for the signal sample.  The kinematic distribution of the control sample separated by true topology is shown in Fig. \ref{fig:ingrid_control_samples}. Stopping and escaping events are considered together as a single \CConepi (events with a muon and a single charged pion track) sample in this analysis (Sample X). Similar to the ND280 control samples, the INGRID data shows a deficit of interactions producing a pion compared to the nominal MC prediction.

\section{Analysis strategy}
\label{sec:fit}

\subsection{Binned likelihood fit}
This analysis uses an unregularized binned maximum likelihood fit similar to the analyses in Refs. \cite{PhysRevD.93.112012, PhysRevD.98.032003, PhysRevD.101.112001, PhysRevD.101.112004, PhysRevD.103.112009}, to fit a set of signal and control samples to provide a data-driven background constraint, to unfold the detector effects, and to extract the number of selected signal events in the analysis bins. The ND280 and INGRID samples are fit simultaneously to extract the \CCzeropi cross section for each detector and produce a correlated result. For the purposes of the analysis, ND280 and INGRID events occupy different bins but are otherwise treated similarly. The analysis framework has been significantly improved compared to previous T2K \CCzeropi results (specifically Refs. \cite{PhysRevD.93.112012, PhysRevD.98.032003, PhysRevD.101.112001, PhysRevD.101.112004}), for example including an improved treatment of the MC statistical uncertainty and principal component analysis to reduce the dimensionality of the fit.

This method varies the input MC using a set of fit parameters for both signal and background events to find the best agreement to the data, and the values and corresponding errors of these parameters at the best-fit point are then used for the cross-section extraction. The primary parameters of interest in the fit are the ``template parameters" $c_i$ which scale the total number of signal events in each kinematic truth bin $i$ (seventy in total for this analysis), and are completely free parameters with no prior constraint. The rest of the parameters are the systematic (or nuisance) parameters that describe variations to the flux, detector, and neutrino interaction model (described in Sec. \ref{subsec:uncertainties}). Separate flux and detector parameters (including correlations when available) are included for ND280 and INGRID, while both detectors use the same neutrino interaction model parameters.

The best-fit parameters are found by minimizing the negative log-likelihood ratio (also approximated as the chi-square), and is split into a statistical and systematic contribution as follows:
\begin{equation}
\label{eq:llh}
    \chi^{2} \approx -2 \log \mathcal{L} = -2 \log \mathcal{L}_\text{stat} -2 \log \mathcal{L}_\text{syst},
\end{equation}
where
\begin{equation}
\label{eq:llh_stat}
\begin{split}
     & -2\log \mathcal{L}_\text{stat} = \\
     &2\sum_j^{\text{reco bins}} \left( \beta_j N^{\text{MC}}_j - N^{\text{obs}}_j + N^{\text{obs}}_j \log \frac{N^{\text{obs}_j}}{\beta_j N^{\text{MC}}_j} + \frac{\beta^2_j - 1}{2\sigma^2_j} \right)
\end{split}
\end{equation}
and
\begin{equation}
\label{eq:llh_syst}
    -2 \log \mathcal{L}_\text{syst} = (\vec{p} - \vec{p}_\text{prior})\mathbf{V}_\text{syst}^{-1}(\vec{p} - \vec{p}_\text{prior}).
\end{equation}

Equation \ref{eq:llh_stat} is the modified statistical log-likelihood ratio following the Barlow--Beeston method \cite{BARLOW1993219, Conway:1306523} for including the uncertainty of finite MC simulation. $N^{\text{MC}}_j$ and $N^{\text{obs}}_j$ are the number of simulated and observed events for each reconstructed bin $j$. The Barlow--Beeston scaling parameter for each bin $\beta_j$ is given by the following:
\begin{equation}
\label{eq:beta}
    \beta_j = \frac{1}{2} \left( -(N^{\text{MC}}_j \sigma^2_j - 1) + \sqrt{(N^{\text{MC}}_j \sigma^2_j - 1)^2 + 4N^{\text{MC}}_j \sigma^2_j} \right)
\end{equation}
where $\sigma^2_j$ is the relative variance of the number of MC events $N^{\text{MC}}_j$ in the bin. In the limit of infinite MC simulation, $\sigma^2_j \rightarrow 0$ and $\beta_j \rightarrow 1$ giving the standard Poisson log-likelihood ratio. Equation \ref{eq:llh_syst} is a Gaussian penalty term to account for the contribution from varying the systematic parameters $\vec{p}$ during the fit compared to their fixed prior values $\vec{p}_\text{prior}$ and uncertainty. A covariance matrix $\mathbf{V}_\text{syst}$ is used to describe the prior uncertainty and correlations between the parameters.

The input MC simulation for a reconstructed bin $j$ is the sum of weighted signal and background events, and can be expressed as:
\begin{equation}
    \label{eq:mc_events}
    N^{\text{MC}}_j = \sum^{\text{true bins}}_i \left( c^{}_{i} w^{\text{sig}}_{ij}(\vec{p}) N^{\text{sig}}_{ij} + w^{\text{bkg}}_{ij}(\vec{p}) N^{\text{bkg}}_{ij} \right)
\end{equation}
where $N^{\text{sig}}_{ij}$ and $N^{\text{bkg}}_{ij}$ are the signal and background events for truth kinematic bin $i$ and reconstructed bin $j$ as predicted by the MC simulation, $c_i$ are the signal template parameters, and $w_{ij}$ are the weights as a function of the systematic parameters $\vec{p}$, and depend on the truth and reconstructed bins $i$ and $j$.

\subsection{Systematic uncertainties}
\label{subsec:uncertainties}

There are three types of systematic uncertainties considered for this analysis and included in the fit as parameters: flux, detector, and neutrino interaction model uncertainties.

\begin{figure}[hbt]
    \centering
    \includegraphics[width=0.4\textwidth,angle=0]{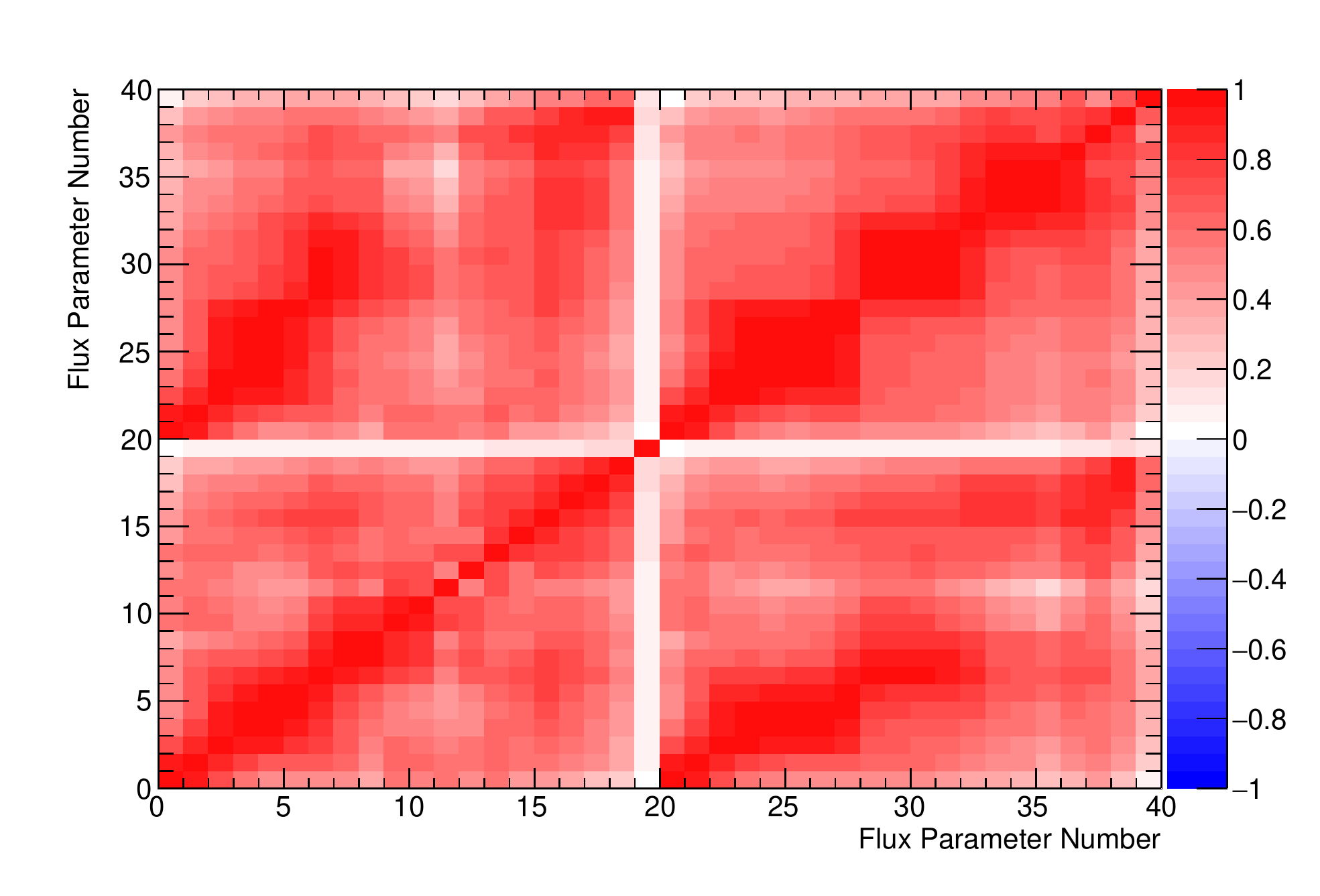}
    \caption{Input flux correlation matrix binned in neutrino energy for both ND280 and INGRID. The flux is highly correlated both across the energy spectrum and between the detectors.}
    \label{fig:flux_correlation}
\end{figure}

The neutrino flux uncertainty is parameterized as scale factors in forty total bins of true neutrino energy with separate flux parameters (or bins) for ND280 and INGRID. Only the $\nu_\mu$ flavor is considered for this analysis due to the small contribution of the $\bar\nu_\mu$, $\nu_e$, and $\bar\nu_e$ flavors. These parameters use the same energy binning scheme and can only affect events for their respective detector (the flux bin edges can be found in Appendix \ref{app:flux_binning}). They have a prior constraint described by a covariance matrix, which includes the correlations between energy bins and between the fluxes at each detector. As shown in Fig. \ref{fig:flux_correlation}, the fluxes at ND280 and INGRID are highly correlated a priori. The high energy bin (10 to 30 GeV) for ND280 is less correlated than the lower energy bins due to an increase in the hadron uncertainties for that bin (mostly from kaon decay). Since the number of events in the analysis corresponding to this energy range is small, it has little effect on the analysis. For a given true neutrino energy bin, identical weights are given to signal and background events. The flux uncertainty is dominated by the hadronic multiplicity and decay modeling, along with other contributions, such as uncertainties in the horn current and alignment.

The detector uncertainty is parameterized as scale factors on the event rate for each reconstructed bin with separate detector parameters for ND280 and INGRID. They have a prior constraint described by a covariance matrix including the correlations between the reconstructed bins for a given detector, however ND280 and INGRID use separate matrices and are completely uncorrelated in the fit. In principle, several detector uncertainties could be considered correlated between them, for example using the same pion secondary interaction modeling in \textsc{geant4} for the detector simulation, but this was out of scope for this analysis. Independent and dedicated control samples for each detector are used to evaluate the detector uncertainties based on data-MC agreement. The largest contribution to the detector uncertainty is the pion secondary interaction modeling. Since the fit includes a detector parameter for each reconstructed bin, this adds up to many hundreds of parameters. Principal component analysis is used to reduce the total number of detector parameters by more than half by transforming the parameters to their eigenspace and removing the parameters that contribute less information according to their eigenvalues such that 99\% of the total information in the covariance matrix is retained.

The neutrino interaction uncertainty is included in the fit using a set of twenty-one parameters that are designed to weight events based on aspects of the neutrino interaction model for both signal and background samples, including final state interactions (similar to Ref. \cite{PhysRevD.101.112001}). A table listing the neutrino interaction parameter names and their priors can be found in Appendix \ref{app:xsec_parameters}, and are described as follows. Variations to the signal model are included through changing the shape of the CCQE cross section by varying the axial mass $M^{QE}_{A}$, and with two parameters to alter the overall normalization of and shape of the 2p2h model. The resonant pion model has two shape parameters, the axial mass $M^{RES}_{A}$ and the axial form factor at zero momentum transfer $C^{A}_{5}$, and a normalization for the non-resonant background $I_{1/2}$. Additionally, two normalization parameters for \CConepi events are included to give the fit additional freedom to adjust the pion background and prevent over-fitting of the flux parameters. For deep inelastic scattering events, a custom shape parameter (DIS multi-pi shape) is used to give greater freedom at lower neutrino energy along with two normalization parameters for DIS and multi-pi events. The other major event topologies (coherent and neutral current) are each given a normalization parameter. Finally a set of six parameters are included to allow the pion FSI model to vary within the fit, separated by different reaction channels (absorption, production, charge exchange, and scattering) and pion momentum range. The prior uncertainty and correlations between parameters are encoded in a single covariance matrix. ND280 and INGRID share the same neutrino interaction parameters but the event weights are calculated for each detector separately.

\subsection{Cross-section extraction}

The flux-integrated differential cross section as a function of true muon kinematics $x = p_\mu \, \cos(\theta_\mu)$ for each detector is calculated using the following:
\begin{equation}
    \label{eq:xsec}
    \frac{\mathrm{d}\sigma}{\mathrm{d}x_i} = \frac{\hat{N}^{\text{sig}}_i}{\epsilon_i \Phi N_{\text{nucleons}} \Delta x_i}
\end{equation}
where $\hat{N}^{\text{sig}}_i$ is the best-fit number of selected signal events in truth bin $i$ summed across all samples, $\epsilon_i$ is the bin-by-bin efficiency correction, $\Phi$ is the integral of the neutrino flux evaluated at the best-fit parameters, $N_{\text{nucleons}}$ is the number of target nucleons in the fiducial volume, and $\Delta x_i$ is the bin width. The bin edges for the extracted cross section can be found in Appendix \ref{app:binning}.

\begin{figure*}[hbt]
    \centering
    \subfloat{\includegraphics[width=0.35\textwidth,angle=0]{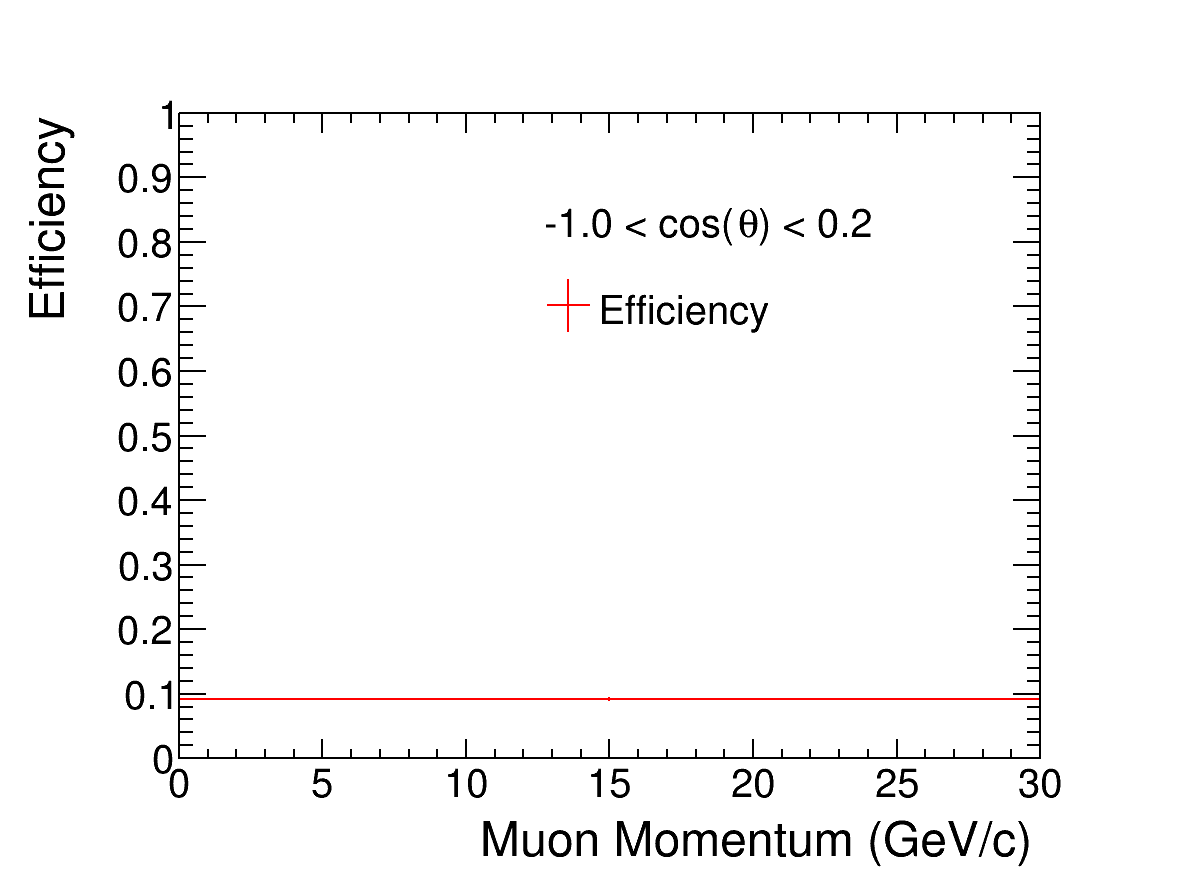}}
    \subfloat{\includegraphics[width=0.35\textwidth,angle=0]{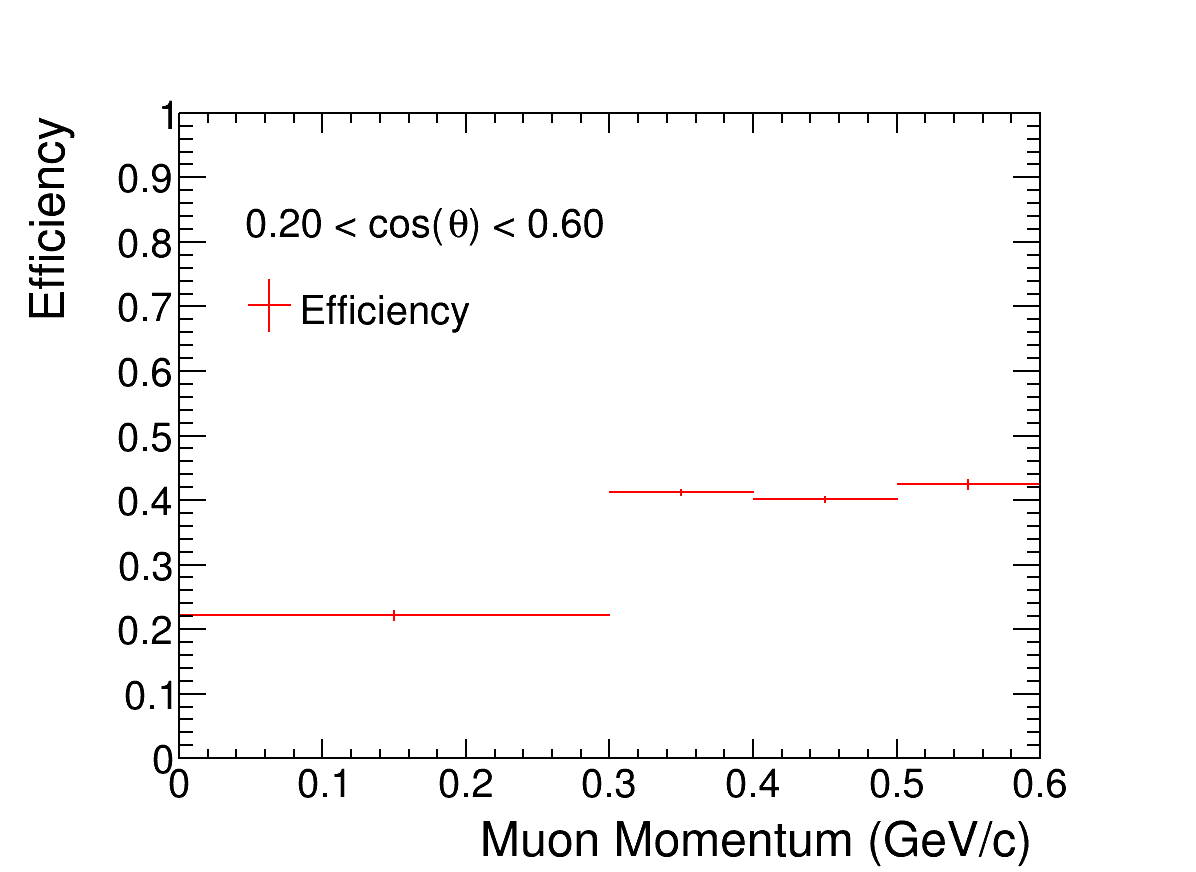}}
    \subfloat{\includegraphics[width=0.35\textwidth,angle=0]{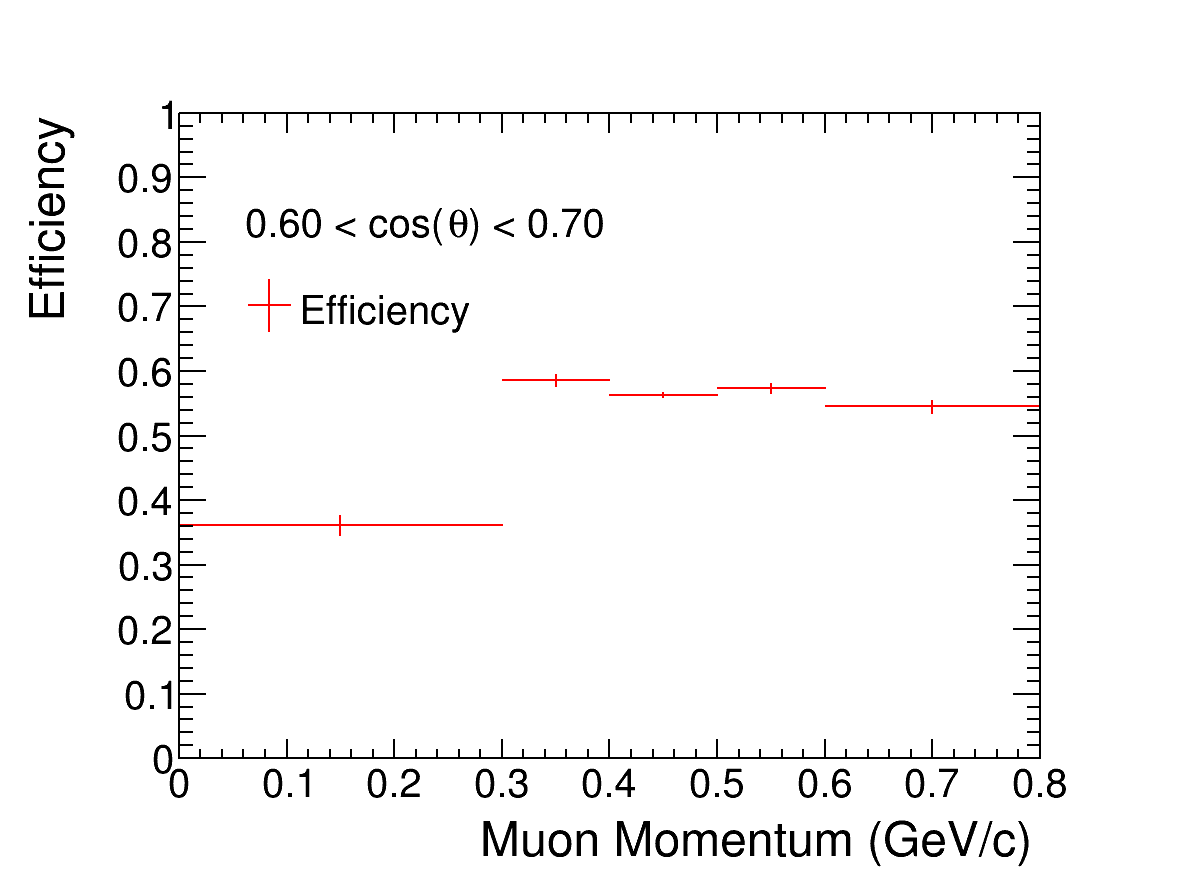}}
    \hfill
    \subfloat{\includegraphics[width=0.35\textwidth,angle=0]{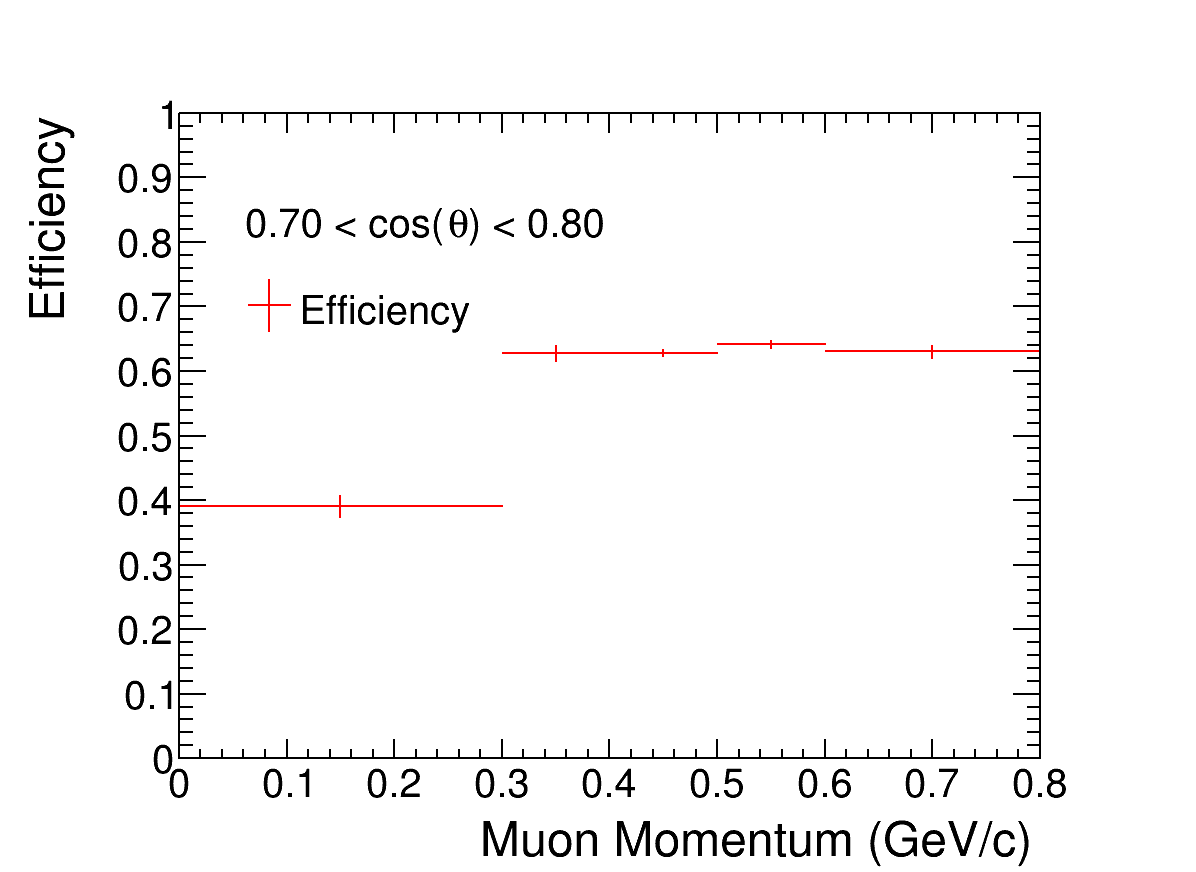}}
    \subfloat{\includegraphics[width=0.35\textwidth,angle=0]{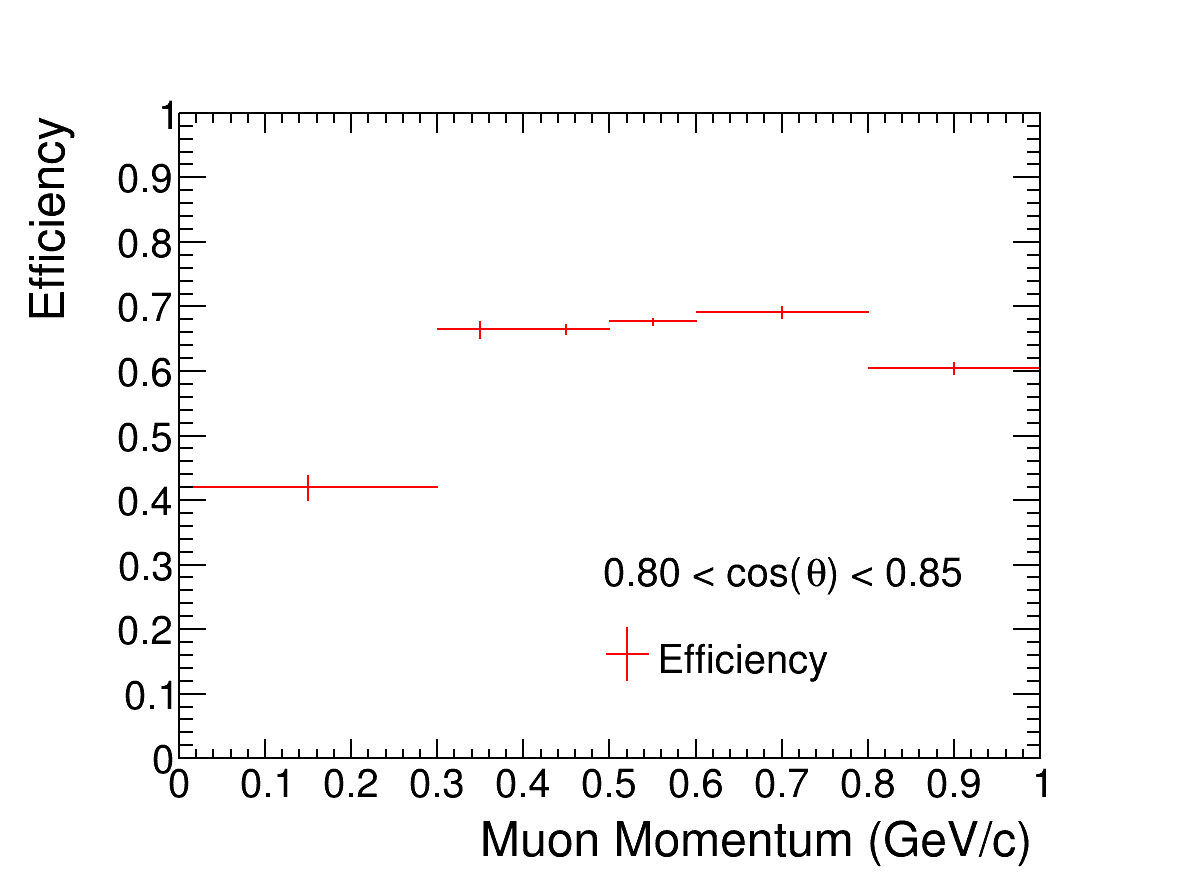}}
    \subfloat{\includegraphics[width=0.35\textwidth,angle=0]{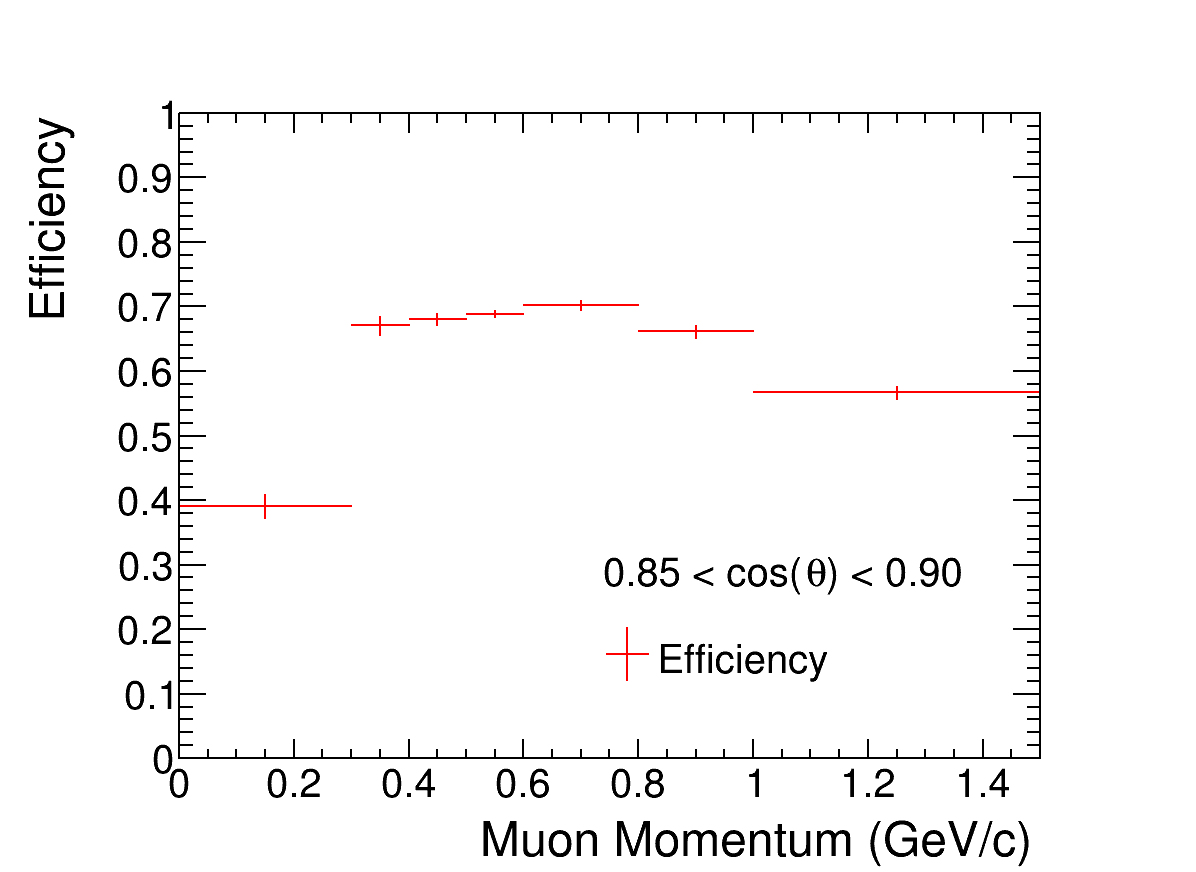}}
    \hfill
    \subfloat{\includegraphics[width=0.35\textwidth,angle=0]{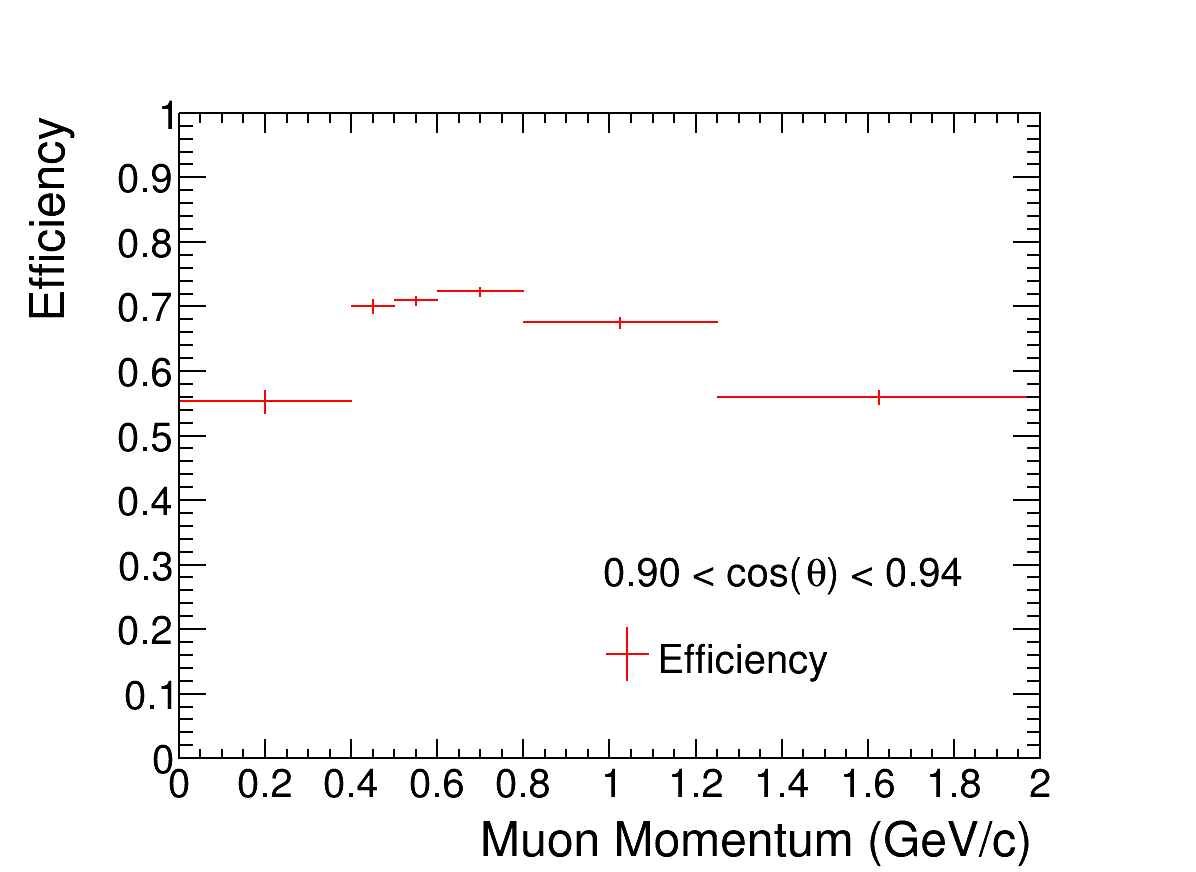}}
    \subfloat{\includegraphics[width=0.35\textwidth,angle=0]{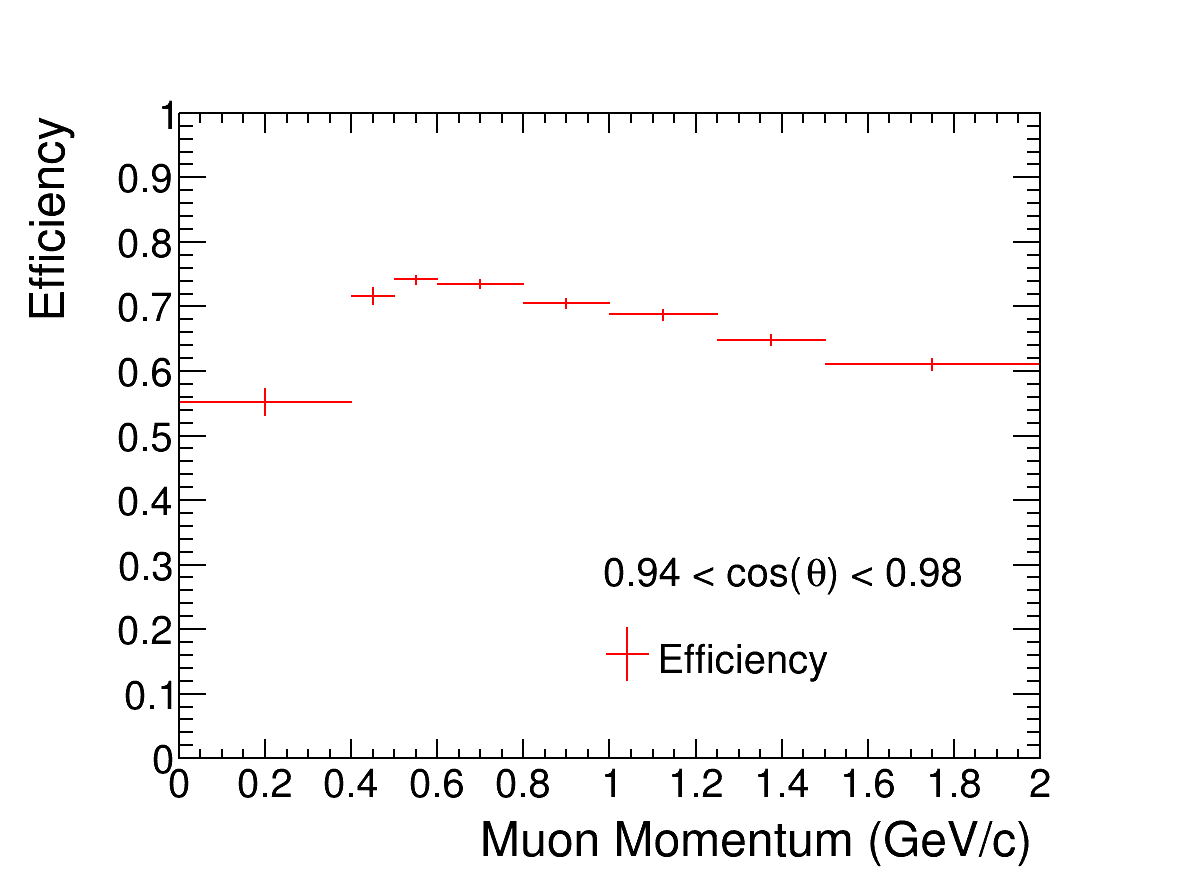}}
    \subfloat{\includegraphics[width=0.35\textwidth,angle=0]{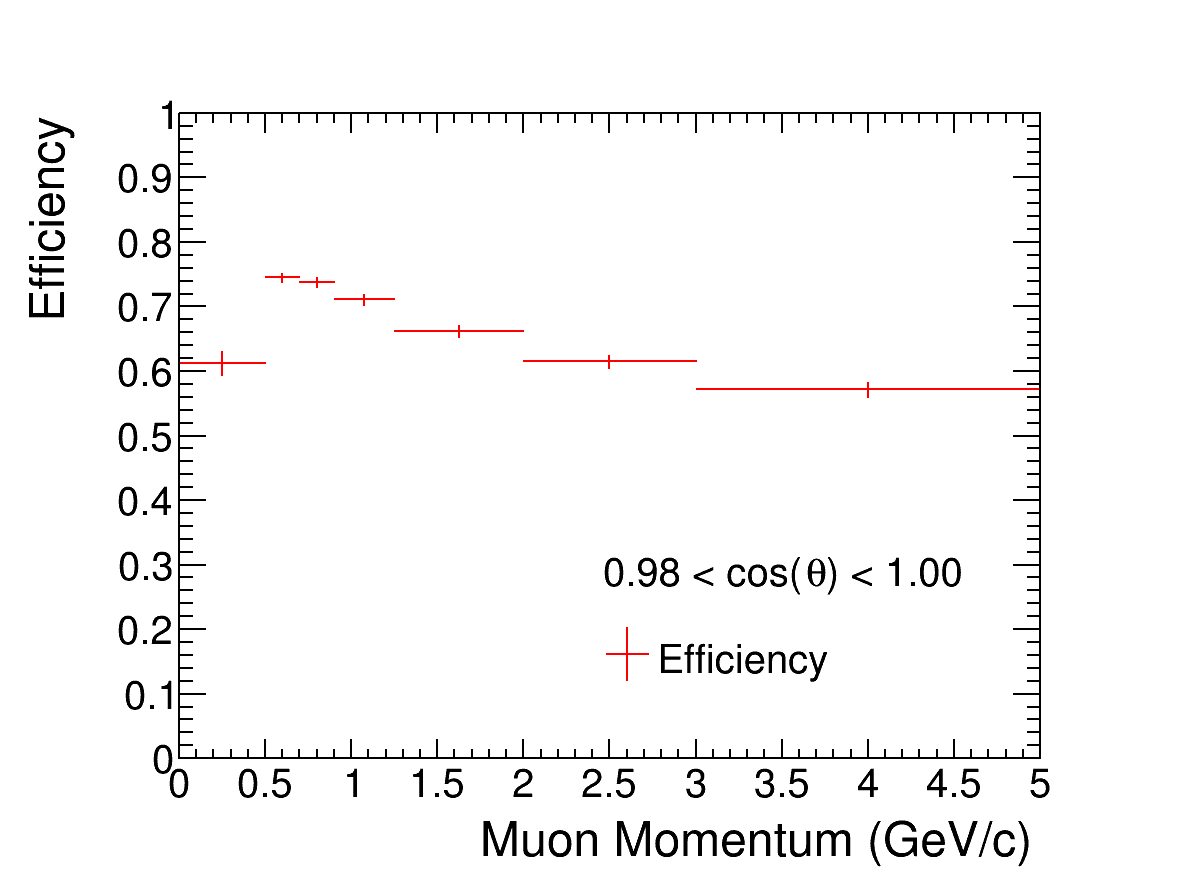}}
    \hfill
    \caption{Selection efficiency with post-fit uncertainty for the ND280 cross-section bins as function of true muon momentum in muon angle bins. Note that the final bin extending to 30 GeV/\textit{c} has been omitted for clarity.}
    \label{fig:eff_nd280}
\end{figure*}

\begin{figure*}[hbt]
    \centering
    \subfloat{\includegraphics[width=0.35\textwidth,angle=0]{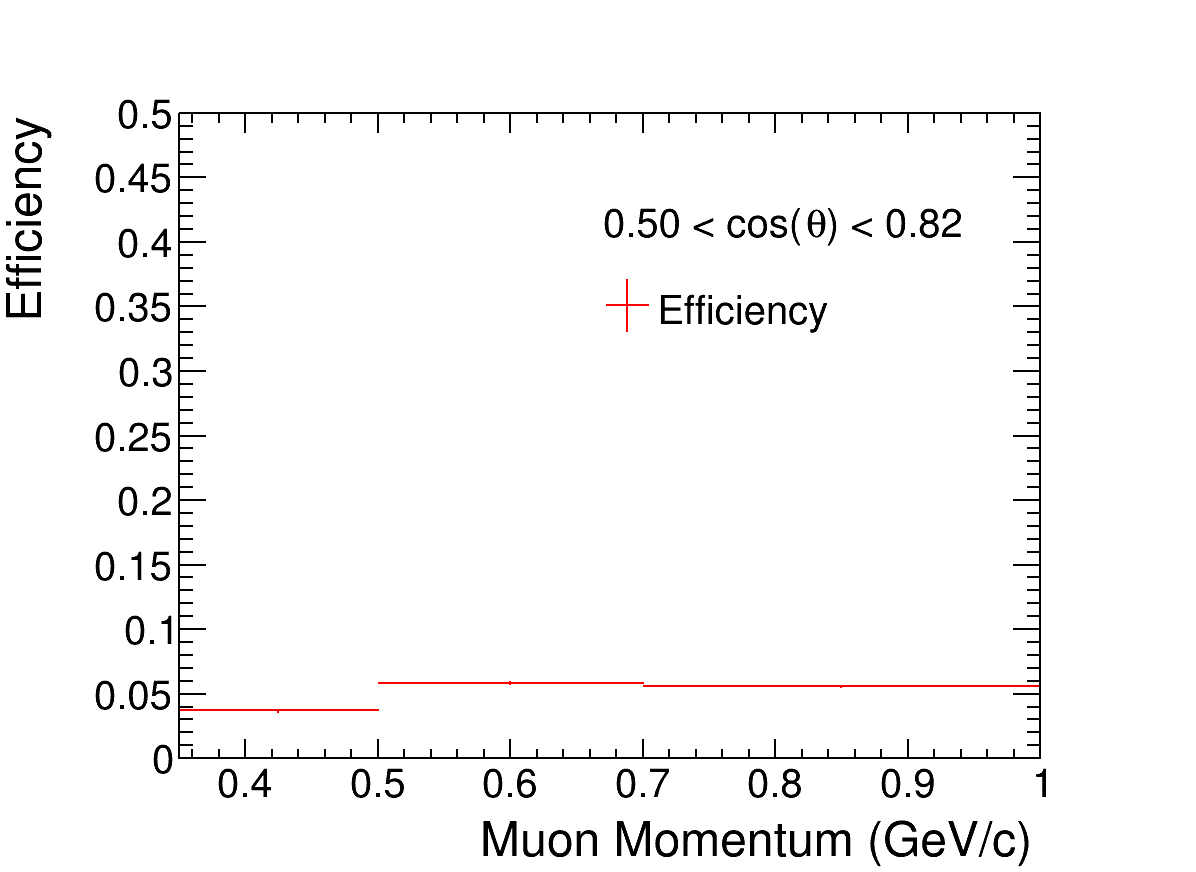}}
    \subfloat{\includegraphics[width=0.35\textwidth,angle=0]{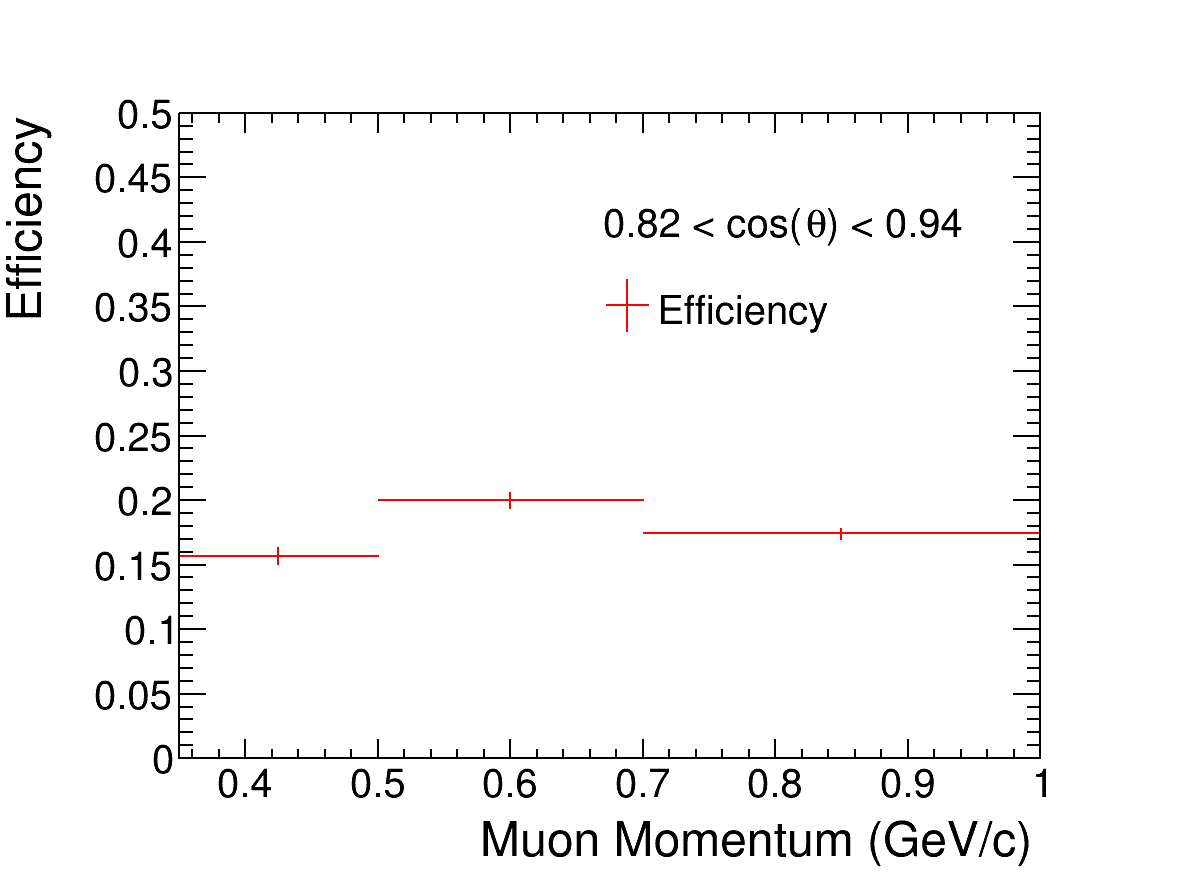}}
    \subfloat{\includegraphics[width=0.35\textwidth,angle=0]{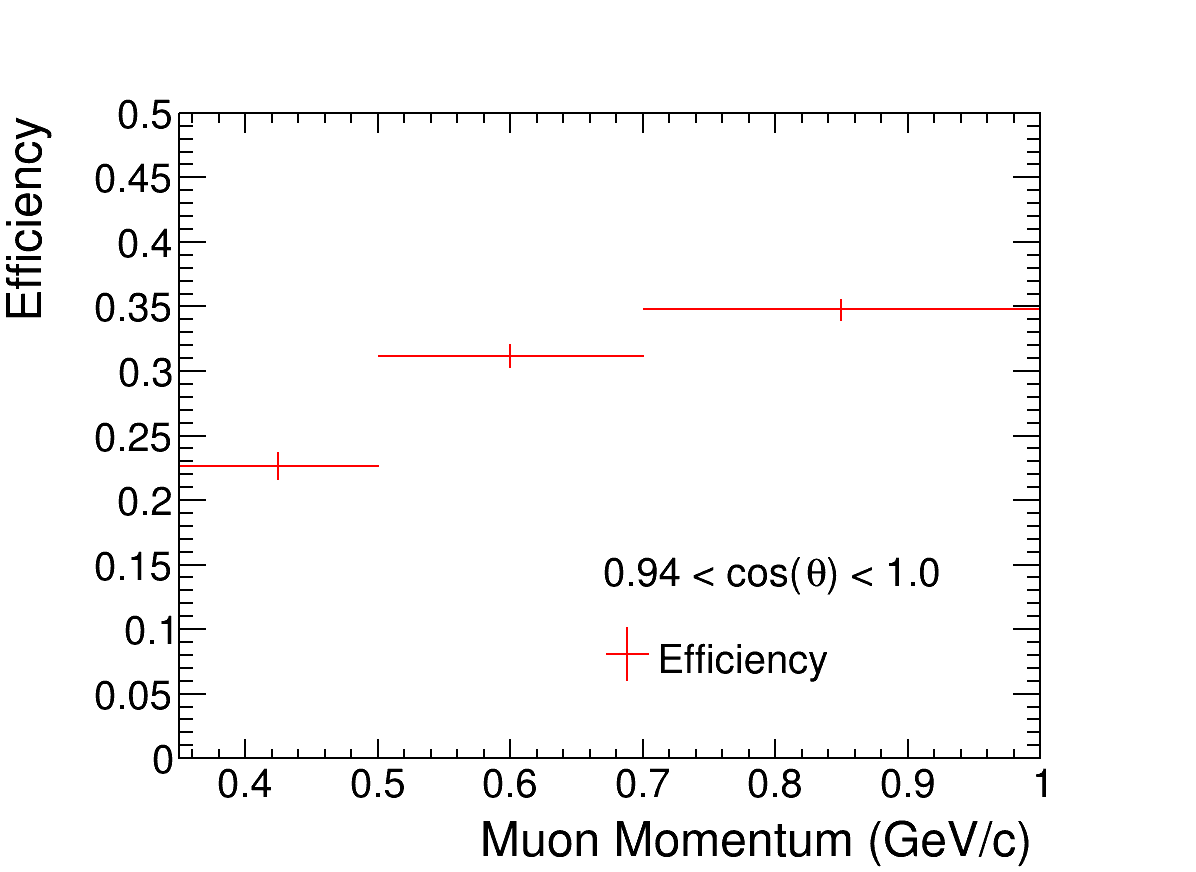}}
    \hfill
    \caption{Selection efficiency with post-fit uncertainty for the INGRID cross-section bins as function of true muon momentum in muon angle bins. Note that the final bin extending to 30 GeV/\textit{c} has been omitted for clarity.}
    \label{fig:eff_ingrid}
\end{figure*}

The cross-section uncertainty is calculated by numerically propagating the post-fit uncertainty for the fit parameters assuming they follow a multivariate Gaussian distribution. The post-fit covariance matrix is Cholesky decomposed and used to create correlated random variations (or ``throws") of the fit parameters that follow the same multivariate distribution as the covariance matrix. This procedure is repeated $10^4$ times to sample the likelihood space encoded in the post-fit covariance matrix. For each thrown variation of the fit parameters, all the events are reweighted using the thrown parameter values and the cross section is recalculated as in Eq. \ref{eq:xsec}. Additionally for each variation, the integrated flux is recalculated, the selection efficiency is allowed to vary based on the thrown parameters, and the number of targets is varied independently for each detector. The efficiency for each cross-section bin and its uncertainty for ND280 and INGRID are shown in Figs. \ref{fig:eff_nd280} and \ref{fig:eff_ingrid} respectively. In general for both ND280 and INGRID the efficiency increases at forward angle and higher momentum as muons leave longer tracks. The number of target nucleons and its uncertainty for ND280 are $5.53 \times 10^{29} \pm 0.67\%$, and INGRID are $1.76 \times 10^{29} \pm 0.38\%$, and includes other elements in addition to carbon and hydrogen present in the fiducial volumes of each detector. The post-fit $\nu_\mu$ flux integral and uncertainty is $\mathrm{2.29} \times 10^{13} \; \mathrm{cm}^{-2}$ and 6.0\% for ND280 and $\mathrm{3.14} \times 10^{13} \; \mathrm{cm}^{-2}$ and 6.1\% for INGRID. The resulting distribution of cross-section values represent the plausible variations of the fit according to the post-fit uncertainties and correlations. Finally, the cross-section ($\mathrm{d}\sigma / \mathrm{d}x$) uncertainties are calculated using the ensemble of random throws and are parameterized using a covariance matrix, assuming the uncertainties are Gaussian distributed.

\begin{figure*}[hbt]
    \centering
    \includegraphics[width=0.35\textwidth,angle=0]{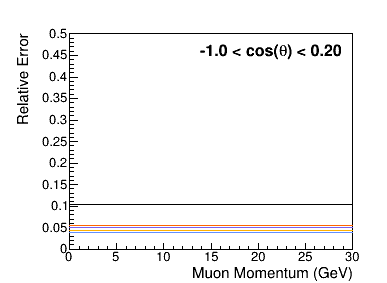}
    \includegraphics[width=0.35\textwidth,angle=0]{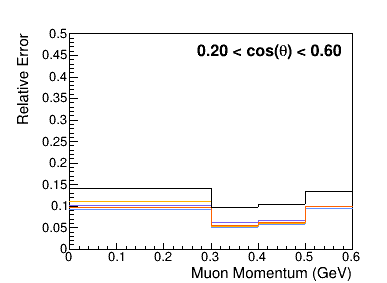}
    \includegraphics[width=0.35\textwidth,angle=0]{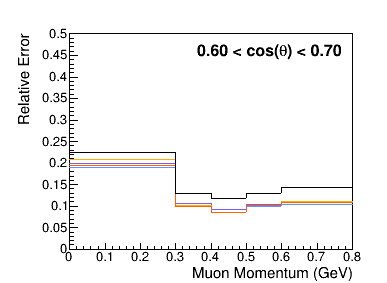}
    \includegraphics[width=0.35\textwidth,angle=0]{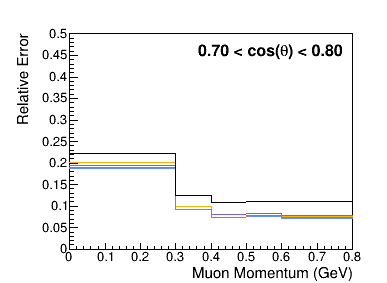}
    \includegraphics[width=0.35\textwidth,angle=0]{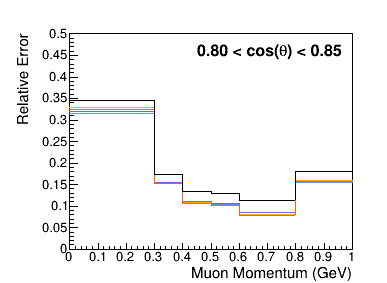}
    \includegraphics[width=0.35\textwidth,angle=0]{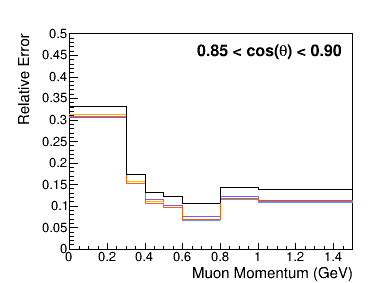}
    \includegraphics[width=0.35\textwidth,angle=0]{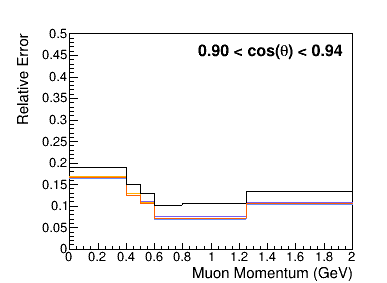}
    \includegraphics[width=0.35\textwidth,angle=0]{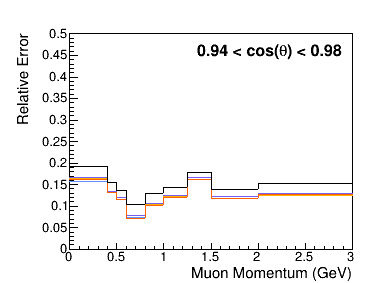}
    \includegraphics[width=0.35\textwidth,angle=0]{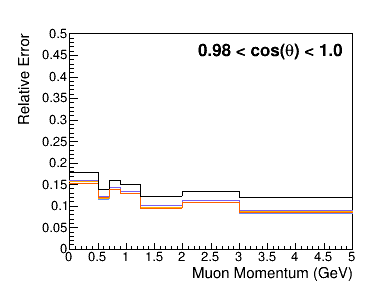}
    \includegraphics[width=0.35\textwidth,angle=0]{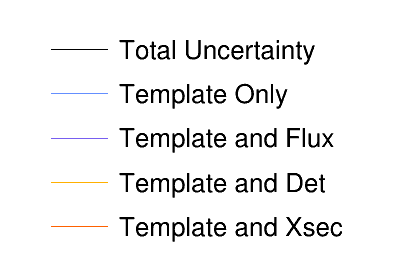}
    \caption{Estimated total systematic uncertainty separated by parameter class for the ND280 cross-section bins as function of true muon momentum in muon angle bins. Note that the final bin extending to 30 GeV/\textit{c} has been omitted for clarity.}
    \label{fig:syst_nd280}
\end{figure*}

\begin{figure*}[hbt]
    \centering
    \includegraphics[width=0.35\textwidth,angle=0]{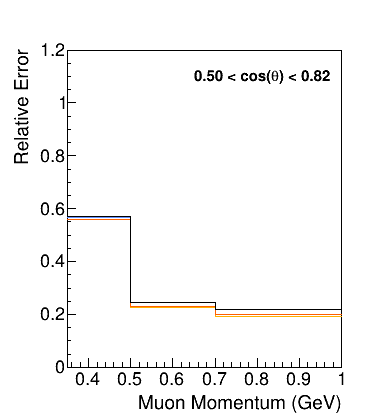}
    \includegraphics[width=0.35\textwidth,angle=0]{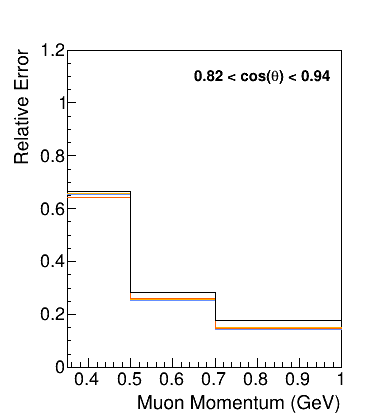}
    \includegraphics[width=0.35\textwidth,angle=0]{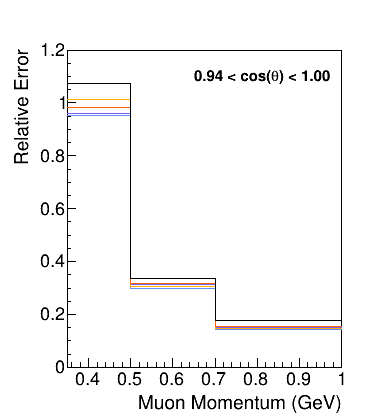}
    \includegraphics[width=0.35\textwidth,angle=0]{Figures/plot_legend_syst.png}
    \caption{Estimated total systematic uncertainty separated by parameter class for the INGRID cross-section bins as function of true muon momentum in muon angle bins. Note that the final bin extending to 30 GeV/\textit{c} has been omitted for clarity.}
    \label{fig:syst_ingrid}
\end{figure*}

A set of fits were performed to estimate the total systematic uncertainty and the contribution from each systematic parameter class (flux, neutrino interaction, and detector) on the measured cross-section bins. A fit using only the template parameters is used as a baseline for the uncertainty (and corresponds approximately to the statistical uncertainty), and additional fits were performed that include each systematic parameter class to estimate its impact (in addition to the template parameters). Each fit used the nominal Monte Carlo prediction as the ``data" so that the best-fit point is at the nominal value for every parameter; this guarantees each fit has the same best-fit point and allows for a more accurate comparison. The results are shown for the analysis cross-section bins for ND280 and INGRID in Figs. \ref{fig:syst_nd280} and \ref{fig:syst_ingrid} respectively, and show roughly equal contributions from the flux, neutrino interaction, and detector parameters to the total uncertainty. Additionally, the uncertainty for the low momentum bins in general is higher than mid to higher momentum bins for a given angle bin. This procedure cannot be applied in the same way to data as each fit will have a different best-fit point resulting in a different total uncertainty, preventing an equal comparison between the fits.

\clearpage
\subsection{Validation}
\label{subsec:validation}

The cross-section extraction was validated using a series of mock data studies, where a known simulated data set was used as input to the fit and the performance of the extraction procedure was studied. These mock data sets cover a wide range of alterations, for example using data-driven modifications such as the deficit seen at low $Q^2$ by \minerva \cite{PhysRevD.99.012004}, modifications to the resonant pion production model, or changes to the flux model. For each of these mock data sets, the cross-section extraction was able to recover the expected true cross section to within the $1\sigma$ uncertainties (often matching the true cross section nearly exactly), showing a robust procedure. The overall $\chi^2$ agreement between the extracted and true cross section is also calculated as follows:
\begin{equation}
    \label{eq:chisq}
    \chi^2 = \sum^{N}_{ij} \left(\frac{\mathrm{d}\sigma^{\text{meas}}}{\mathrm{d}x_i} - \frac{\mathrm{d}\sigma^{\text{true}}}{\mathrm{d}x_i} \right) \mathbf{V}^{-1}_{ij} \left(\frac{\mathrm{d}\sigma^{\text{meas}}}{\mathrm{d}x_j} - \frac{\mathrm{d}\sigma^{\text{true}}}{\mathrm{d}x_j} \right)
\end{equation}
where $N$ is the number of cross-section bins, $x_i$ and $x_j$ are the $i$th and $j$th kinematic bin respectively, and $\mathbf{V}$ is the cross-section covariance matrix.

The post-fit p-value was calculated for the fit to data (or mock data) as another metric to check the validity of the result. First, numerous statistical and systematically varied samples of the MC prediction were produced and then fit to build a post-fit log-likelihood distribution from Eq. \ref{eq:llh}. The p-value is the fraction of the simulated likelihood distribution that is greater than the post-fit log-likelihood of the fit in question. A value of 0.05 for the p-value has been chosen as the threshold to require further investigation of the result.

\section{Results}
\label{sec:results}

\begin{figure*}[hbt]
    \centering
    \includegraphics[width=0.35\textwidth,angle=0]{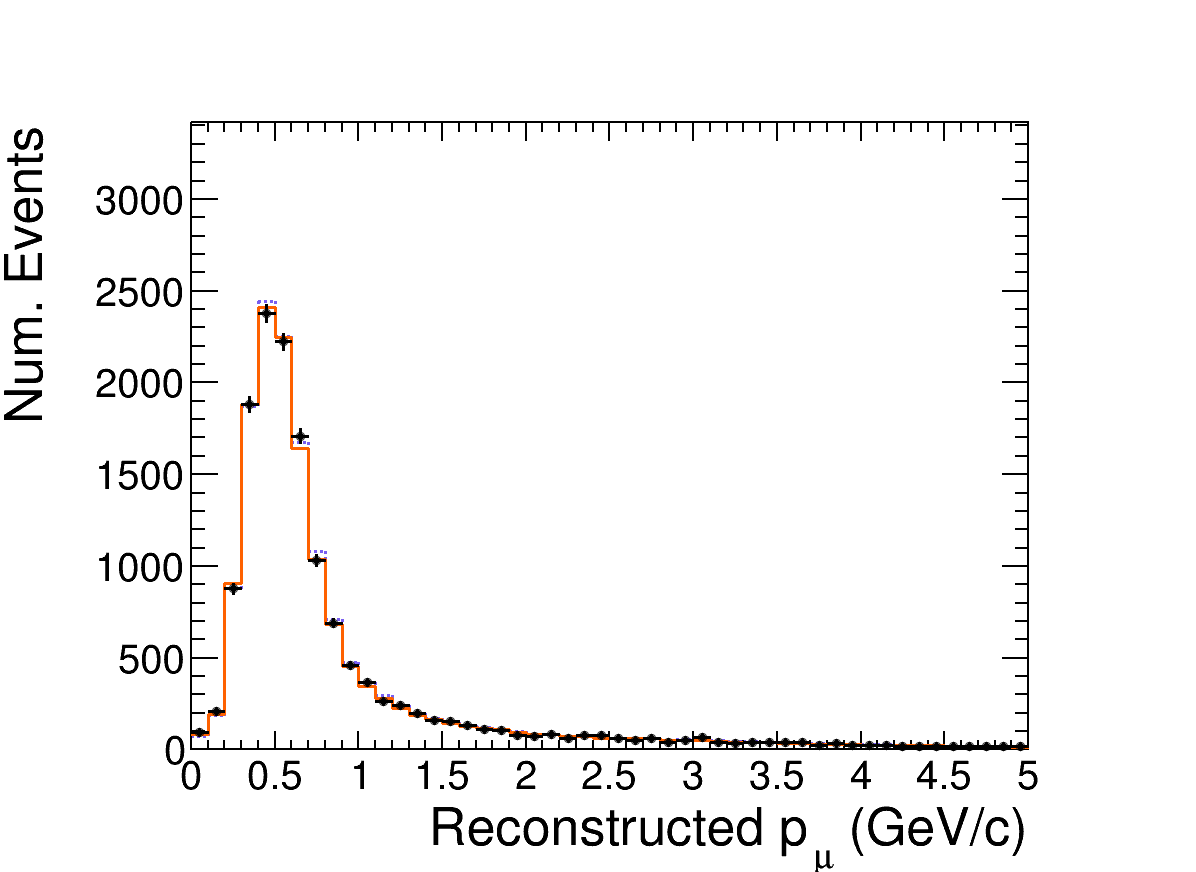}
    \includegraphics[width=0.35\textwidth,angle=0]{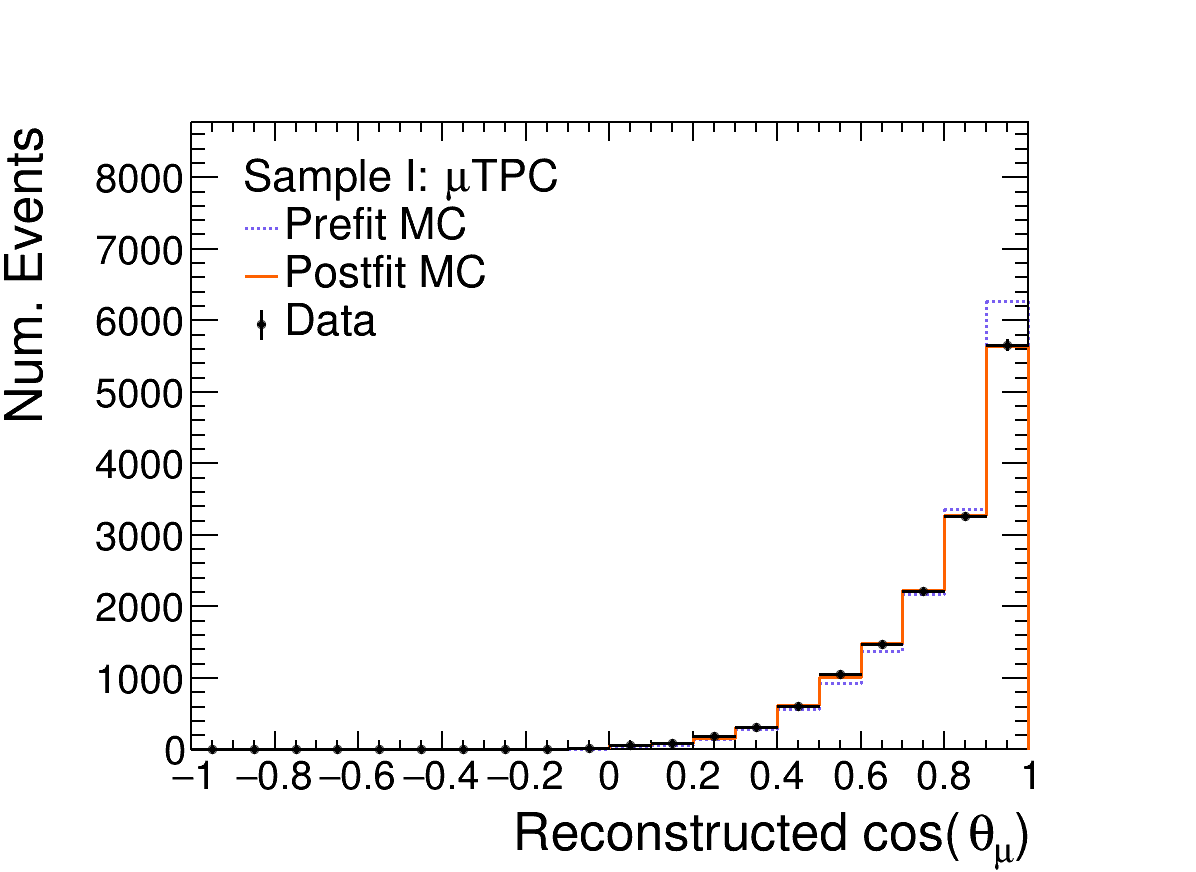}
    \includegraphics[width=0.35\textwidth,angle=0]{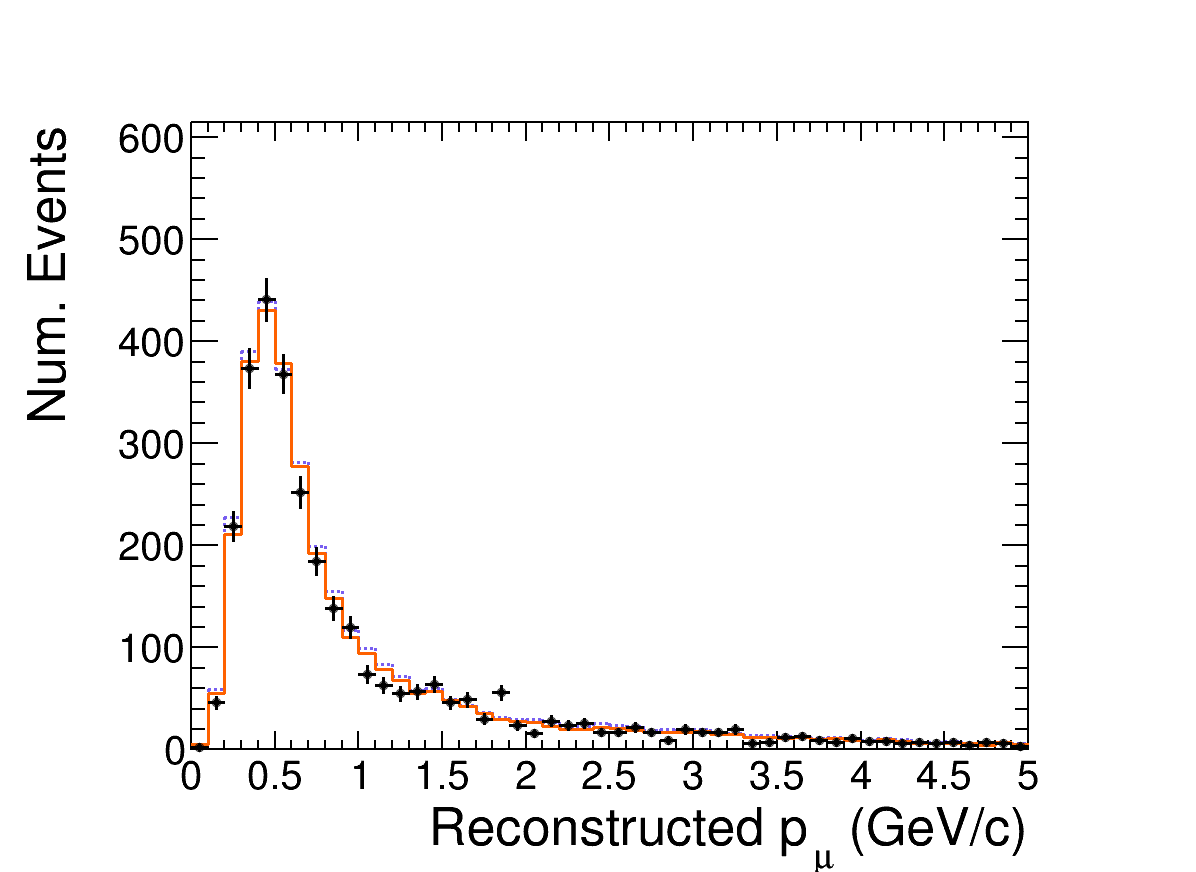}
    \includegraphics[width=0.35\textwidth,angle=0]{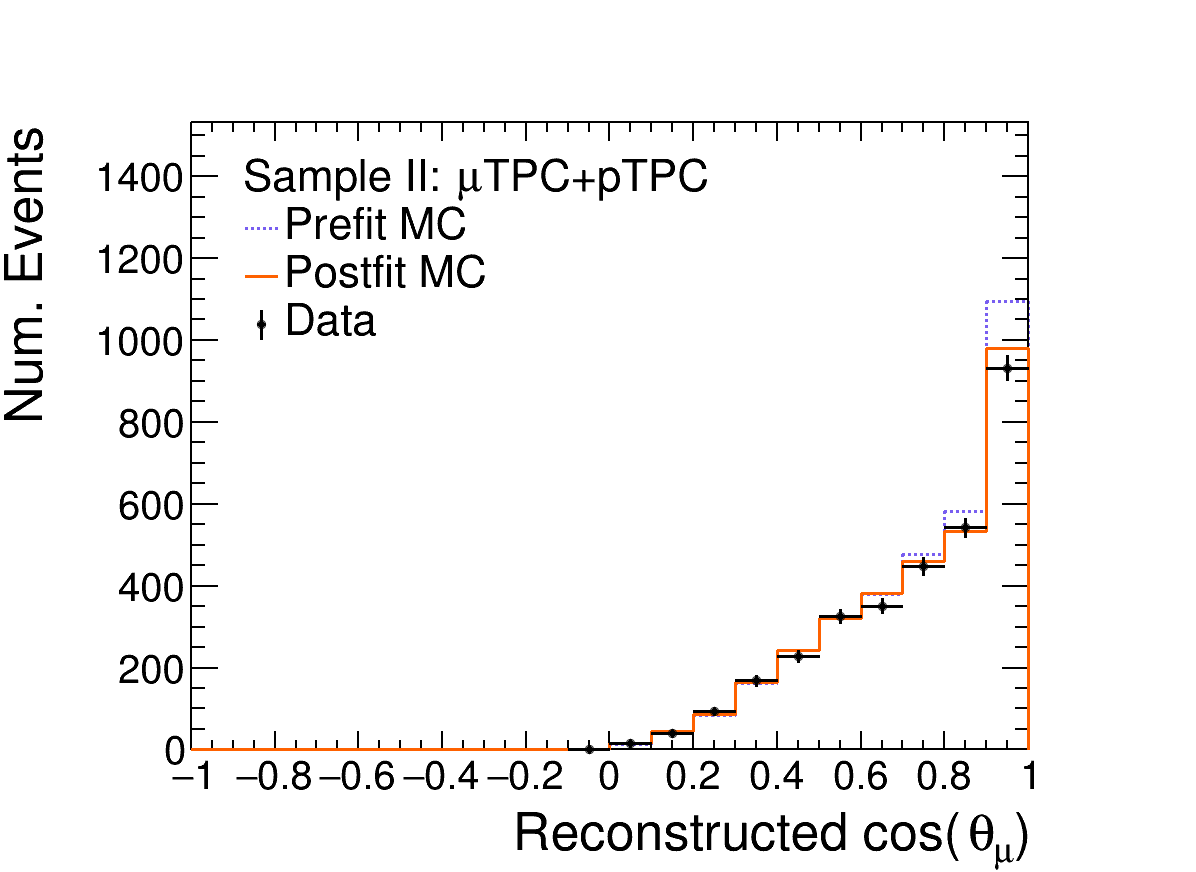}
    \includegraphics[width=0.35\textwidth,angle=0]{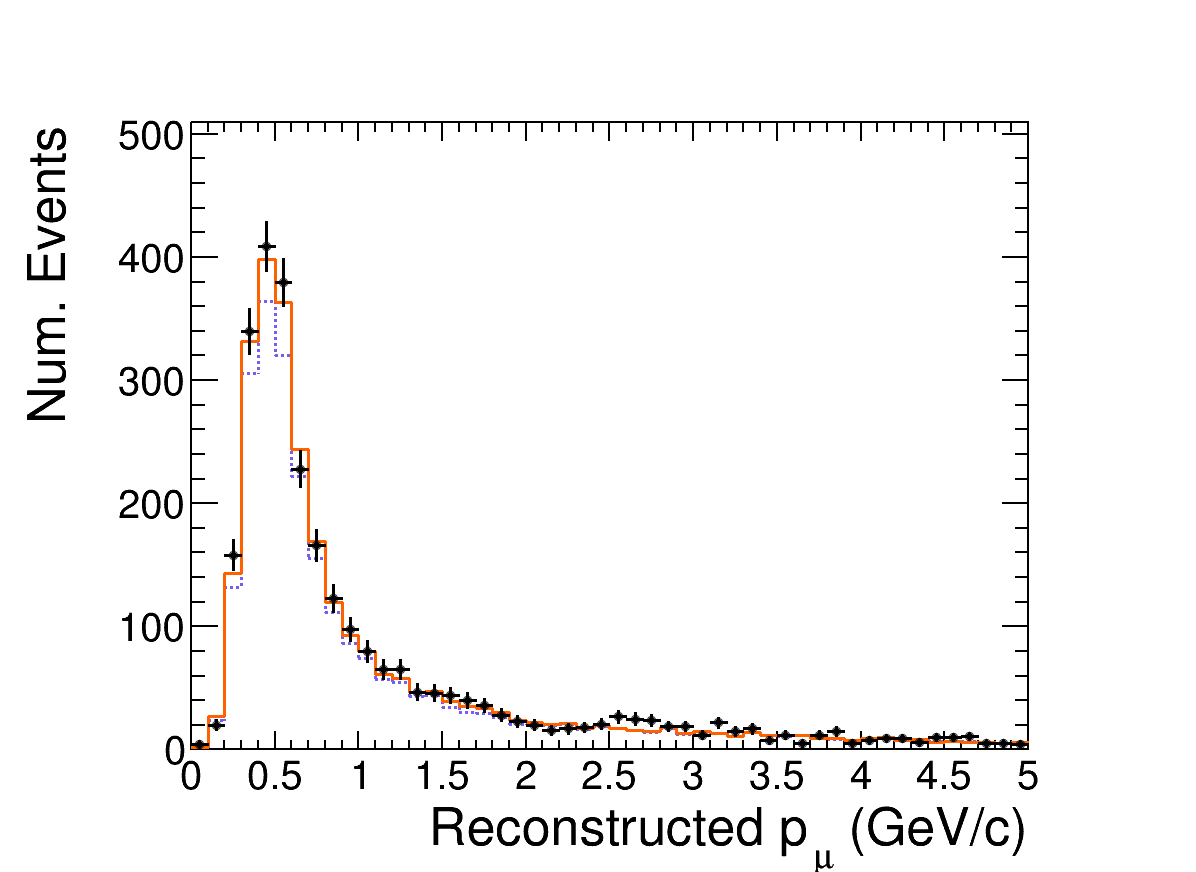}
    \includegraphics[width=0.35\textwidth,angle=0]{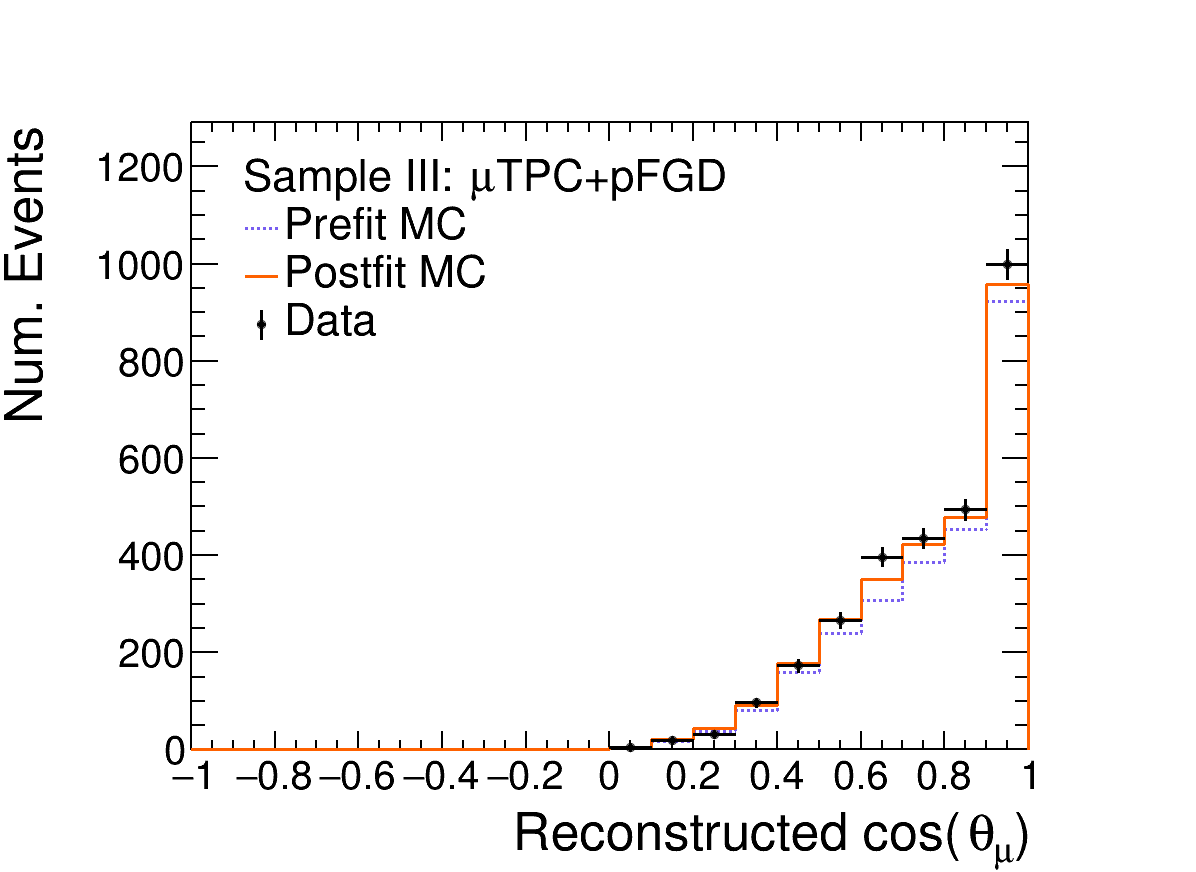}
    \includegraphics[width=0.35\textwidth,angle=0]{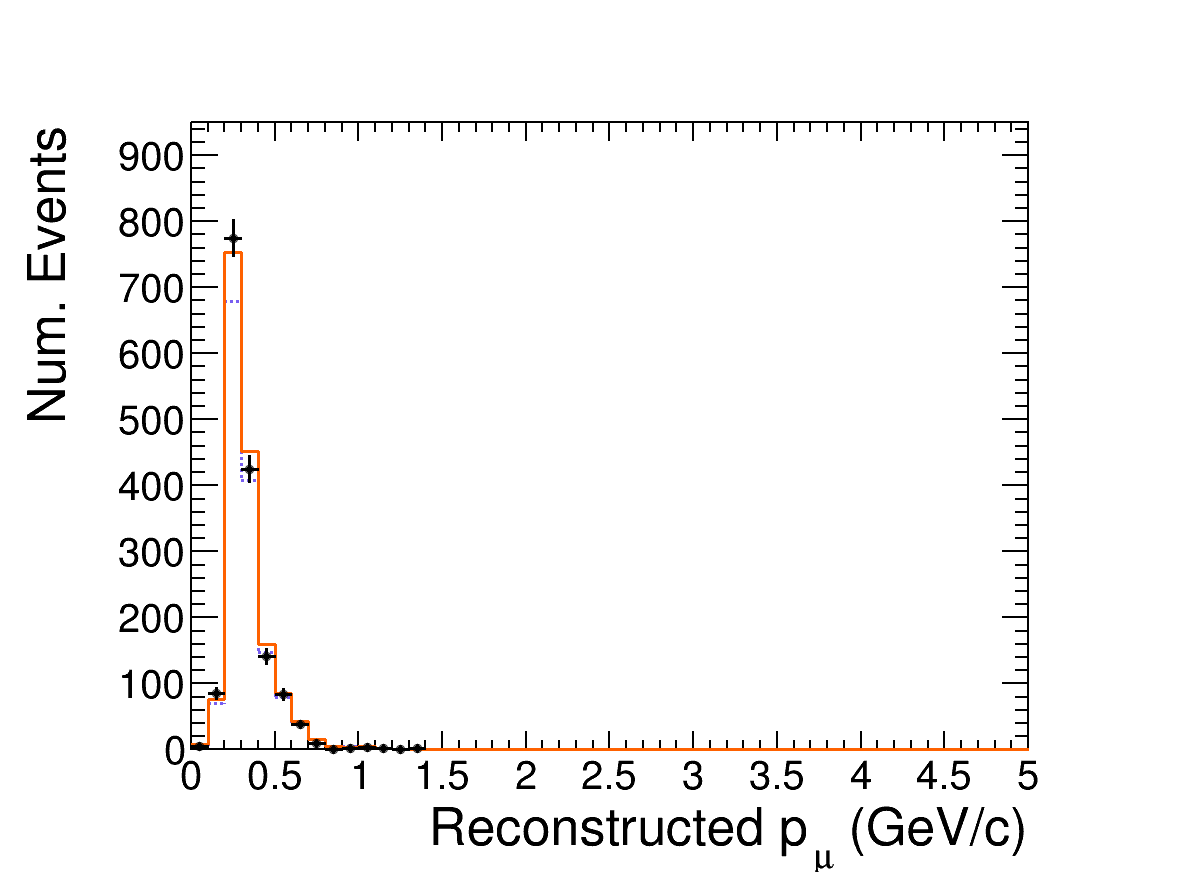}
    \includegraphics[width=0.35\textwidth,angle=0]{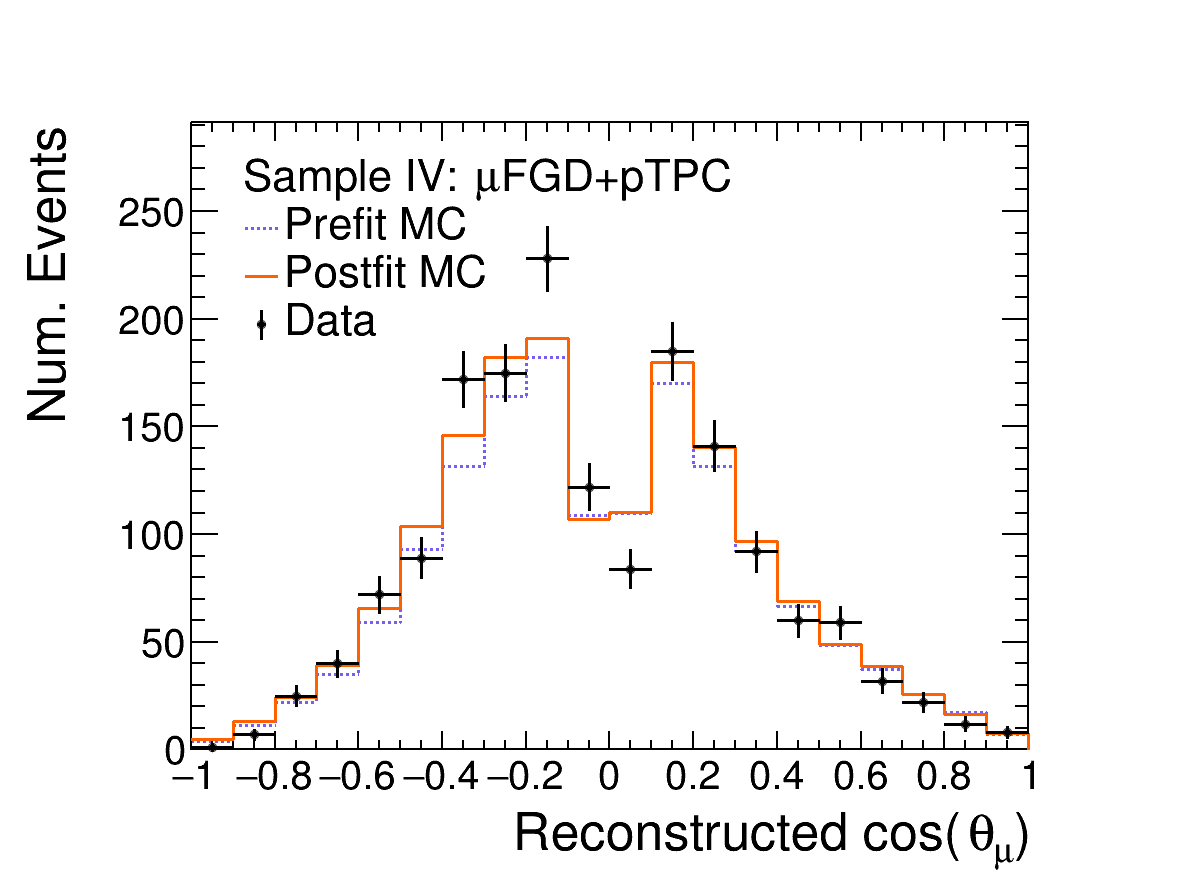}
    \includegraphics[width=0.35\textwidth,angle=0]{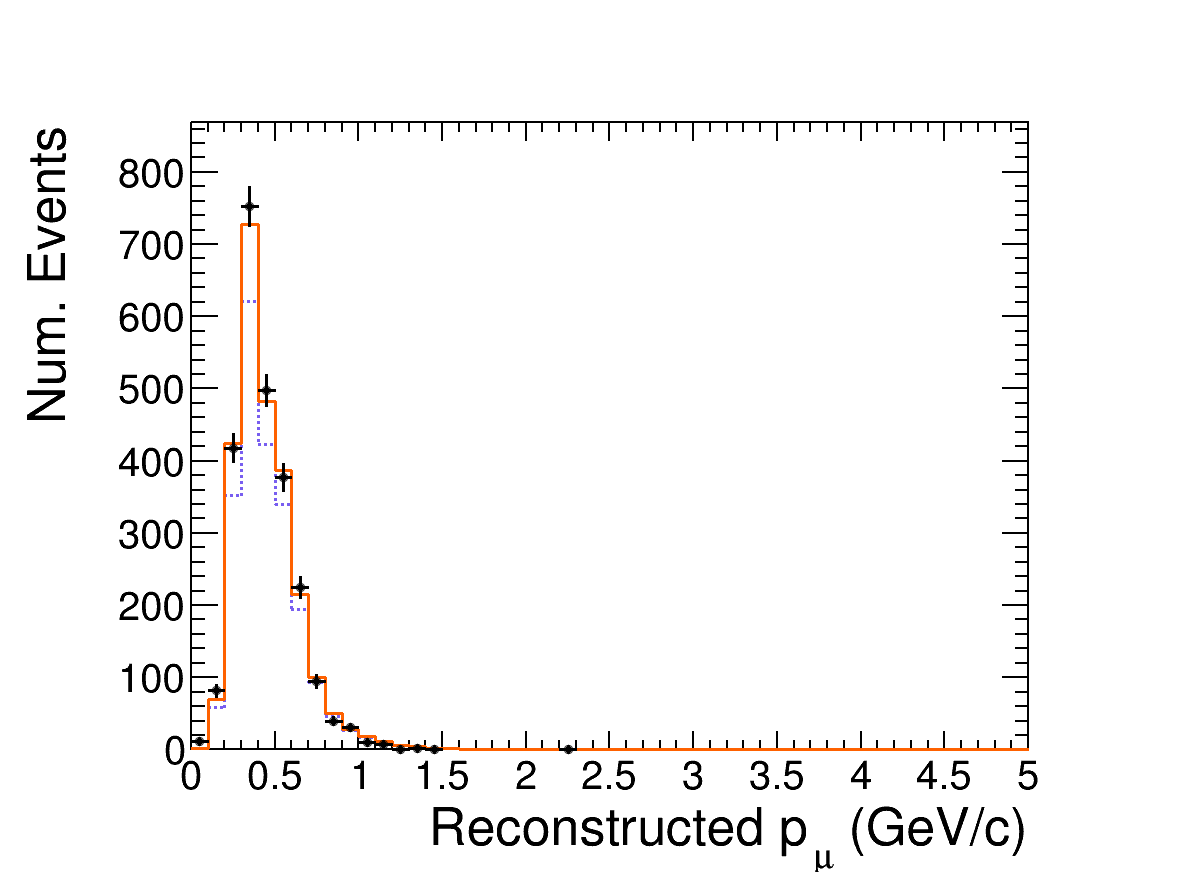}
    \includegraphics[width=0.35\textwidth,angle=0]{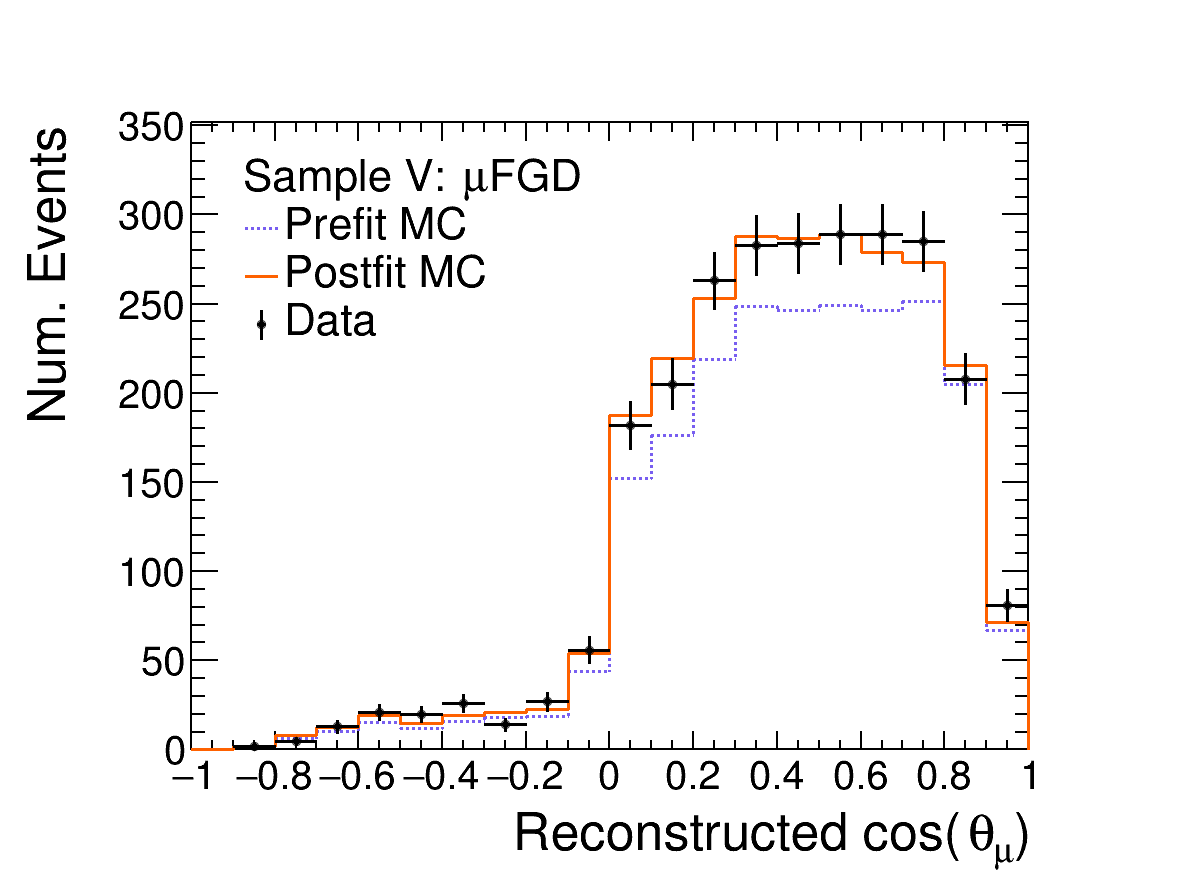}
    \caption{Event distribution for measured data and the pre-/post-fit MC prediction in reconstructed muon momentum and angle for the ND280 signal samples.}
    \label{fig:nd280_signal_prepost}
\end{figure*}

\begin{figure*}[hbt]
    \centering
    \includegraphics[width=0.35\textwidth,angle=0]{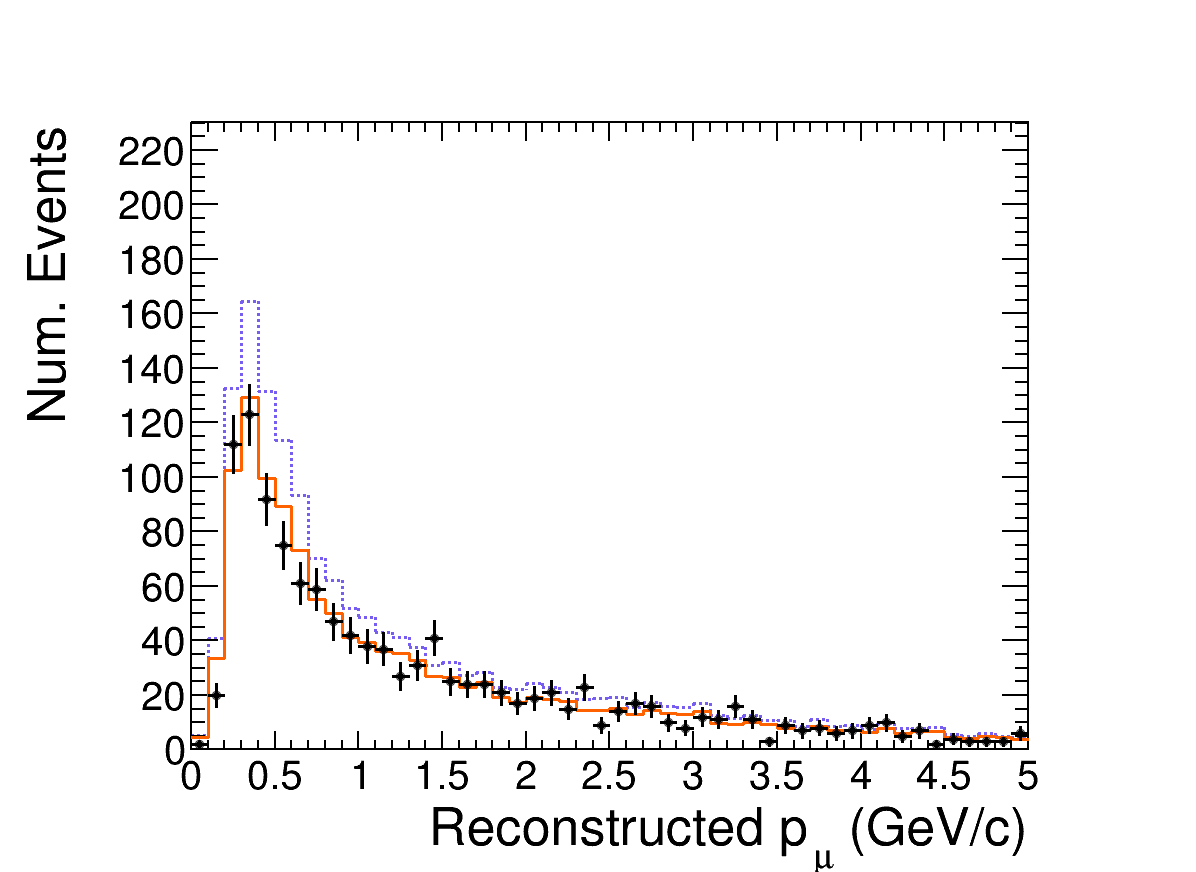}
    \includegraphics[width=0.35\textwidth,angle=0]{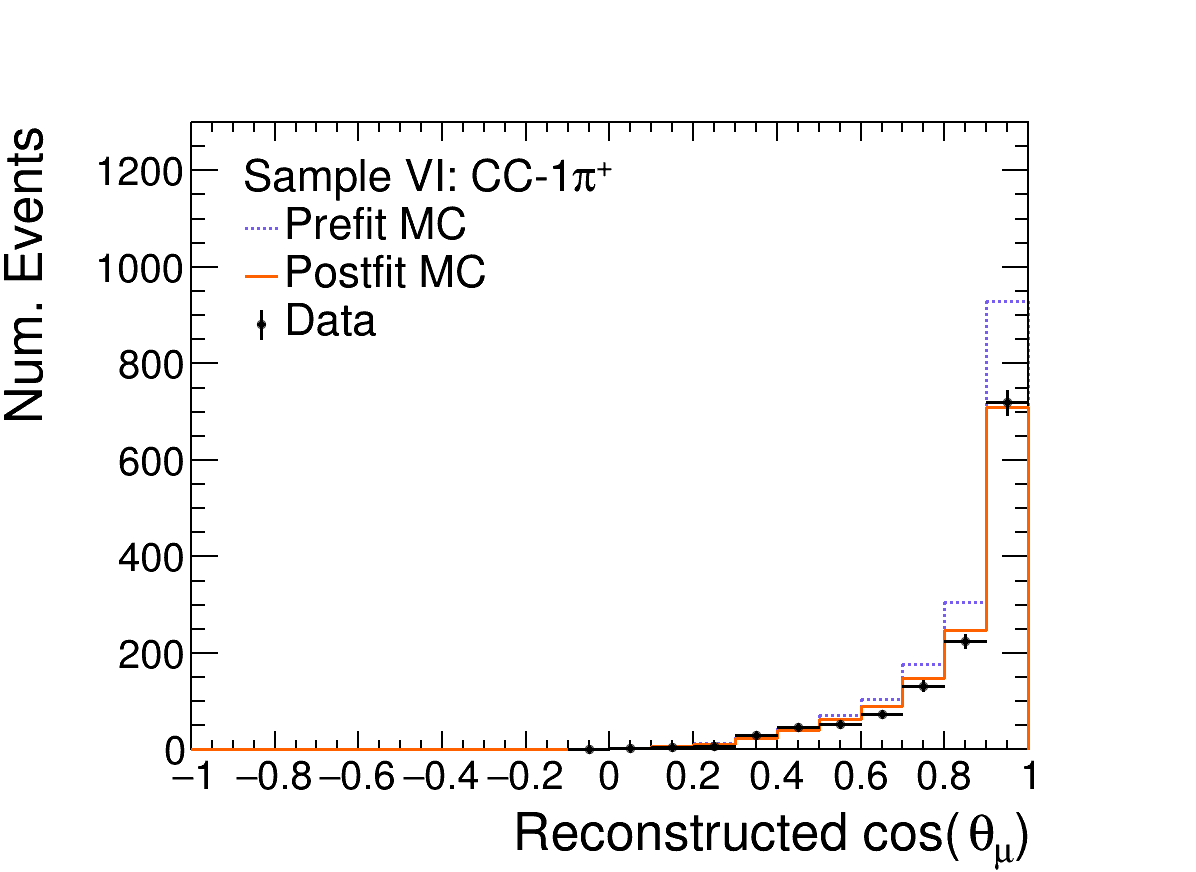}
    \includegraphics[width=0.35\textwidth,angle=0]{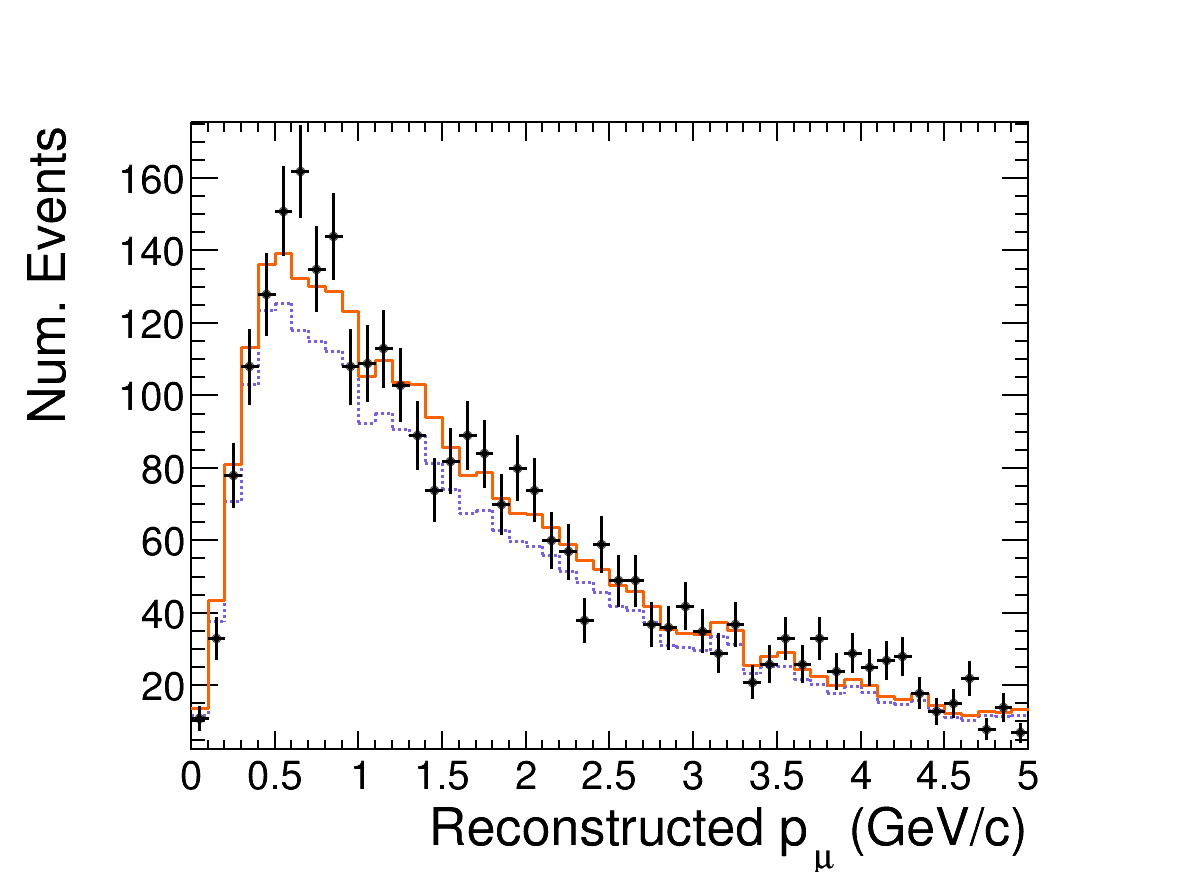}
    \includegraphics[width=0.35\textwidth,angle=0]{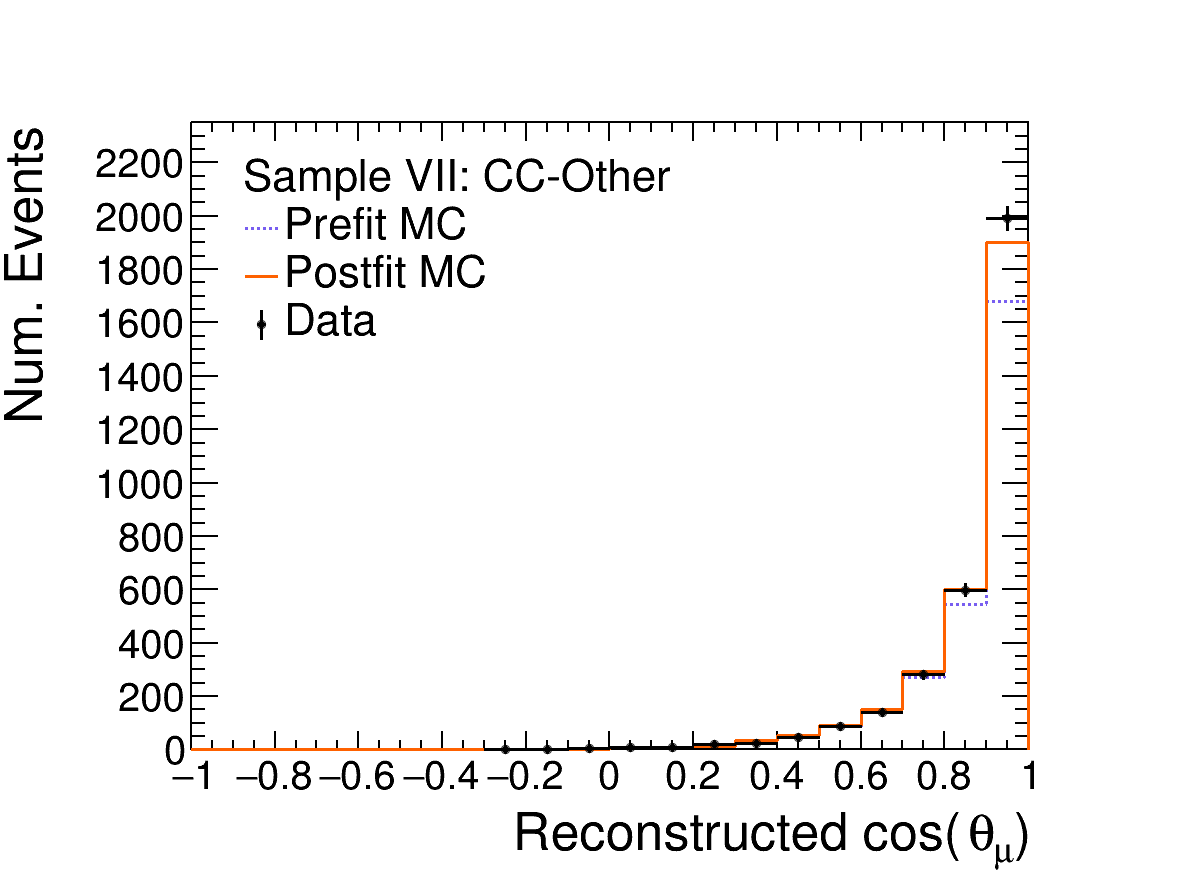}
    \includegraphics[width=0.35\textwidth,angle=0]{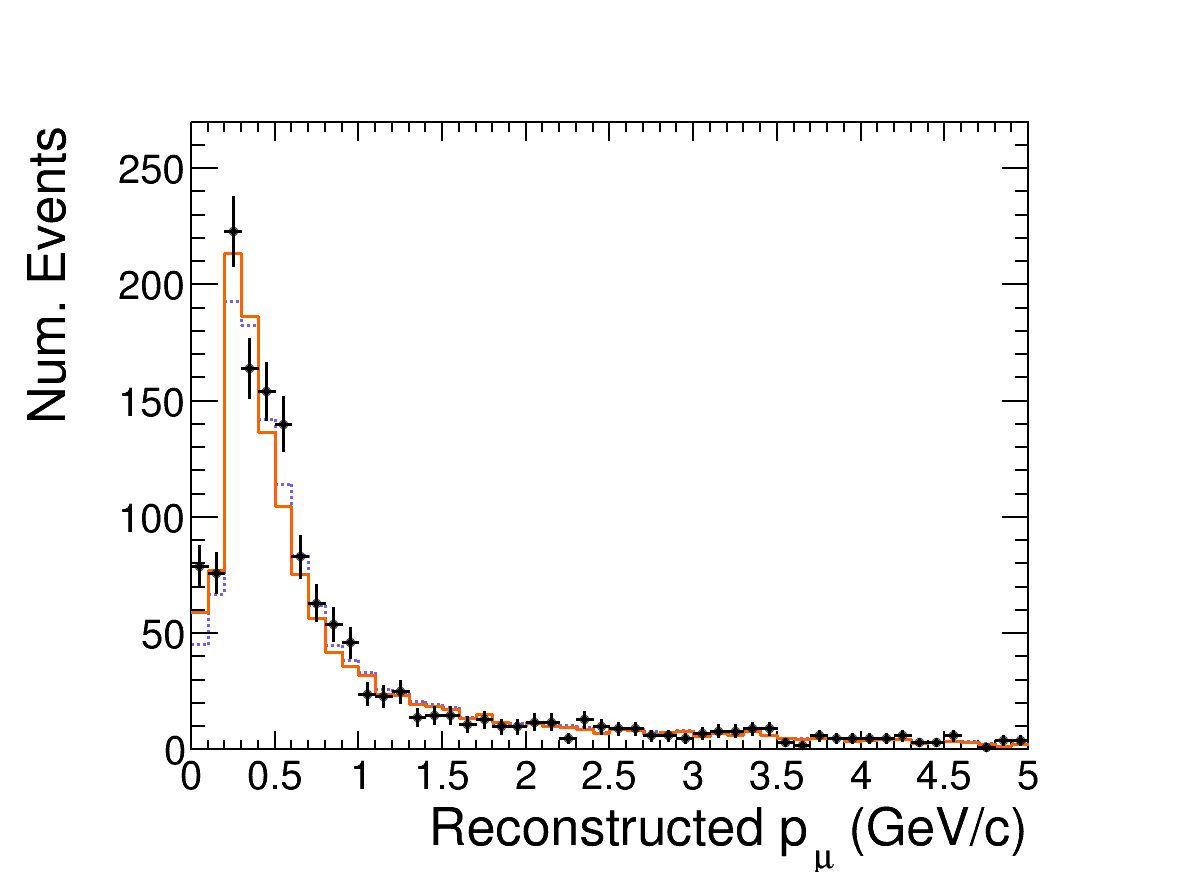}
    \includegraphics[width=0.35\textwidth,angle=0]{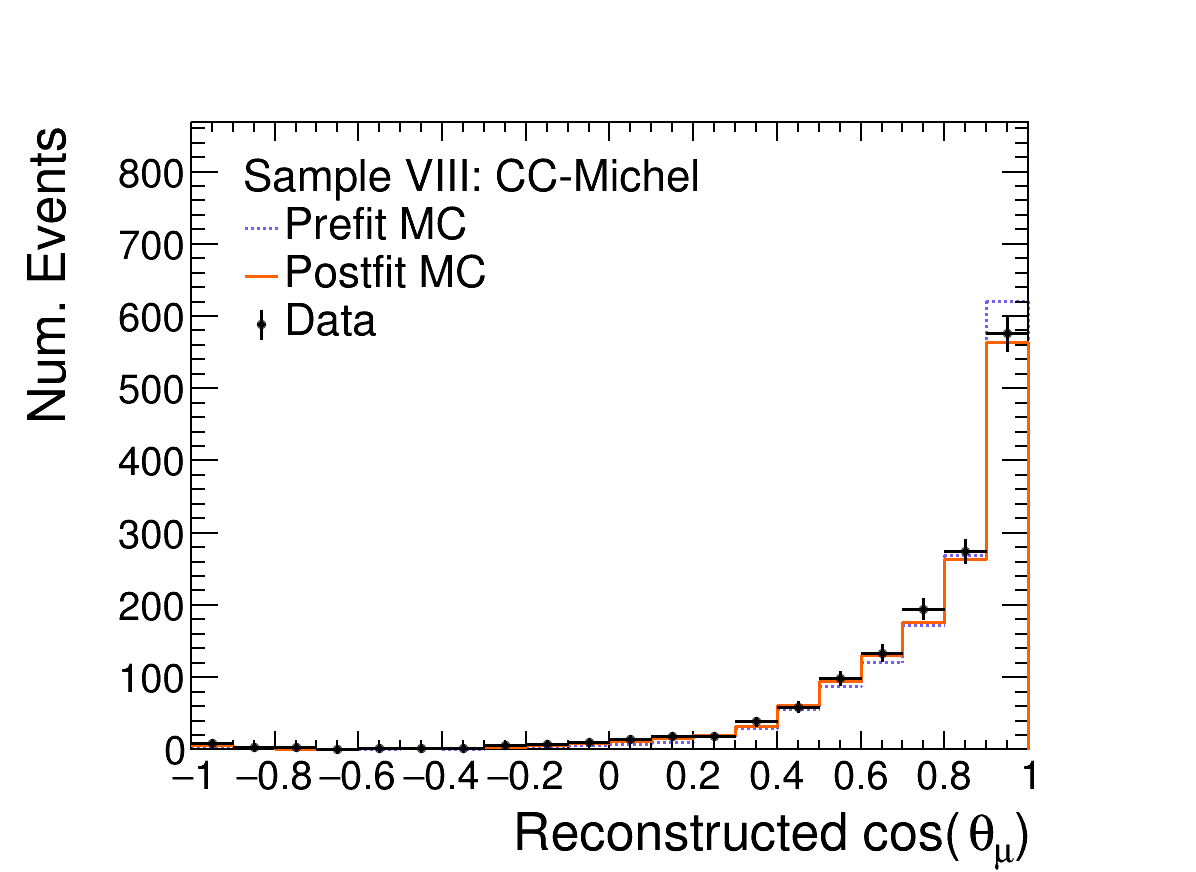}
    \caption{Event distribution for measured data and the pre-/post-fit MC prediction in reconstructed muon momentum and angle for the ND280 control samples.}
    \label{fig:nd280_control_prepost}
\end{figure*}

\begin{figure*}[hbt]
    \centering
    \includegraphics[width=0.35\textwidth,angle=0]{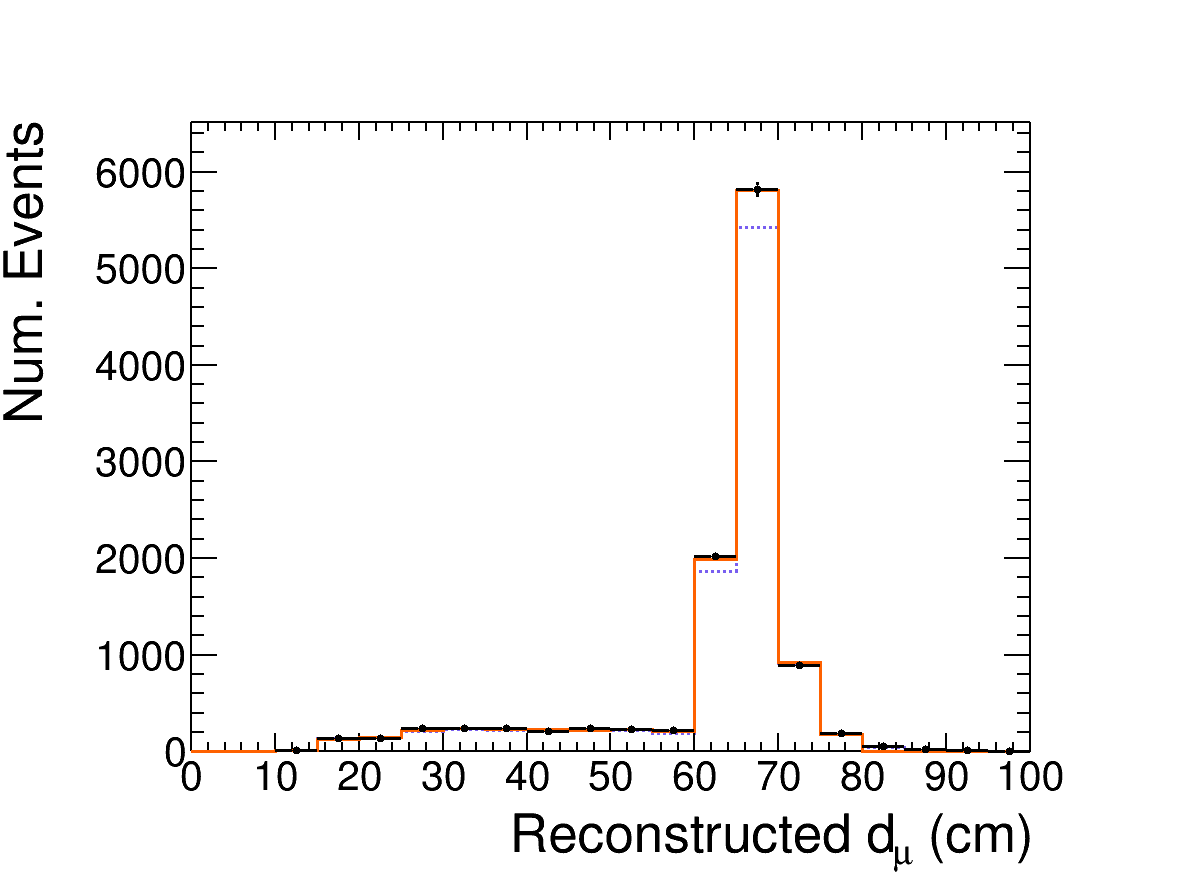}
    \includegraphics[width=0.35\textwidth,angle=0]{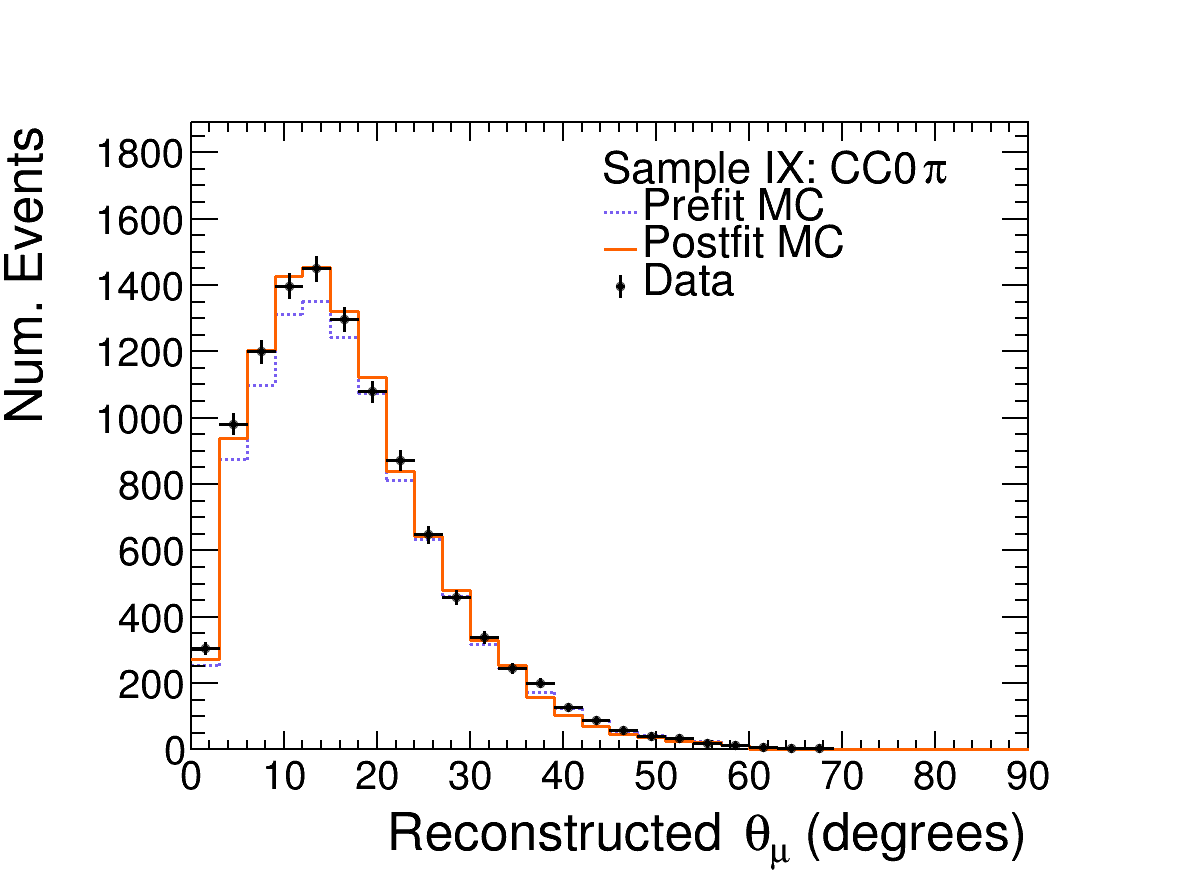}
    \includegraphics[width=0.35\textwidth,angle=0]{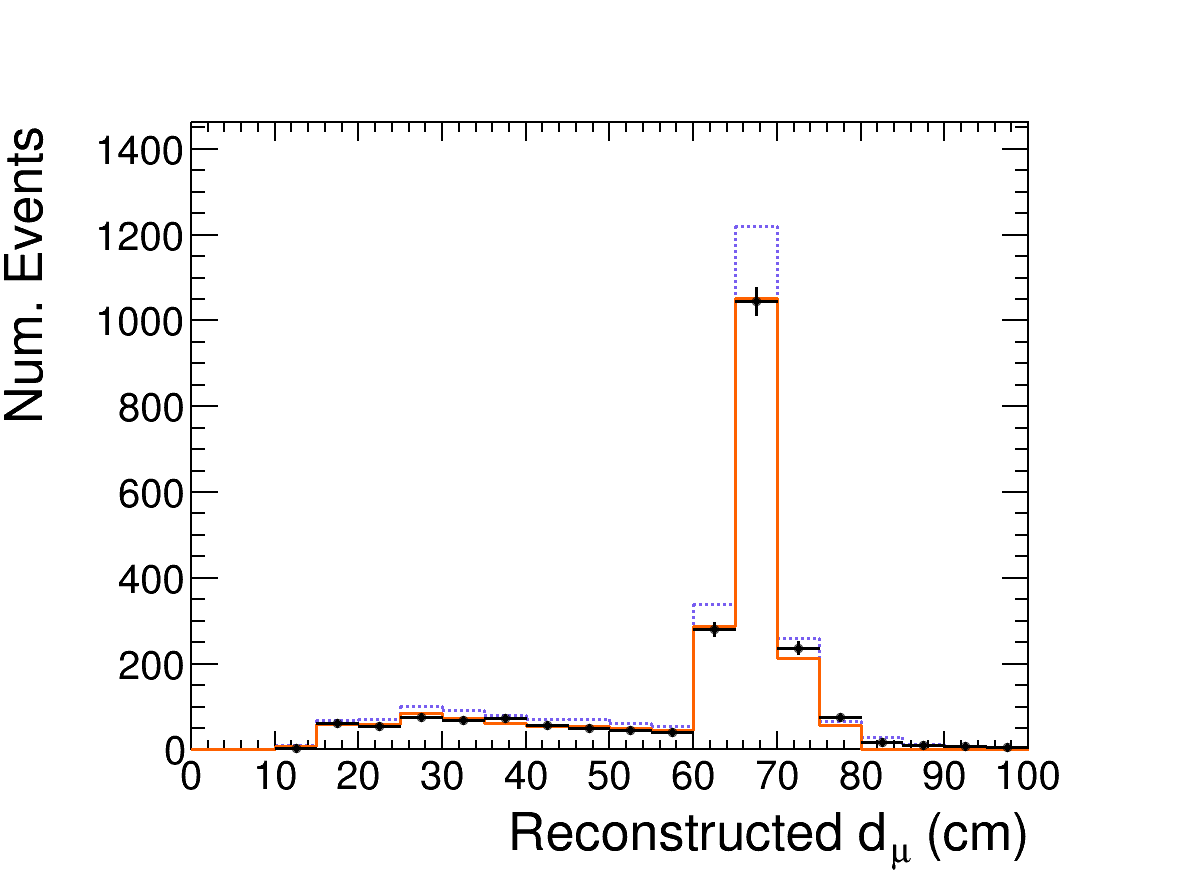}
    \includegraphics[width=0.35\textwidth,angle=0]{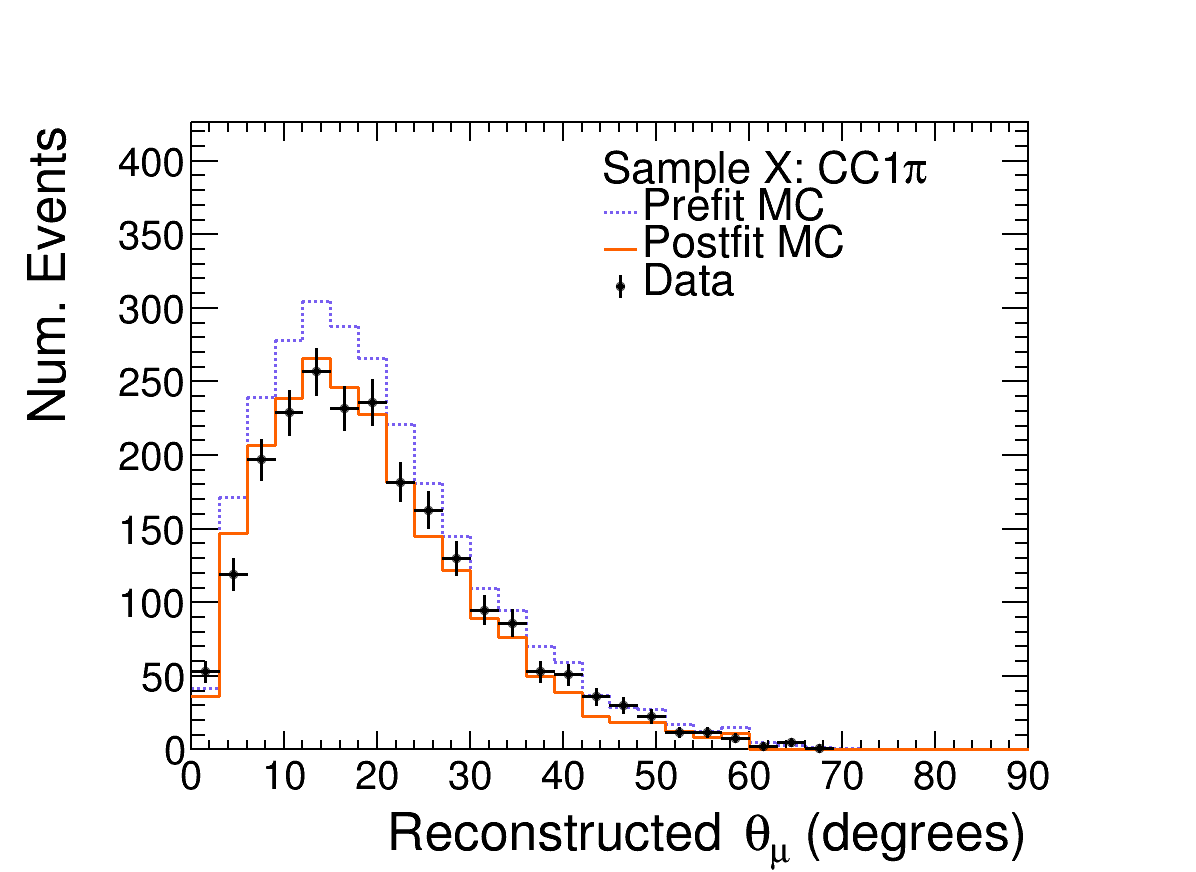}
    \caption{Event distribution for measured data and the pre-/post-fit MC prediction in reconstructed muon distance and angle for the INGRID signal sample (top) and control sample (bottom).}
    \label{fig:ingrid_prepost}
\end{figure*}

The distributions of the reconstructed events used to calculate the cross section are shown in Figs. \ref{fig:nd280_signal_prepost} and \ref{fig:nd280_control_prepost} for ND280 and Fig. \ref{fig:ingrid_prepost} for INGRID as a function of the muon kinematics compared to the nominal and post-fit MC predictions with data plus statistical errors overlaid. In general the fit is able to adjust the fit parameters as described in Sec. \ref{sec:fit} to match the observed data in the signal samples for both detectors within expected statistical fluctuations.

\begin{figure*}[hbt]
    \centering
    \includegraphics[width=0.35\textwidth,angle=0]{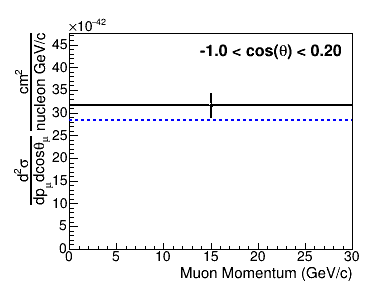}
    \includegraphics[width=0.35\textwidth,angle=0]{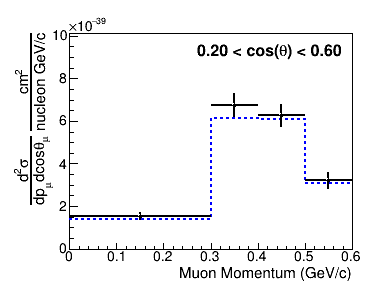}
    \includegraphics[width=0.35\textwidth,angle=0]{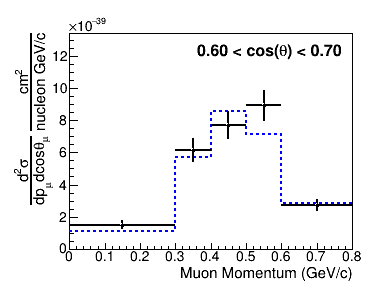}
    \includegraphics[width=0.35\textwidth,angle=0]{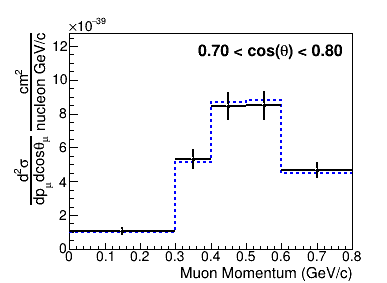}
    \includegraphics[width=0.35\textwidth,angle=0]{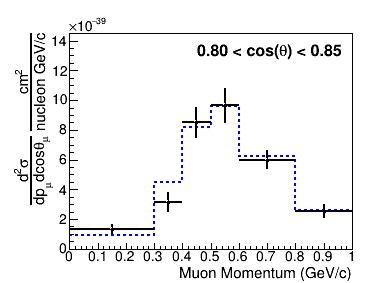}
    \includegraphics[width=0.35\textwidth,angle=0]{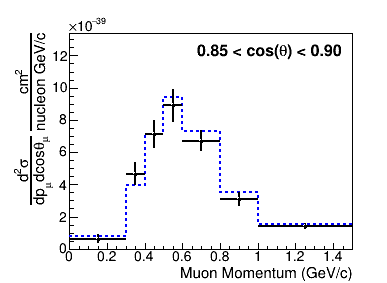}
    \includegraphics[width=0.35\textwidth,angle=0]{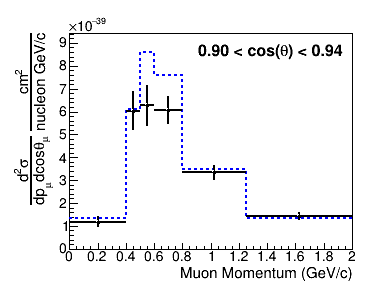}
    \includegraphics[width=0.35\textwidth,angle=0]{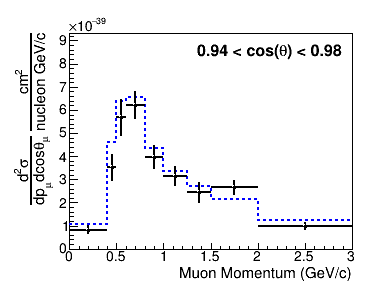}
    \includegraphics[width=0.35\textwidth,angle=0]{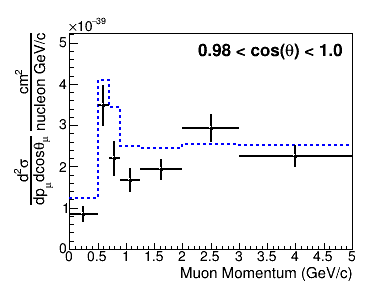}
    \includegraphics[width=0.35\textwidth,angle=0]{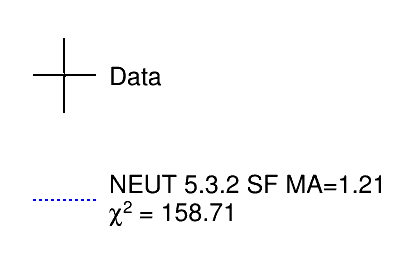}
    \caption{Extracted ND280 cross section as a function of muon momentum in angle bins compared to the nominal NEUT MC prediction. Note that the final bin extending to 30 GeV/\textit{c} has been omitted for clarity.}
    \label{fig:data_result_nd280}
\end{figure*}

\begin{figure*}[hbt]
    \centering
    \includegraphics[width=0.35\textwidth,angle=0]{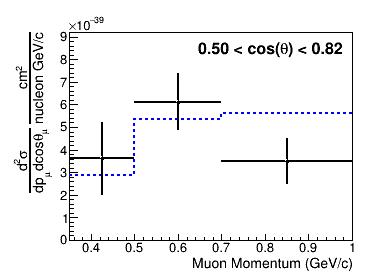}
    \includegraphics[width=0.35\textwidth,angle=0]{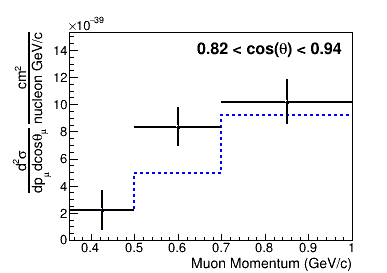}
    \includegraphics[width=0.35\textwidth,angle=0]{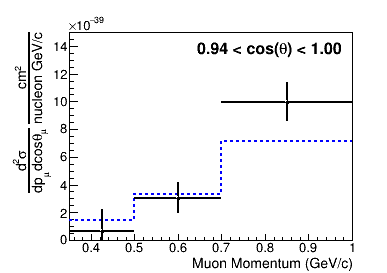}
    \includegraphics[width=0.35\textwidth,angle=0]{Figures/plot_legend_data.png}
    \caption{Extracted INGRID cross section as a function of muon momentum in angle bins compared to the nominal NEUT MC prediction. Note that the final bin extending to 30 GeV/\textit{c} has been omitted for clarity.}
    \label{fig:data_result_ingrid}
\end{figure*}

Nearly all systematic parameters had a post-fit value within their $1\sigma$ prior uncertainty, with the normalization parameters for DIS and multi-$\pi$ events pulled further than $1\sigma$ to accommodate the difference between the nominal MC and the data in the control samples as discussed previously in Sec. \ref{sec:strategy}. Additionally, the 2p2h shape parameter was pulled to the boundary corresponding to pushing the distribution of 2p2h events toward smaller momentum and energy transfer. The p-value for the fit to data compared to the nominal input model was calculated to be approximately 0.01, and was investigated further to verify the robustness of the result. The extracted cross sections for ND280 and INGRID are shown in Figs. \ref{fig:data_result_nd280} and \ref{fig:data_result_ingrid} respectively, and includes an additional uncertainty to account for possible missing freedom in the fit indicated by the poor p-value. A discussion of the poor p-value is provided in the following section (Sec. \ref{subsec:p-value}). Additional plots of the extracted cross section including the last momentum bin extending to 30 GeV/\textit{c} are shown in Appendix \ref{app:log_plots}.

\subsection{Discussion on the poor p-value}
\label{subsec:p-value}

As part of the data fit validation a p-value was calculated for the post-fit result as described in Sec. \ref{subsec:validation} to gauge the compatibility of the result with the nominal input model. The overall p-value for the original fit to data was calculated to be approximately 0.01, indicating that, relative to the input model, the observed data was an unlikely fluctuation. Based on the post-fit likelihood for each sample and the systematic contribution, the two main hypotheses were the pion modelling for background events (including the separation of samples for tracked versus Michel-tagged pions) and the high momentum bins considered in the fit. Several tests and different configurations of the fit were considered to see the impact on the result and to evaluate the need for an additional uncertainty.

To test the freedom of the pion model, two additional interaction model parameters were included to introduce more freedom at low pion and muon momentum where the majority of tension is present. One parameter was allowed to further alter the overall normalization of out-of-fiducial volume events (which are more likely at lower muon momentum), and the other changes the kinematics of the intermediate $\Delta$ decay in resonant pion interactions, which modifies the outgoing pion spectrum. In the fit to data, the rate of out-of-fiducial events was increased and the $\Delta$-decay parameter was moved to increase the rate of low-momentum pions (which correlates with low-momentum muons). However the total post-fit likelihood only improved by about 1\% compared to the data fit without these extra parameters, showing little sensitivity to the change in pion kinematics.

Next an alternative configuration of the fit combines the ND280 \CConepiplus and CC-Michel control samples (Samples VI and VIII respectively) in an attempt to quantify the role of the control samples on the background model, inspired by how these samples were used in previous T2K analyses \cite{PhysRevD.101.112001, PhysRevD.101.112004}. Overall the fit with the merged control samples gives similar results to the original fit, but with slightly better post-fit likelihood for the systematic contribution and a p-value of 0.024. The post-fit likelihood for the signal samples are nearly identical to the original showing little impact from the merged versus separated pion control samples. The merged sample performs slightly better compared to being separated, highlighting the tension between these samples as a driver of the poor p-value.

For the final test of the pion model, the fit was tested using only a single pion control sample (ND280 \CConepiplus, ND280 CC-Michel, INGRID \CConepi) at a time. The post-fit likelihood for the other samples largely were the same and the systematic contribution to the likelihood was about 10\% smaller (which is expected as fewer bins should in general easier to describe using the same systematic parameters) for each test case, indicating a robust result with respect to the pion control samples. The little to no effect on the signal samples across these tests is primarily due to the low contamination of pion events and the relative fraction of events in the signal versus control samples.

Another alternative configuration removes the high momentum bins extending to 30 GeV/\textit{c} for each ND280 sample (except the backward bin) as these bins were contributing a disproportional amount to the total $\chi^2$ relative to what would be expected from only statistical fluctuations. This was performed as an additional closure test to check that there was no pathological behavior regarding these bins that could be driving the p-value result. The fit performs moderately better as expected when removing problematic bins with a p-value of 0.065. Given the modest increase of the p-value and since the analysis is less sensitive to this kinematic region, these bins were considered not to be a problem.

\begin{figure}[hbt]
    \centering
    \includegraphics[width=0.40\textwidth,angle=0]{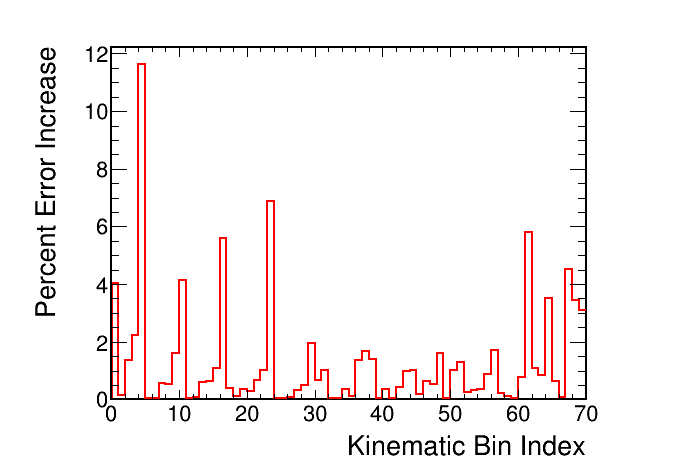}
    \caption{Percent error increase for each cross-section bin (flattened as a 1D array) from the the additional studies for the small p-value.}
    \label{fig:error_increase}
\end{figure}

The goal of these tests was to see if a large difference in p-value or post-fit likelihood was noticeable and if it required an additional uncertainty. The consistency of the post-fit results for the original fit to data and the variety of alternative fit configurations is evidence of a robust result for the analysis, and based on these additional investigations the small p-value is not an indication of a biased result and is driven by the measured data. Furthermore the slight variation in extracted cross section from the different tests were all well covered by the uncertainty on the original result. However, to be conservative, an additional uncorrelated uncertainty is added across all bins to cover the average difference between the original fit and the fit with merged ND280 control samples and the fit where the highest ND280 momentum bins were removed as these showed the largest difference in the extracted cross section. This additional uncertainty uses a similar method to the PDG to calculate how much the errors needed to be scaled to cover the different central values at 68\% confidence level by construction \cite{PDG}. The percent error increase for each kinematic bin is shown in Fig. \ref{fig:error_increase} where most bins had a ~2\% or less increase in error and the large increases generally correspond to the highest momentum bins.

\subsection{Model comparisons}
In this section, the measured cross sections are compared to a small selection of neutrino interaction models. The agreement between the measurement and models is quantified via the $\chi^2$ (relative to the number of degrees of freedom) as in Eq. \ref{eq:chisq} where $\sigma^{\text{true}}$ is replaced by the model prediction $\sigma^{\text{model}}$. The model predictions were produced by generating a sufficiently large number of neutrino interactions on hydrocarbon using the T2K on- and off-axis flux for each model. Generated events that satisfy the \CCzeropi signal definition are selected to calculate the flux integrated cross section for each detector. The comparisons between this result and following models were facilitated by the \textsc{nuisance} software package \cite{NUISANCE}.
\begin{enumerate}[(i)]
    \item \textsc{neut} 5.5.0 \cite{Hayato:2009zz, HAYATO2002171} : several different NEUT configurations were used: two using the Benhar \textit{et al.} \cite{BENHAR1994493} spectral function approach for the nuclear ground state with different values of the axial mass ($M_A^{QE} = 1.03$ and $M_A^{QE} = 1.21$ $\mathrm{GeV}/c^2$), and one using the Nieves \textit{et al.} model \cite{PhysRevC.83.045501} for the nuclear ground state and quasi-elastic interactions (which is based on a local Fermi gas). All NEUT models use the same pion production model, multi-nucleon model from Nieves \textit{et al.} \cite{PhysRevC.83.045501}, and FSI interaction model as described earlier in Sec. \ref{subsec:sim}. This is a newer version of NEUT compared to the nominal MC production (NEUT 5.3.2) used in the analysis.
    \item \textsc{genie} 3.0.6 \cite{Andreopoulos:2009rq, Andreopoulos:2015wxa}: two different GENIE configurations were used: the ``G18\_01a'' tune, which includes the Bodek--Ritchie modified relativistic Fermi gas (BRRFG) for the ground state nuclear model, the GENIE empirical multi-nucleon model \cite{GENIE_MEC}, the Rein--Sehgal model for pion production, and the hA model for FSI; the ``G18\_10b'' tune, which uses the local Fermi gas (LFG) for the nuclear model, the Nieves \textit{et al.} multi-nucleon model, the Berger--Sehgal model for pion production, and the hN model for FSI. The cross-section models for both of these configurations are tuned using a preliminary version of the fit to bubble chamber data as described in Ref. \cite{GENIE:2021zuu}.
    \item \textsc{nuwro} 21.09 \cite{GOLAN2012499}: several NuWro configurations were used: one using a spectral function approach for the nuclear ground state with the Nieves multi-nucleon model, and multiple using a local Fermi gas (LFG) for the nuclear ground state with a different available multi-nucleon model: SuSAv2 \cite{PhysRevD.94.093004}, Nieves \textit{et al.}, or Martini \textit{et al.} \cite{PhysRevC.80.065501}. All configurations use the same models for pion production and FSI interactions and set $M_A^{QE} = 1.03$ $\mathrm{GeV}/c^2$.
\end{enumerate}
Tab. \ref{tab:chisq_table} contains the $\chi^2$ for the joint result and the $\chi^2$ considering a single detector. The joint $\chi^2$ for a given model comparison will be slightly different from the sum of the individual $\chi^2$ values due to the correlations between the detectors. Overall the generator predictions do not describe the data well according to $\chi^2 / N$ values ranging from approximately 1.5 to 3.0 for $N=70$ degrees of freedom (measured bins) with 58 ND280 bins and 12 INGRID bins. The larger $\chi^2$ values for ND280 compared to INGRID is primarily due to ND280 having a finer binning than INGRID. A unique aspect of this measurement is the ability to compare how a given model does for ND280 and INGRID individually and how the full result with correlations between ND280 and INGRID is better or worse than the naive sum. For example, the two GENIE models used in this paper show opposite behavior: one model describes ND280 better than the other but does worse describing INGRID and vice versa.

\begin{figure*}[hbt]
    \centering
    \includegraphics[width=0.35\textwidth,angle=0]{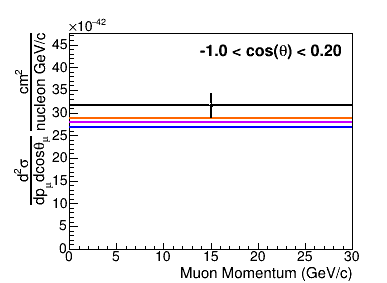}
    \includegraphics[width=0.35\textwidth,angle=0]{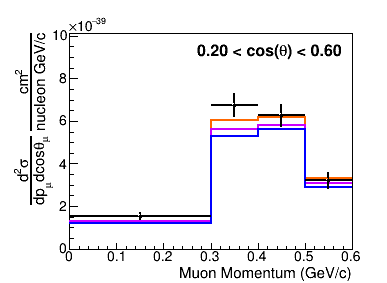}
    \includegraphics[width=0.35\textwidth,angle=0]{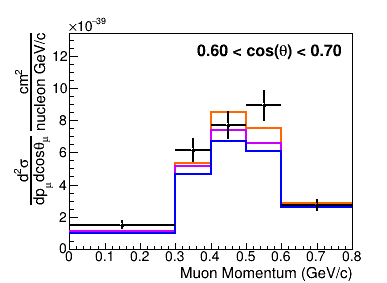}
    \includegraphics[width=0.35\textwidth,angle=0]{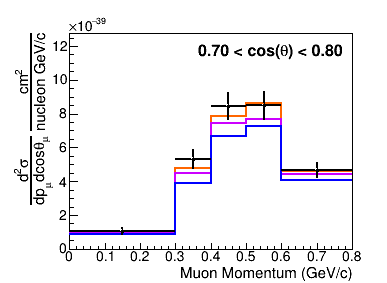}
    \includegraphics[width=0.35\textwidth,angle=0]{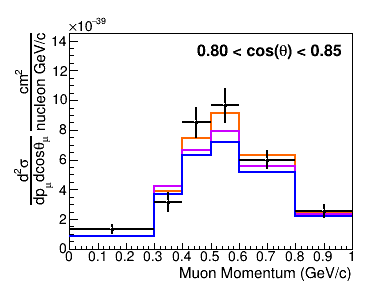}
    \includegraphics[width=0.35\textwidth,angle=0]{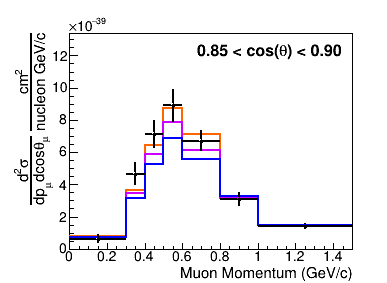}
    \includegraphics[width=0.35\textwidth,angle=0]{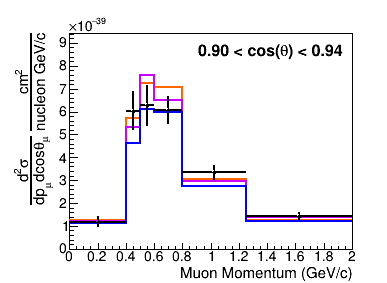}
    \includegraphics[width=0.35\textwidth,angle=0]{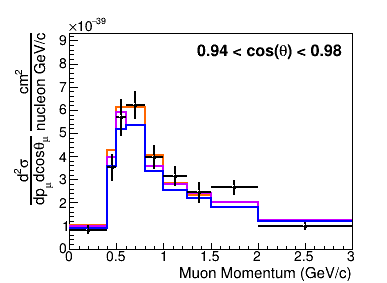}
    \includegraphics[width=0.35\textwidth,angle=0]{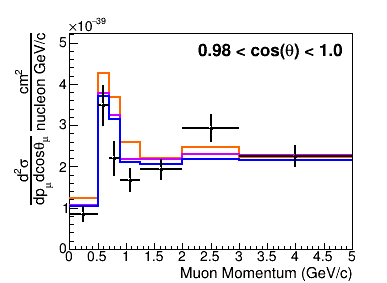}
    \includegraphics[width=0.35\textwidth,angle=0]{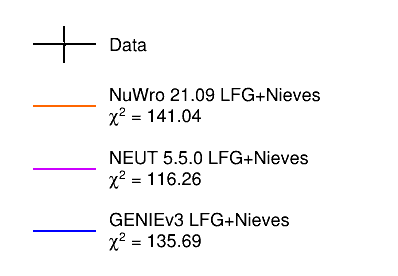}
    \caption{Extracted ND280 cross section as a function of muon momentum in angle bins compared to NEUT, GENIE, and NuWro all using a similar model. Note that the final bin extending to 30 GeV/\textit{c} has been omitted for clarity.}
    \label{fig:generator_comp_nd280}
\end{figure*}

\begin{figure*}[hbt]
    \centering
    \includegraphics[width=0.35\textwidth,angle=0]{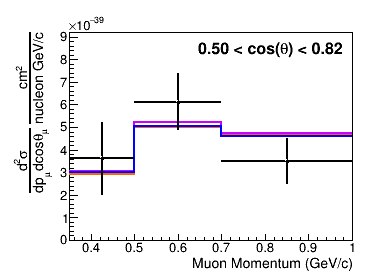}
    \includegraphics[width=0.35\textwidth,angle=0]{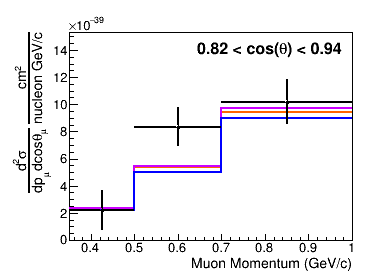}
    \includegraphics[width=0.35\textwidth,angle=0]{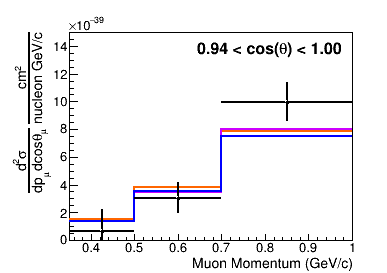}
    \includegraphics[width=0.35\textwidth,angle=0]{Figures/plot_legend_generator_comp.png}
    \caption{Extracted INGRID cross section as a function of muon momentum in angle bins compared to NEUT, GENIE, and NuWro all using a similar model. Note that the final bin extending to 30 GeV/\textit{c} has been omitted for clarity.}
    \label{fig:generator_comp_ingrid}
\end{figure*}

Figures \ref{fig:generator_comp_nd280} and \ref{fig:generator_comp_ingrid} show the data compared to each generator's implementation of a LFG nuclear ground state plus the Nieves \textit{et al.} multi-nucleon model. The pion production models will be roughly similar, however the FSI treatment is different between each prediction. The generators mostly differ in the normalization for the ND280 bins at the middle momentum range around the T2K flux peak energy of 0.6 GeV, however all show very similar INGRID predictions. For this particular set of models and generator versions, NEUT performs notably better than GENIE and NuWro.

\begin{figure*}[hbt]
    \centering
    \includegraphics[width=0.35\textwidth,angle=0]{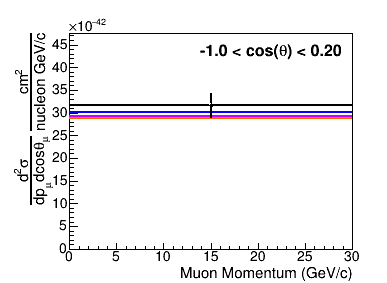}
    \includegraphics[width=0.35\textwidth,angle=0]{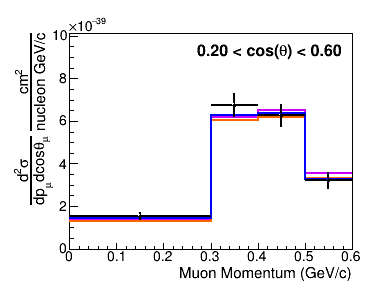}
    \includegraphics[width=0.35\textwidth,angle=0]{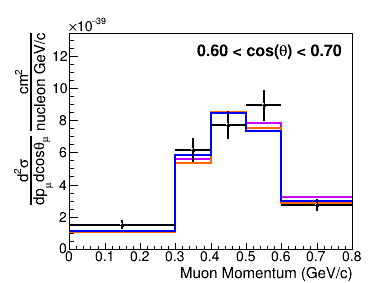}
    \includegraphics[width=0.35\textwidth,angle=0]{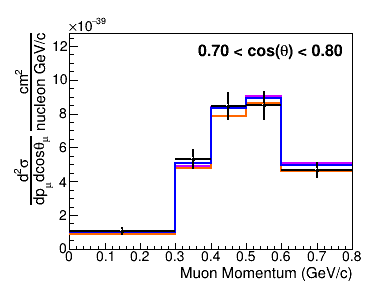}
    \includegraphics[width=0.35\textwidth,angle=0]{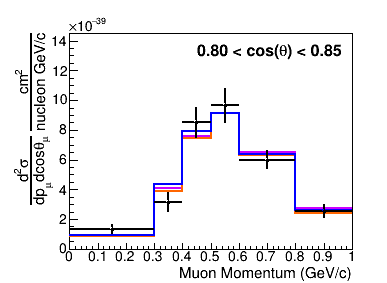}
    \includegraphics[width=0.35\textwidth,angle=0]{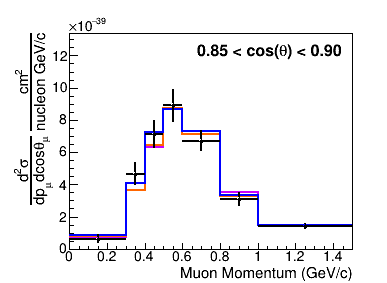}
    \includegraphics[width=0.35\textwidth,angle=0]{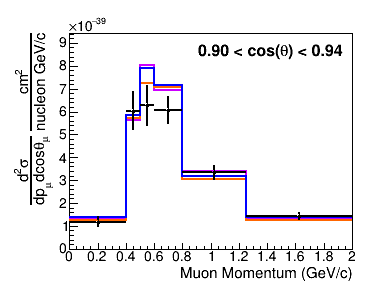}
    \includegraphics[width=0.35\textwidth,angle=0]{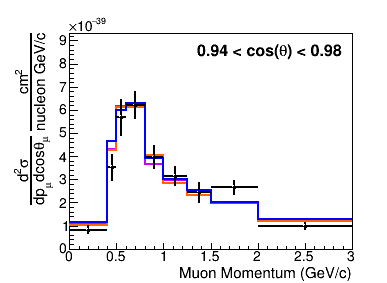}
    \includegraphics[width=0.35\textwidth,angle=0]{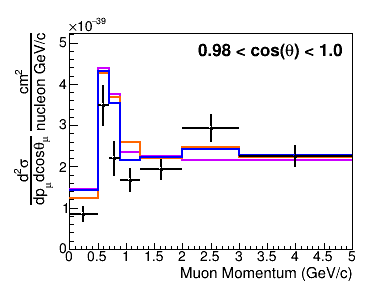}
    \includegraphics[width=0.35\textwidth,angle=0]{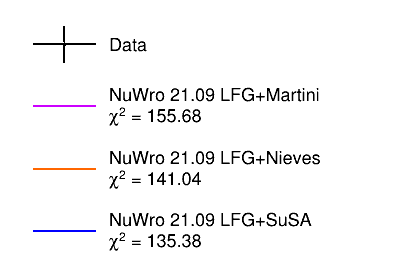}
    \caption{Extracted ND280 cross section as a function of muon momentum in angle bins compared to several different multi-nucleon predictions using NuWro and the same LFG ground state. Note that the final bin extending to 30 GeV/\textit{c} has been omitted for clarity.}
    \label{fig:multinucleon_comp_nd280}
\end{figure*}

\begin{figure*}[hbt]
    \centering
    \includegraphics[width=0.35\textwidth,angle=0]{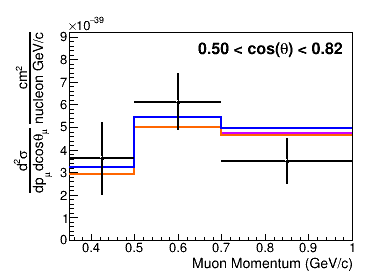}
    \includegraphics[width=0.35\textwidth,angle=0]{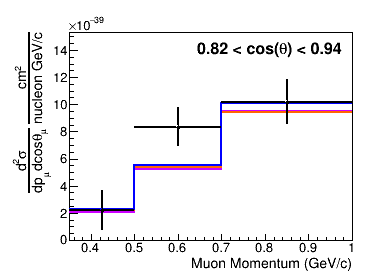}
    \includegraphics[width=0.35\textwidth,angle=0]{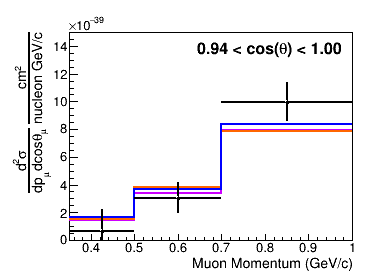}
    \includegraphics[width=0.35\textwidth,angle=0]{Figures/plot_legend_2p2h_comp.png}
    \caption{Extracted INGRID cross section as a function of muon momentum in angle bins compared to several different different multi-nucleon predictions using NuWro and the same LFG ground state. Note that the final bin extending to 30 GeV/\textit{c} has been omitted for clarity.}
    \label{fig:multinucleon_comp_ingrid}
\end{figure*}

Figures \ref{fig:multinucleon_comp_nd280} and \ref{fig:multinucleon_comp_ingrid} show the data compared to several different multi-nucleon predictions using NuWro with a LFG ground state and the same parameters for all other aspects of the generation. The predictions are very similar between the different multi-nucleon models as implemented in NuWro, with a slight preference for SuSAv2.

\begin{table}[htb]
    \centering
    \begin{tabular}{l|r|r|r}
         Model & ND280 & INGRID & Joint \\
         \hline
         Nominal MC (NEUT)  & 136.34 & 18.21 & 158.71  \\ 
         NEUT LFG+Nieves    & 106.46 & 11.46 & 116.26  \\
         NEUT SF+Nieves $M_A=1.03$ & 194.88 & 14.36 & 209.18  \\
         NEUT SF+Nieves $M_A=1.21$ & 158.71 &  9.98 & 170.93  \\
         NuWro SF+Nieves    & 122.74 & 15.68 & 137.02  \\
         NuWro LFG+Nieves   & 125.88 & 12.75 & 141.04  \\
         NuWro LFG+SuSAv2   & 121.57 & 11.13 & 135.38  \\
         NuWro LFG+Martini  & 138.86 & 12.46 & 155.68  \\
         GENIE BRRFG+EmpMEC & 141.40 & 12.80 & 156.05  \\
         GENIE LFG+Nieves   & 125.50 & 14.45 & 135.69  \\
         \hline
    \end{tabular}
    \caption{Agreement between this result and the various model comparisons as measured by the $\chi^2$ for both the joint result and when compared to each detector individually. ND280 has 58 cross-section bins and INGRID has 12 cross-section bins for a combined 70 total bins.}
    \label{tab:chisq_table}
\end{table}

\subsection{Comparison to previous result}

\begin{figure*}[hbt]
    \centering
    \includegraphics[width=0.35\textwidth,angle=0]{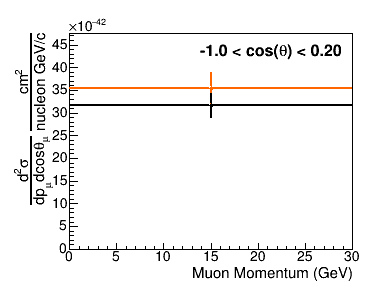}
    \includegraphics[width=0.35\textwidth,angle=0]{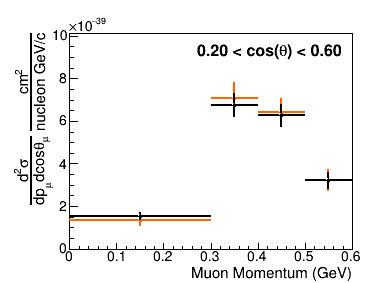}
    \includegraphics[width=0.35\textwidth,angle=0]{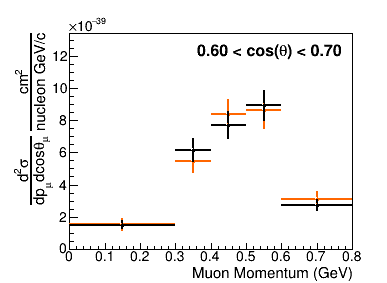}
    \includegraphics[width=0.35\textwidth,angle=0]{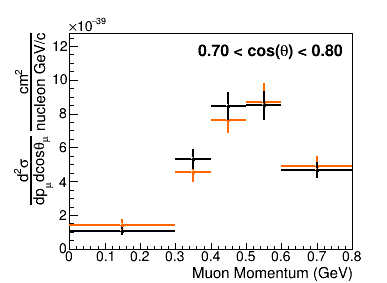}
    \includegraphics[width=0.35\textwidth,angle=0]{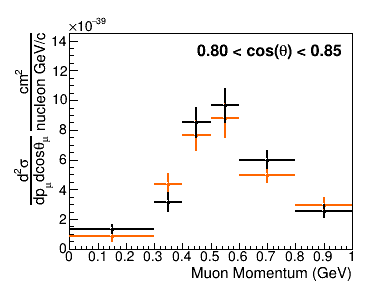}
    \includegraphics[width=0.35\textwidth,angle=0]{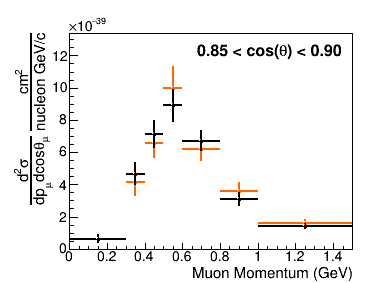}
    \includegraphics[width=0.35\textwidth,angle=0]{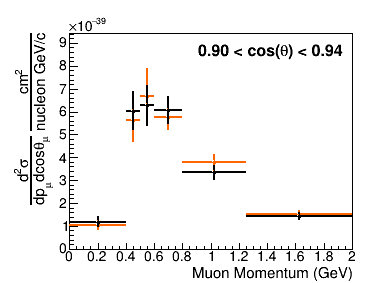}
    \includegraphics[width=0.35\textwidth,angle=0]{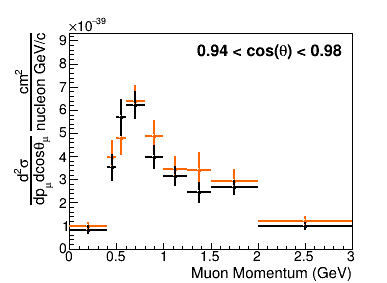}
    \includegraphics[width=0.35\textwidth,angle=0]{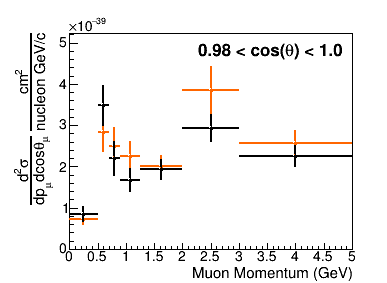}
    \includegraphics[width=0.35\textwidth,angle=0]{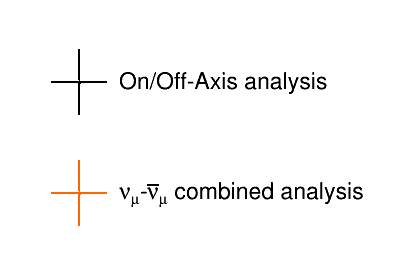}
    \caption{Extracted ND280 cross section from this analysis compared to the neutrino analysis in Ref. \cite{PhysRevD.101.112001}. The final momentum bin extending to 30 GeV/\textit{c} has been omitted for clarity.}
    \label{fig:t2k_result_comparison}
\end{figure*}

This analysis uses the same binning for the ND280 samples as the \CCzeropi analysis from Ref. \cite{PhysRevD.101.112001}, allowing for a direct comparison between the results. The main differences are the inclusion of more data for this result (T2K Run 8), increasing the neutrino-mode sample statistics by approximately a factor of two, and the configuration of the fits, where this analysis did a neutrino-only joint fit of on- and off-axis data and Ref. \cite{PhysRevD.101.112001} did a joint anti-neutrino and neutrino fit with only off-axis data. Both results are shown in Fig. \ref{fig:t2k_result_comparison} with the final bin extending to 30 GeV/\textit{c} omitted for clarity. The majority of the bins agree within their $1\sigma$ error bars, and show a trend for this result to report a smaller cross section at medium to higher muon momentum (above 0.8 GeV/\textit{c}) that is more pronounced at more forward-going angles. Additionally, the high fluctuation in the cross section seen in the 2.0 to 3.0 GeV/\textit{c} momentum bin in the most forward angle bin ($0.98 < \cos{\theta_{\mu}} < 1.00$) is now smaller and closer in value to the neighboring bins compared to previous T2K \CCzeropi results.

\section{Conclusions}

This paper presents the first measurement of neutrino interactions without pions in the final state using multiple energy spectra at T2K with the on- and off-axis near detectors. The analysis was performed using a joint maximum likelihood fit with signal and control samples from both detectors to minimize the background uncertainties and perform the unfolding from reconstructed to truth variables. The results include the cross-section measurement at each detector and the correlation between them, providing additional information compared to the individual measurements. Generator models continue to struggle to describe the data, and for the comparisons performed in this paper, the NEUT implementation of a LFG ground state plus the Nieves \textit{et al.} multi-nucleon model has the smallest $\chi^{2} / N \sim 1.66$, which is still very poor agreement.

This analysis is the next step in combined measurements at T2K and further opens up the possibility for more complex combinations of analyses. Only neutrino-mode data was considered for this first analysis using multiple energy spectra, but future analyses will include the anti-neutrino data. Additionally, future versions of this analysis will include the T2K replica target measurements from NA61/SHINE \cite{Berns:2018tap} for the flux modeling, and updates of the neutrino interaction model.

Since this analysis was finalized, the WAGASCI \cite{Kin_2017} and BabyMIND \cite{Antonova_2017} detectors were added to the T2K near detector hall at an off-axis angle of 1.5 degrees and have started taking data. WAGASCI/BabyMIND data could be used to extend this analysis to use three different energy spectra, and provide additional statistics for interactions on both hydrocarbon and water. The upcoming J-PARC accelerator upgrade \cite{jparc_beam_tdr} will increase the beam power providing a higher rate of data taking. Finally, the ND280 upgrade \cite{ndup_tdr} will increase the detector capabilities, providing increased angle coverage, better low momentum tracking, and additional target mass.

The data release for this analysis is hosted at Ref. \cite{data_release}. It contains the best-fit cross-section results, the nominal MC prediction, the associated covariance matrix, the analysis binning, and the flux histograms.

\section*{Acknowledgements}

We thank the J-PARC staff for superb accelerator performance. We thank the CERN NA61/SHINE Collaboration for providing valuable particle production data. We acknowledge the support of MEXT,   JSPS KAKENHI  and bilateral programs, Japan; NSERC, the NRC, and CFI, Canada; the CEA and CNRS/IN2P3, France; the DFG, Germany; the INFN, Italy; the Ministry of Science and Education and the National Science Centre, Poland; the RSF, RFBR and the Ministry of Science and Higher Education, Russia; MICINN and ERDF funds and CERCA program, Spain; the SNSF and SERI, Switzerland; the STFC and UKRI, UK; and the DOE, USA. We also thank CERN for the UA1/NOMAD magnet, DESY for the HERA-B magnet mover system, the BC DRI Group, Prairie DRI Group, ACENET, SciNet, and CalculQuebec consortia in the Digital Research Alliance of Canada, and GridPP in the United Kingdom, and the CNRS/IN2P3 Computing Center in France. In addition, the participation of individual researchers and institutions has been further supported by funds from the ERC (FP7), “la Caixa” Foundation, the European Union’s Horizon 2020 Research and Innovation Programme under the Marie Sklodowska-Curie grant; the JSPS, Japan; the Royal Society, UK; French ANR; and the DOE Early Career programme, USA.
This work was supported in part through computational resources and services provided by the Institute for Cyber-Enabled Research at Michigan State University.
For the purposes of open access, the authors have applied a Creative Commons Attribution licence to any Author Accepted Manuscript version arising.

\bibliography{apssamp}

\clearpage
\appendix
\section{Analysis binning}
\label{app:binning}

\begin{table}[!htbp]
    \begin{center}
        \begin{tabular}{c|r|l}
            \# bins & $\cos(\theta_{\mu})$ bin & $p_{\mu}$ bin edges (GeV/\textit{c}) \\
            \hline
            1 & -1.00, 0.20 & 0, 30 \\
            5 &  0.20, 0.60 & 0, 0.3, 0.4, 0.5, 0.6, 30 \\
            6 &  0.60, 0.70 & 0, 0.3, 0.4, 0.5, 0.6, 0.8, 30 \\
            6 &  0.70, 0.80 & 0, 0.3, 0.4, 0.5, 0.6, 0.8, 30 \\
            7 &  0.80, 0.85 & 0, 0.3, 0.4, 0.5, 0.6, 0.8, 1.0, 30 \\
            8 &  0.85, 0.90 & 0, 0.3, 0.4, 0.5, 0.6, 0.8, 1.0, 1.5, 30 \\
            7 &  0.90, 0.94 & 0, 0.4, 0.5, 0.6, 0.8, 0.8, 1.25, 2.0, 30 \\
            10&  0.94, 0.98 & 0, 0.4, 0.5, 0.6, 0.8, 0.8, 1.25, 1.5, 2.0, \\
              &             & 3.0, 30 \\
            8 &  0.98, 1.00 & 0, 0.5, 0.7, 0.9, 1.25, 2.0, 3.0, 5.0, 30 \\
        \end{tabular}
        \caption{ND280 analysis binning for extracted cross section.}
        \label{tab:nd280_binning}
    \end{center}
\end{table}

\begin{table}[!htbp]
    \begin{center}
        \begin{tabular}{c|r|l}
            \# bins & $\cos(\theta_{\mu})$ bin & $p_{\mu}$ bin edges (GeV/\textit{c}) \\
            \hline
            4 &  0.50, 0.82 & 0.35, 0.50, 0.70, 1.0, 30 \\
            4 &  0.82, 0.94 & 0.35, 0.50, 0.70, 1.0, 30 \\
            4 &  0.94, 1.00 & 0.35, 0.50, 0.70, 1.0, 30 \\
        \end{tabular}
        \caption{INGRID analysis binning for extracted cross section.}
        \label{tab:ingrid_binning}
    \end{center}
\end{table}

\section{Flux energy binning}
\label{app:flux_binning}

\begin{table}[!htbp]
    \begin{center}
        \begin{tabular}{ c|c|c }
            \hline
            ND280 & INGRID & Energy bin (GeV)\\
            \hline
            $f_0$    & $f_{20}$ & 0.0 - 0.1\\
            $f_1$    & $f_{21}$ & 0.1 - 0.2\\
            $f_2$    & $f_{22}$ & 0.2 - 0.3\\
            $f_3$    & $f_{23}$ & 0.3 - 0.4\\
            $f_4$    & $f_{24}$ & 0.4 - 0.5\\
            $f_5$    & $f_{25}$ & 0.5 - 0.6\\
            $f_6$    & $f_{26}$ & 0.6 - 0.7\\
            $f_7$    & $f_{27}$ & 0.7 - 0.8\\
            $f_8$    & $f_{28}$ & 0.8 - 1.0\\
            $f_9$    & $f_{29}$ & 1.0 - 1.2\\
            $f_{10}$ & $f_{30}$ & 1.2 - 1.5\\
            $f_{11}$ & $f_{31}$ & 1.5 - 2.0\\
            $f_{12}$ & $f_{32}$ & 2.0 - 2.5\\
            $f_{13}$ & $f_{33}$ & 2.5 - 3.0\\
            $f_{14}$ & $f_{34}$ & 3.0 - 3.5\\
            $f_{15}$ & $f_{35}$ & 3.5 - 4.0\\
            $f_{16}$ & $f_{36}$ & 4.0 - 5.0\\
            $f_{17}$ & $f_{37}$ & 5.0 - 7.0\\
            $f_{18}$ & $f_{38}$ & 7.0 - 10.0\\
            $f_{19}$ & $f_{39}$ & 10.0 - 30.0\\
            \hline
        \end{tabular}
        \caption{The neutrino energy binning used for the flux systematic parameters. Both the ND280 and INGRID flux parameters use the same energy binning, but are treated as separate parameters in the fit.}
       \label{tab:flux_binning}
    \end{center}
\end{table}

\section{Neutrino interaction parameters}
\label{app:xsec_parameters}
\begin{table}[!htbp]
    \begin{center}
        \begin{tabular}{ l|l|r|r }
            \hline
            Parameter & Type & Prior & Error \\
            \hline
            $M_A^{QE}$ & Signal shape  & 1.21 & 0.3 \\
            2p2h $\nu$ norm. & Signal normalization & 1.0  & 1.0 \\
            2p2h $\nu$ shape & Signal shape & 1.0  & 1.0 \\
            $M_A^{Res}$  & Bkg shape & 0.95 & 0.15 \\
            $C_A^5$    & Bkg shape   & 1.01 & 0.12 \\
            Bkg Resonant ($I_{1/2}$) & Bkg normalization  & 1.3 & 0.2 \\
            DIS Multiple pion & Bkg shape & 1.0 & 0.4 \\
            CC-1$\pi$ $E_\nu < $ 2.5 GeV & Bkg normalization & 1.0 & 0.5 \\
            CC-1$\pi$ $E_\nu > $ 2.5 GeV & Bkg normalization & 1.0 & 0.5 \\
            CC DIS           & Bkg normalization & 1.0 & 0.5 \\
            CC Multi-$\pi$   & Bkg normalization & 1.0 & 0.5 \\
            CC Coherent on C & Bkg normalization & 1.0 & 1.0 \\
            NC Coherent      & Bkg normalization & 1.0 & 0.3 \\
            NC Other         & Bkg normalization & 1.0 & 0.3 \\
            CC $\nu_{e}$     & Bkg normalization & 1.0 & 0.03 \\
            \hline
            FSI Inelastic, LE & Bkg shape & 1.0 & 0.41 \\
            FSI $\pi$ absorbtion & Bkg shape & 1.1 & 0.41 \\
            FSI Charge exchange, LE & Bkg shape & 1.0 & 0.57 \\
            FSI Inelastic, HE & Bkg shape & 1.8 & 0.34 \\
            FSI $\pi$ production & Bkg shape & 1.0 & 0.50 \\
            FSI Charge exchange, HE & Bkg shape & 1.8 & 0.28 \\
            \hline
        \end{tabular}
        \caption{Cross-section modeling parameters used in this analysis along with their type, prior, and prior fractional uncertainty.}
        \label{tab:xsec_par}
    \end{center}
\end{table}

\section{Additional data plots}
\label{app:log_plots}
\begin{figure*}[hbt]
    \centering
    \includegraphics[width=0.35\textwidth,angle=0]{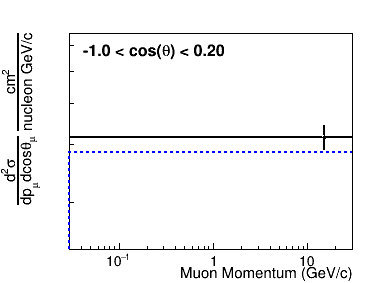}
    \includegraphics[width=0.35\textwidth,angle=0]{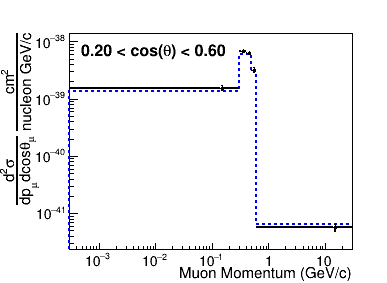}
    \includegraphics[width=0.35\textwidth,angle=0]{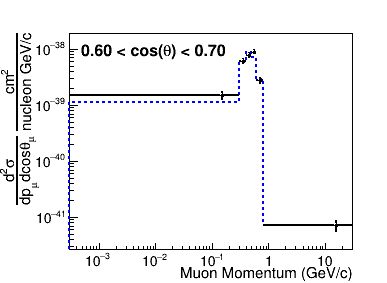}
    \includegraphics[width=0.35\textwidth,angle=0]{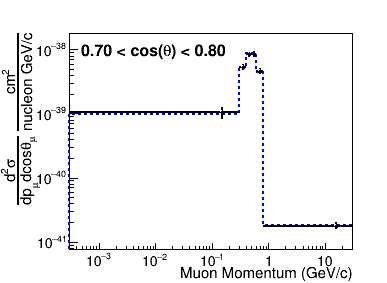}
    \includegraphics[width=0.35\textwidth,angle=0]{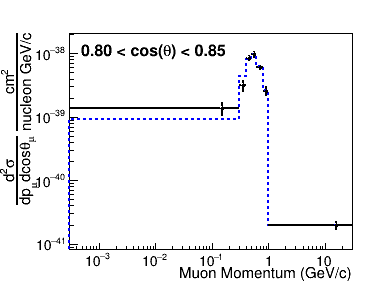}
    \includegraphics[width=0.35\textwidth,angle=0]{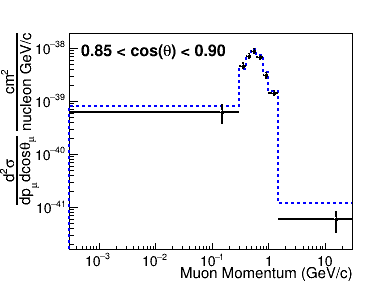}
    \includegraphics[width=0.35\textwidth,angle=0]{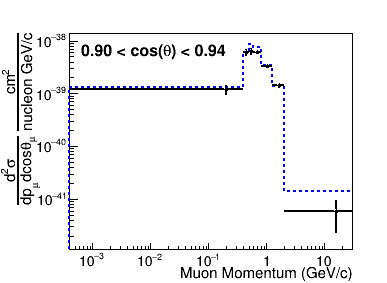}
    \includegraphics[width=0.35\textwidth,angle=0]{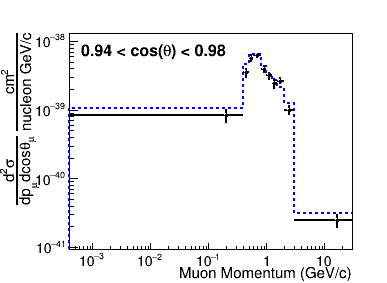}
    \includegraphics[width=0.35\textwidth,angle=0]{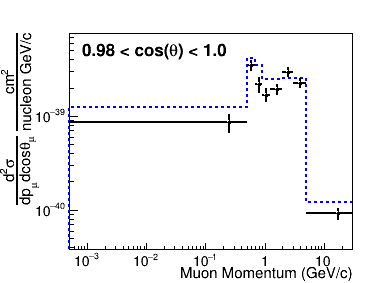}
    \includegraphics[width=0.35\textwidth,angle=0]{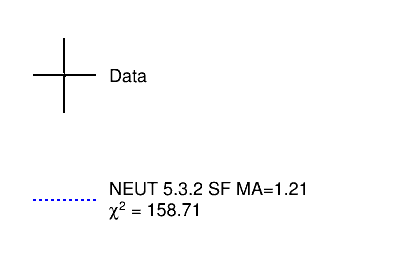}
    \caption{Extracted ND280 cross section as a function of muon momentum in angle bins compared to the nominal NEUT MC prediction including the final momentum bin.}
    \label{fig:data_result_nd280_log}
\end{figure*}

\begin{figure*}[hbt]
    \centering
    \includegraphics[width=0.35\textwidth,angle=0]{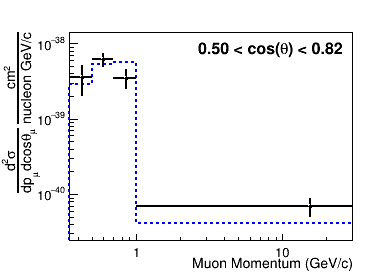}
    \includegraphics[width=0.35\textwidth,angle=0]{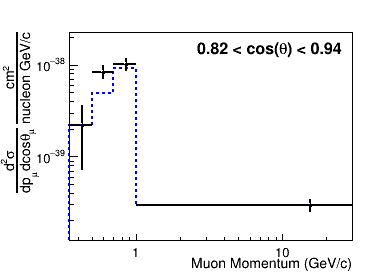}
    \includegraphics[width=0.35\textwidth,angle=0]{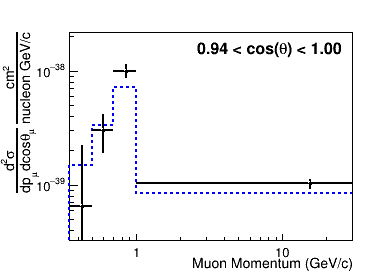}
    \includegraphics[width=0.35\textwidth,angle=0]{Figures/plot_data_log_legend.png}
    \caption{Extracted INGRID cross section as a function of muon momentum in angle bins compared to the nominal NEUT MC prediction including the final momentum bin.}
    \label{fig:data_result_ingrid_log}
\end{figure*}
\end{document}